\documentclass[11pt,3p,review,authoryear]{elsarticle}

\usepackage[dvipsnames]{xcolor}
\usepackage{soul,framed} 
\usepackage{mathtools}
\usepackage{lscape}
\colorlet{shadecolor}{yellow}
\usepackage{natbib}
\usepackage[ruled,vlined]{algorithm2e}
\usepackage{amsmath}
\usepackage{amsfonts}
\usepackage{amssymb}
\usepackage{subcaption}
\usepackage{hyperref}
\usepackage{lscape}
\usepackage{multirow}
\usepackage{pifont}

\usepackage{array}

\usepackage[utf8]{inputenc}
\usepackage{placeins}

\usepackage{blindtext}
\usepackage{dirtytalk}
\usepackage{adjustbox}
\usepackage{pdfpages} 
\usepackage{pgffor} 
\usepackage{xr}
\makeatletter
\AtBeginDocument{\let\LS@rot\@undefined}
\cleardoublepage\makeatother

\usepackage[shortlabels]{enumitem}
\usepackage{mdwtab}
\usepackage{eqparbox}
\usepackage{url}
\newtheorem{definition}{Definition}
\newtheorem{theorem}{Theorem}
\newtheorem{proof}{Proof}
\newtheorem{lemma}{Lemma}
\newtheorem{corollary}{Corollary}

\newtheorem{remark}{Remark}
\newtheorem{assumption}{Assumption}
\usepackage{mathtools}
\usepackage{dsfont}
\usepackage{bbold}
\usepackage[english]{babel} 
\usepackage{blindtext}
\usepackage{epstopdf}
\DeclareGraphicsExtensions{.png,.pdf,.jpg}
\usepackage{float}
\usepackage{hyperref}
\hypersetup{
    colorlinks=true,
    linkcolor=blue,
    filecolor=magenta,      
    urlcolor=cyan,
    pdftitle={Overleaf Example},
    pdfpagemode=FullScreen,
    }
\usepackage[utf8]{inputenc}
\usepackage{blindtext}

\begin{document}

\begin{frontmatter}

\title{Neural ARFIMA model for forecasting BRIC exchange rates with long memory}


\author{
Donia Besher \textsuperscript{1},
Madhurima Panja \textsuperscript{1},
Shovon Sengupta \textsuperscript{1},
Tanujit Chakraborty\textsuperscript{1,2}
\footnote[3]{\textit{Corresponding author}: \textit{Mail}: tanujit.chakraborty@sorbonne.ae}\\
{\scriptsize \textsuperscript{1} SAFIR, Sorbonne University Abu Dhabi, UAE.},\\
{\scriptsize \textsuperscript{2} Sorbonne Center for Artificial Intelligence, Sorbonne University, Paris, France.}}

\begin{abstract}
{Exchange rate forecasting remains a challenging problem, particularly for emerging economies, where the observed time series exhibit pronounced long-memory dependence, nonlinear dynamics, and sensitivity to macro-financial drivers. Classical models such as ARFIMA capture long-range persistence but fail to adequately represent nonlinear relationships, while modern machine learning approaches often neglect the underlying long-memory structure in macroeconomic series. To address this gap, we propose a Neural AutoRegressive Fractionally Integrated Moving Average (NARFIMA) model that integrates ARFIMA-based long-memory modeling with neural networks for nonlinear function approximation, while incorporating exogenous macroeconomic and uncertainty indicators. The framework provides a unified approach for capturing persistence, nonlinear dynamics, and external shocks. We establish asymptotic stationarity of the NARFIMA process and develop conformal prediction intervals for distribution-free uncertainty quantification. Empirical results for BRIC exchange rates show that NARFIMA consistently outperforms a broad range of forecasting benchmarks across multiple horizons, underscoring the importance of explicitly modeling long-memory dependence in exchange rate dynamics. 
The \texttt{narfima} \textbf{R} package provides an implementation of our approach.
}

\end{abstract}

\end{frontmatter}

\section{Introduction}
{Exchange rates significantly influence macroeconomic outcomes, shaping trade balances, capital flows, inflationary pressures, and financial stability. Consequently, they remain a central focus for policymakers, central banks, and market participants in an increasingly interconnected global economy \citep{eichengreen1998exchange, hausmann1999financial, stoica2016exchange}. Their role in assessing countries' financial stability has long been established \citep{taylor2001role}, with accurate forecasts proving indispensable for guiding monetary policy, designing capital controls, and implementing macro-prudential measures \citep{wieland2013forecasting, lubik2007central}. For commodity-dependent economies, where exchange rate fluctuations directly impact inflation forecasts and broader economic stability, reliable projections are particularly vital \citep{rossi2013exchange}. Within this context, spot exchange rates, reflecting current market prices for immediate currency delivery, are especially important as they directly influence trade prices, international capital flows, external debt obligations, and monetary policy decisions \citep{pilbeam2015forecasting}. Unlike aggregate measures such as the real effective exchange rate, spot rates respond swiftly to short-term market conditions, policy shifts, and uncertainty shocks, with their continuous trading facilitating rapid assimilation of new information \citep{bartsch2019economic}. This makes spot markets an ideal setting for analyzing the transmission of oil price shocks and policy uncertainty to exchange rate dynamics across multiple horizons. Given the rising global prominence of the BRIC economies\footnote{As of January 1, 2024, the BRICS has expanded to include eleven members (Brazil, Russia, India, China, South Africa, Argentina, Egypt, Ethiopia, Iran, Saudi Arabia, and the United Arab Emirates). This expansion has increased the collective share of global GDP. Further details can be found at: \url{https://www.weforum.org/stories/2024/11/brics-summit-geopolitics-bloc-international/}.}, collectively accounting for roughly 37.3\% of world gross domestic product (GDP) based on purchasing power parity (PPP) as of 2024 and a rapidly growing share of global trade \citep{o2011growth, hopewell2017brics, SENGUPTA2025953, nasir2018implications}, understanding and accurately forecasting spot exchange rates in these economies has become increasingly critical for macroeconomic planning, financial stability, and policy coordination. Although emerging economies such as the BRIC nations have become increasingly influential in the global economy, they remain vulnerable to global shocks, often experiencing rapid and severe currency fluctuations than developed nations. This increased susceptibility stems from greater exposure to external shocks, fragile financial markets, and risk of sudden reversals in capital flows, a phenomenon termed ``flight-to-quality'' \citep{bernanke1994financial, calvo2005sudden}. 
}

Translating these macroeconomic vulnerabilities into predictive models presents a fundamental challenge, primarily because exchange rate series simultaneously exhibit two important statistical characteristics: long memory and nonlinearity. Long memory refers to the persistence of dependence across observations, where shocks continue to influence future values over extended periods \citep{granger1980introduction, beran2017statistics}. Because long memory processes retain distant historical information, this persistence remains highly relevant even across long forecasting horizons \citep{Hosking1981fractional}. This behavior is well-documented in exchange rates and broader macroeconomic aggregates, underscoring the necessity of explicitly accounting for long-range dependence during model construction \citep{baillie1996memory, doukhan2002theory,cont2005financial}. Simultaneously, exchange rates frequently exhibit nonlinear dynamics arising from changing economic conditions, interactions among macroeconomic variables, and evolving investor behavior \citep{fernandes1998nonlinearity, kilian2003exchange, lee2013nonlinear}. These nonlinear mechanisms imply that the relationship governing exchange rate trajectories cannot be fully characterized through fixed linear structures, further complicating the forecasting task. This forecasting complexity is further compounded by external drivers, as exchange rate dynamics are influenced by a broad range of macroeconomic and uncertainty-related factors. Prior research has studied variables such as economic policy uncertainty \citep{balcilar2016does}, equity market volatility \citep{mun2008emv}, monetary policy uncertainty \citep{mueller2017exchange}, oil price growth rates \citep{beckmann2020relationship}, and interest rate differentials \citep{saracc2016impact}. These drivers introduce additional nonlinear interactions that further amplify the forecasting complexity. 

Modeling the joint presence of long memory and nonlinearity presents a significant forecasting quest, since existing methodologies only address one side of the problem. Previous studies have widely adopted traditional time series forecasting approaches, such as linear autoregressive fractionally integrated moving average (ARFIMA) models with Gaussian disturbances, to capture long-memory behavior in exchange rates via fractional differencing \citep{Hosking1981fractional, karemera2006assessing}. However, these models have limited capabilities in addressing the nonlinearities and structural complexities inherent in exchange rate dynamics. Conversely, purely neural network-based methods excel at capturing complex nonlinear relationships \citep{hornik1989approximators}, but remain limited in their ability to capture long-range dependence. Even recurrent architectures such as Long Short Term Memory (LSTM) \citep{hochreiter1997lstm}, though specifically designed to mitigate short-term memory, fail to learn highly persistent long-range dependencies in practice \citep{bengio1993dependence, zhao2020memory}. This failure stems from a lack of the theoretical foundations required to model long memory dynamics and ensure desirable asymptotic properties \citep{hornik1989approximators}. Deep learning models also face a practical limitation in macroeconomic settings despite their flexibility. These data-intensive models require large amount of observations, whereas macroeconomic variables are typically observed at monthly or quarterly frequencies, resulting in sample sizes that are insufficient for robust model training for deep neural networks \citep{li2020deeplearning, chapman2023macroeconomic}.  
Consequently, the inability of existing methodologies to simultaneously model long memory and nonlinearity represents an important methodological gap in exchange rate forecasting and reinforces the need for flexible yet theoretically grounded forecasting framework.

To address these methodological limitations, we propose the Neural AutoRegressive Fractionally Integrated Moving Average (NARFIMA) model, an ensemble approach that integrates the strengths of ARFIMA with autoregressive neural networks (ARNNs). The mechanism of the proposed NARFIMA model operates through a structured two-stage procedure. In the initial phase, 
a linear ARFIMA model is fitted with lagged observations of the exchange rate series, auxiliary uncertainty measures, and macroeconomic drivers to produce in-sample residuals. These ARFIMA model residuals capture the unexplained variations in the exchange rate dataset after accounting for long-memory effects and key economic drivers. In the second step, the residuals are combined with the exchange rate data, uncertainty measures, and macroeconomic drivers, and are then modeled using an ARNN architecture to capture nonlinear dependencies and underlying structural patterns. 
This ensemble framework can capture long-term memory, nonlinear dependencies, and complex interactions between exchange rates and multiple economic drivers. 
Furthermore, the asymptotic stationarity and geometric ergodicity conditions obtained through Markov chain analysis confirm the theoretical robustness of the NARFIMA model. These theoretical properties have several economic implications and policy relevance, as central banks and financial institutions rely on forecasting models that are asymptotically stable and reliable.
Empirical evaluations conducted on the BRIC economies' exchange rates demonstrate that the NARFIMA model substantially enhances predictive accuracy and robustness, surpassing state-of-the-art statistical and machine learning forecasting frameworks. 
Its superior performance underscores the practical relevance and reliability of our proposed approach. 
Experimental evaluation suggests that NARFIMA is an effective forecasting technique, particularly suited for analyzing the complex exchange rate dynamics under varying policy uncertainties and macroeconomic shocks. 
To quantify uncertainty, we integrate conformal prediction with NARFIMA to obtain prediction intervals, which are vital for effective policy design.

{The remainder of this paper is structured as follows.  Section~\ref{sec:literature_review} reviews the literature on exchange rate forecasting methods and motivates the selection of the macroeconomic factors considered as exogenous variables in this study. Section~\ref{Section_Data_and_Methodology} offers a detailed description of the data characteristics.  Section~\ref{Section_Proposed_Model} introduces the proposed NARFIMA methodology, detailing its architecture and statistical properties. 
In Section \ref{Section_Experimental_Evaluation}, we present the empirical causality analysis, performance evaluation, and the results of the statistical significance tests. Uncertainty quantification through conformal prediction and sensitivity analysis is presented in Section \ref{uq}. 
Policy implications and the conclusion are discussed in Sections~\ref{Sec_Policy_Implications} and~\ref{Section_Conclusion}, respectively.}

\section{Literature Review}~\label{sec:literature_review}
\subsection{Methods of Exchange Rate Forecasting}
The complexity of exchange rate dynamics in emerging markets motivated the development of models beyond standard benchmarks. A past study by \citealp{meese1983empirical} demonstrated that simple random walk models often outperform more advanced econometric approaches in out-of-sample forecasts. Numerous efforts have been made to improve the accuracy of exchange rate predictions; the existing literature encompasses a wide range of strategies, including Taylor rule-based fundamentals \citep{molodtsova2009out}, nonlinear methods \citep{kilian2003so}, and Kalman filter-based models \citep{date2025modelling}. More recently, with advances in data-driven techniques and the increasing availability of macroeconomic datasets, the adoption of statistical and machine learning techniques for exchange rate forecasting has surged \citep{plakandaras2015forecasting}. For instance, \citealp{ngan2013forecasting} employed an ARIMA model to capture the exchange rate dynamics in Vietnam, while \citealp{karemera2006assessing} revealed that an ARFIMA framework can outperform the random walk method in forecasting exchange rates for several developed economies. Similarly, \citealp{pilbeam2015forecasting} investigated the effectiveness of the univariate generalized autoregressive conditional heteroskedasticity (GARCH) model in forecasting foreign exchange market volatility. \citealp{galeshchuk2016neural} explored the use of neural networks in forecasting exchange rates across multiple currencies. More recently, deep learning approaches have been applied to forecast exchange rates in emerging economies \citep{abir2024use}; however, these algorithms require high-frequency datasets with extensive training data points. Moreover, they frequently overfit and struggle to capture long-memory characteristics. The majority of these studies have predominantly focused on the exchange rates of developed countries and Organization for Economic Co-operation and Development (OECD) nations, largely overlooking the distinct dynamics of emerging economies and their interactions with macroeconomic drivers. Furthermore, the structural complexities present in BRIC exchange rate series, including long memory, nonlinearity, and non-stationary behavior, are not adequately addressed by any single existing approach, which motivates the development of the NARFIMA model.

\subsection{Macroeconomic Drivers of BRIC Exchange Rates}
External macroeconomic variables carry meaningful predictive power for exchange rates in emerging economies. Among these, uncertainty measures such as economic policy uncertainty (EPU) and geopolitical risks (GPR) have received extensive emphasis in the literature \citep{kumar2024bayesian, salisu2022exchange}. 
From an economic perspective, when domestic uncertainty exceeds foreign uncertainty, domestic investors tend to invest in foreign currency assets, triggering exchange rate movements \citep{balcilar2016does}. Moreover, economic uncertainties affect expectations about costs and returns, which in turn influence both supply and demand in currency markets \citep{benigno2012risk}. Motivated by these theoretical foundations, recent empirical studies have focused on modeling the interplay between EPU and exchange rate dynamics. For instance, \citealp{zhou2020can} demonstrated the enhanced forecasting performance of GARCH-Mixed Data Sampling models incorporating Sino-US EPU in predicting Chinese exchange rate volatility, while \citealp{benigno2012risk} explored how monetary, inflation, and productivity uncertainties influence real exchange rates. Further studies have consistently confirmed the robust predictive power of EPU in emerging markets across short and long horizons \citep{colombo2013economic, sin2015economic, juhro2018can, abid2020economic}. Alongside EPU, other financial uncertainty indices such as the US Equity Market Volatility (EMV) and US Monetary Policy Uncertainty (MPU) measure distinct aspects of the economic environment that potentially impact currency markets. Empirical evidence from \citealp{mueller2017exchange} indicates that US MPU significantly affects currency risk premia, with emerging market currencies exhibiting greater sensitivity due to “reach for yield” behavior. Additionally, \citealp{istrefi2018subjective} documents an inverse relationship between US MPU and economic activity, where increased US MPU typically prompts dollar appreciation driven by safe-haven demand during financial stress. 
In addition to uncertainty measures, other external economic factors, such as the relationship between oil prices and exchange rates, form a critical nexus in international finance, particularly for economies with significant oil exposure. 
For instance, \citealp{beckmann2020relationship} demonstrated that the oil price and exchange rate relationship exhibits time-varying characteristics depending on the nature of the underlying shock driving oil prices. 
The distinction between oil-exporting and oil-importing countries is fundamental, as highlighted by \citealp{chen2024dynamic}, who found that exchange rate-oil price connectedness intensifies during crisis periods, with stronger transmission channels evident in oil-exporting economies. This finding is especially relevant for BRIC countries, which span the spectrum from major oil exporters (Russia, Brazil) to significant importers (China, India). 
Alongside oil price dynamics, \citealp{andrieș2017relationship} examined the relationship 
between interest rates and exchange rates in Romania, revealing that the strength of this relationship varies across different time horizons, 
while \citealp{saracc2016impact} 
found the same conclusion for Turkey.
Their analysis further highlighted that interest rate differentials serve as a more reliable predictor of exchange rate movements than absolute interest rate levels. For BRIC economies, which exhibit varying degrees of capital account openness and monetary policy independence, the interest rate channel represents a critical transmission mechanism through which both domestic and external shocks propagate to exchange rates. Thus, when combined with oil price dynamics and uncertainty measures, they can capture the complex dynamics governing exchange rates in emerging economies. These theoretical and empirical considerations inform the selection of candidate exogenous variables examined in this study, namely global EPU (GEPU), US EMV, US MPU, oil price growth rates, geopolitical risk indices, and country‑specific short‑term interest rate differentials. 

\section{Data Characteristics}\label{Section_Data_and_Methodology}
In this study, we analyze monthly spot exchange rates and several macroeconomic covariates for the BRIC economies from January 1997 (1997-01) to October 2023 (2023-10). Monthly frequency is preferable to daily because exchange rates at higher frequencies are dominated by short-term noise rather than the persistent trends relevant for forecasting. Additionally, the multi-step predictions required for longer horizons would compound estimation errors, further motivating the use of monthly observations. Our analysis employs a rolling window approach with six forecast horizons: short-term (1- and 3-month-ahead), semi-long-term (6- and 12-month-ahead), and long-term (24- and 48-month-ahead) for each country. For short-term forecasting, with 1-month-ahead and 3-month-ahead horizons, the training periods span from 1997-01 to 2023-09 and 1997-01 to 2023-07, while the test horizons are 2023-10 and 2023-08 to 2023-10, respectively. In the case of semi-long-term forecasting, which includes 6-month-ahead and 12-month-ahead horizons, the training periods extend from 1997-01 to 2023-04 and 1997-01 to 2022-10, with corresponding test horizons from 2023-05 to 2023-10 and 2022-11 to 2023-10, respectively. Finally, for long-term forecasting with 24-month-ahead and 48-month-ahead horizons, the training periods range from 1997-01 to 2021-10 and 1997-01 to 2019-10, while the test horizons cover 2021-11 to 2023-10 and 2019-11 to 2023-10, respectively. For the statistical analyses presented in this section, we rely on the training sample for the 48-month-ahead horizon (1997-01 to 2019-10). This sample remains entirely separate from all out-of-sample test periods, thereby avoiding any potential look-ahead bias in the results reported. 

The target variable, spot exchange rate of BRIC countries, is obtained from the Federal Reserve Economic Data (FRED) repository\footnote{\url{https://fred.stlouisfed.org/}.}. Monthly exchange rate values are calculated as the averages of daily exchange rates based on noon buying rates in New York City for foreign currency cable transfers and are seasonally unadjusted. Table \ref{PLOTS:ACF} presents the time series plots of exchange rate dynamics, along with their autocorrelation functions (ACF) plots for the BRIC nations. We observe that all BRIC currencies, except the Chinese yuan, have depreciated against the USD over time. China maintained a currency peg until 2005, and even after shifting away from the peg, its central bank continued active intervention in the exchange rate market. In contrast, the other BRIC nations follow a floating exchange rate system. Additionally, a sharp spike in exchange rates is evident during the global recession, affecting all BRIC countries except China. On the other hand, the ACF plots for the BRIC nations indicate that the autocorrelation decays slowly and stays above the significance bounds for some lags, providing weaker evidence of long memory. Furthermore, to investigate possible structural breakpoints in the exchange rate series, we employ the ordinary least squares (OLS)-based CUSUM test, which helps account for regime shifts \citep{ploberger1992cusum}. The test examines the cumulative sum of recursive residuals to identify any deviations from the model’s stability, indicating potential structural breakpoints. Our implementation of the OLS-based CUSUM test shows that no statistically significant breakpoints were found in the exchange rate series for BRIC countries, as depicted in Table \ref{PLOTS:ACF}. The absence of significant breakpoints supports the conclusion that the exchange rate series has been relatively stable during the observed period, suggesting that regime shifts are unlikely to affect the forecasting models.

\begin{table*}[ht]
    \small 
    \centering
    \caption{Time plots of the training data for exchange rates of BRIC nations, alongside the ACF plots and OLS-based CUSUM test results.}
    \label{PLOTS:ACF}
    \begin{adjustbox}{max width=\textwidth}
    \begin{tabular}{p{6cm} p{4cm} p{8cm}}
        \hline
        \hspace{1.5cm}Training data (01-1997 to 10-2019) & \hspace{2.9cm}ACF Plot & \hspace{3.7cm}OLS-based CUSUM test\\ \hline
        
        \multicolumn{3}{c}{
    \includegraphics[width=1.5\textwidth,height=35mm]{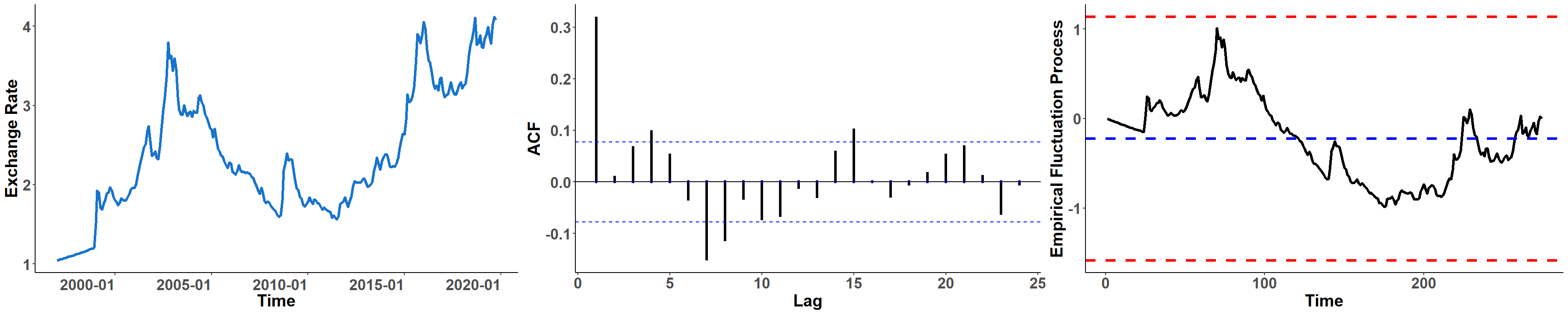} }\\ 
        \multicolumn{3}{c}{\centering \textbf{Brazil}}\\ \hline
        
        \multicolumn{3}{c}{
    \includegraphics[width=1.5\textwidth,height=35mm]{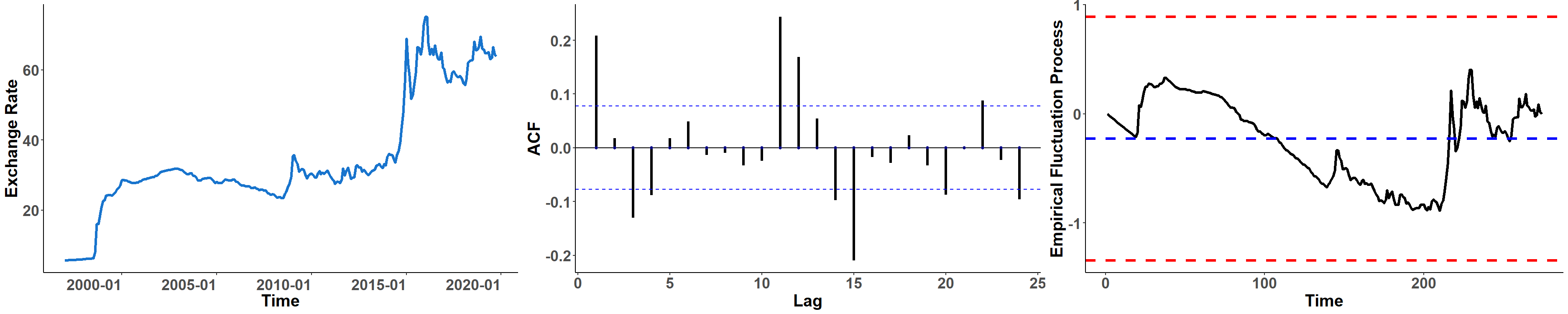}}\\
        \multicolumn{3}{c}{\centering \textbf{Russia}}\\ \hline
        
        \multicolumn{3}{c}{
    \includegraphics[width=1.5\textwidth,height=35mm]{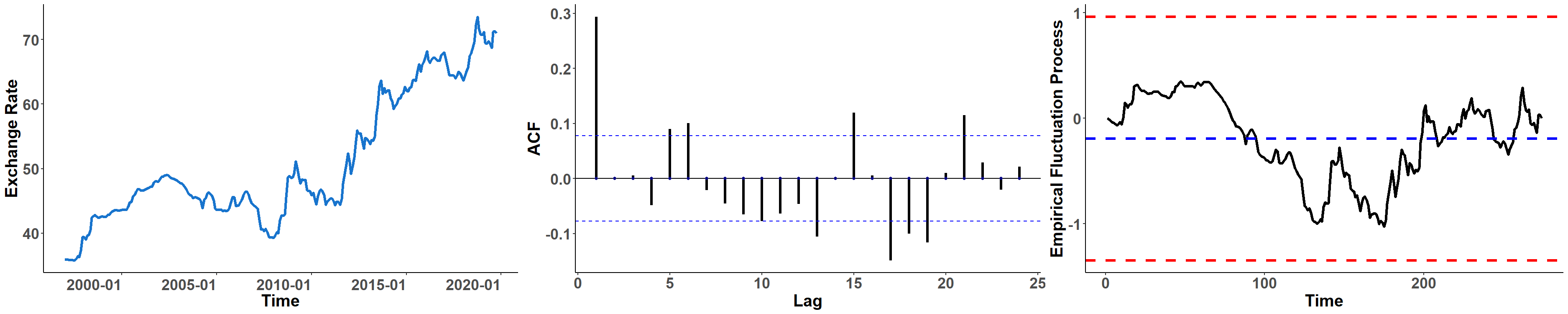}}\\
        \multicolumn{3}{c}{\centering \textbf{India}}\\ \hline
        \multicolumn{3}{c}{
    \includegraphics[width=1.5\textwidth,height=35mm]{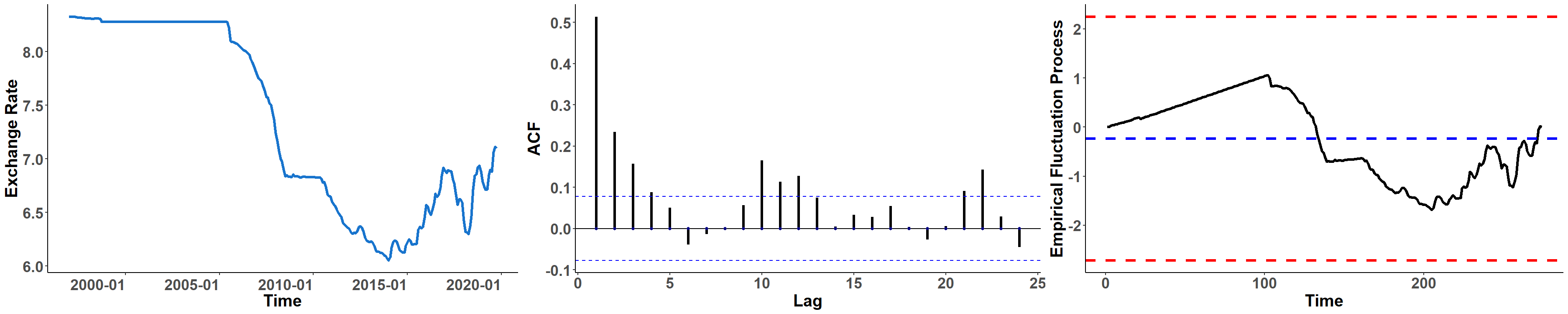}}\\
        \multicolumn{3}{c}{\centering \textbf{China}}\\ \hline 
    \end{tabular} 
    \end{adjustbox}
\end{table*}
 
Among the economic drivers, we consider four widely used news-based uncertainty measures\footnote{The historical dataset for all uncertainty indicators is obtained from \url{https://www.policyuncertainty.com/}.}, namely GEPU, US EMV, US MPU, and GPR. An overview of their definitions and construction is provided in the Appendix~\ref{uncertainty_desc}. Beyond uncertainty indices, we incorporate macroeconomic indicators such as oil prices, interest rates, and inflation. We consider the global price of West Texas Intermediate (WTI) crude oil (USD per barrel) and stabilize its variability by computing oil price growth rates. The resultant series helps capture the effect of oil price fluctuations on exchange rate movements over time. Additionally, we include short-term interest rates, which reflect the cost of borrowing for a duration under 24 hours between financial institutions or for government securities. These rates, measured in percentages for the BRIC economies and the US, serve as a key indicator of short-term financial system liquidity, responding to central bank interventions and market fluctuations. To assess relative monetary policy stance and capital flow dynamics, we compute the short-term interest rate differential (IRD) between the US and each BRIC economy. This country-specific short-term IRD significantly influences exchange rates, as higher BRIC interest rates tend to attract capital inflows, strengthening the domestic currency, while lower rates can lead to depreciation. We also consider consumer price index (CPI) inflation rates for both the US and the BRIC economies, which measure price changes for a fixed basket of goods and services. The CPI inflation differential, reflecting the gap between US and country-specific CPI inflation rates, influences exchange rate movements, as higher domestic inflation indicates currency depreciation and vice versa. All macroeconomic indicators used in this analysis are collected from the FRED repository. 

We compute the summary statistics and analyze several global characteristics of the exchange rate datasets and auxiliary variables. Notably, we focus on seven key time series features: skewness, kurtosis, nonlinearity, long-range dependence, seasonality, stationarity, and outlier detection \citep{hyndman2018forecasting}. We employ Tsay's and Keenan's one-degree tests to assess nonlinearity, while the Kwiatkowski–Phillips–Schmidt–Shin (KPSS) test is used to evaluate stationarity. Seasonal patterns are examined using Ollech and Webel's test, and long-range dependence is analyzed using the Hurst exponent. Moreover, to detect outliers in the dataset, we apply the Bonferroni outlier test using studentized residuals \citep{weisberg1982residuals}. These statistical features, summarized in Appendix~\ref{stat_prop}, reveal that most macroeconomic time series are non-stationary, except for the oil price growth rate series, the CPI inflation series for Brazil and India, and the GPR series for Brazil and Russia. Most of the series do not exhibit seasonality, except for the GEPU and US EMV series. Additionally, nonlinear patterns are present in the majority of the dataset. All time series, except for the exchange rate series of Russia, exhibit significant outliers. The Hurst exponent values are greater than 0.5, suggesting a persistent long-range dependence in exchange rate dynamics and all economic indicators. Furthermore, to provide a more rigorous statistical assessment of long memory, we estimate the fractional differencing parameter $d$ under an ARFIMA specification via Whittle likelihood \citep{whittle1951hypothesis, beran2017statistics}. This approach tests the null hypothesis of a short memory process ($d = 0$) against the alternative of long memory ($d > 0$). Rejection of the null provides robust evidence of long-range dependence, where the magnitude of the estimated parameter quantifies the series' structural persistence. The estimated values of $d$, as reported in Table~\ref{Table_Whittle_Estimator} of Appendix~\ref{stat_prop}, are statistically different from zero at the 1\% significance level across all series, with values ranging from 0.3 to 0.49. This provides strong evidence against short memory and indicates that the series exhibits highly persistent, long-range dependence.  

\section{Neural ARFIMA Model}\label{Section_Proposed_Model}
This section outlines the workflow of the proposed Neural ARFIMA (NARFIMA) model, designed for forecasting exchange rate dynamics in the BRIC economies. The NARFIMA framework integrates the strengths of the AutoRegressive Fractionally Integrated Moving Average with causal exogenous variables (ARFIMAx) to capture long-range dependencies in time series while leveraging a neural network to model complex nonlinear interactions within macroeconomic variables. The proposed model follows a sequential approach, where ARFIMAx is first employed to model exchange rate series based on historical observations and macroeconomic drivers. ARFIMAx captures short-term dependencies through its autoregressive and moving average components, while its fractional differencing mechanism accounts for long-term dependencies in exchange rate dynamics. The inclusion of macroeconomic indicators as exogenous variables further strengthens the model by incorporating external influences, providing a comprehensive representation of the exchange rate dynamics. 

\subsection{Model Formulation}\label{model_formulation}
Given $T$ historical observations of exchange rate series $\left\{y_t; t = 1, 2, \ldots, T\right\}$ and $r$ macroeconomic drivers $\left\{X_{j,1}, X_{j,2}, \ldots, X_{j,T}\right\}_{j = 1}^r$, the ARFIMAx model is formulated as:

\begin{equation}\label{eq_ARFIMAx}
    \left(1 - \sum_{i = 1}^{\tilde{p}} \tilde{\phi}_i \operatorname{B}^i\right) \left(1 - \operatorname{B}\right)^d y_t = \tilde{\mu} + \sum_{j = 1}^r \tilde{\pi}_j \operatorname{B}  X_{j, t} + \left(1 + \sum_{k = 1}^{\tilde{q}} \tilde{\theta}_k \operatorname{B}^k\right)\epsilon_t,
\end{equation}
where $\epsilon_t$ is white noise and $\operatorname{B}$ represents the backshift operator, such that $\operatorname{B}y_t = y_{t-1}$ and $\operatorname{B}X_{j,t} = X_{j, t-1}$. The parameters $\tilde{p}$ and $\tilde{q}$ denote the number of autoregressive and moving average terms, respectively, while $d \in \left(0, 0.5\right)$ represents the fractional differencing parameter responsible for capturing long-memory effects \citep{granger1980introduction}. The coefficients $\left\{\tilde{\phi}, \; \tilde{\theta}, \; \tilde{\pi}, \; \tilde{\mu}\right\}$ correspond to the autoregressive, moving average, exogenous components, and bias term, respectively. The fractional differencing operator, given by:
$$
\left(1 - \operatorname{B}\right)^d = \sum_{v = 0}^{\infty} \frac{\Gamma\left(v - d\right)\operatorname{B}^v}{\Gamma\left(-d\right)\Gamma\left(v+1\right)},
$$ 
where $\Gamma\left(\cdot\right)$ denotes the gamma function, ensures that $y_t$ is transformed into a stationary process. By allowing $d$ to take non-integer values, this operator effectively captures long-term memory effects in time series, making it particularly well-suited for exchange rate datasets with slow decaying autocorrelation. Thus, the predictions $\left\{\hat{y}_t^{ARFIMA}\right\}$ generated from the ARFIMAx model by estimating the coefficients in  Eqn.~\eqref{eq_ARFIMAx} capture the linear trajectory of the exchange rate series and the influence of auxiliary covariates. However, exchange rate series of emerging economies like BRIC often exhibit complex nonlinear structures and dynamic causal interactions, which the linear ARFIMAx model may fail to fully explain. Consequently, the residuals 
$$e_t = y_t - \hat{y}_t^{ARFIMA}$$ of the ARFIMAx framework, captures the unexplained variations in exchange rate dynamics after accounting for linear long-memory effects. These residuals contain learnable structures with nonlinear dependencies and high-frequency fluctuations that cannot be effectively modeled using traditional parametric approaches. To address these limitations, the NARFIMA$(p,q,k)$ framework integrates a feed-forward neural network to model complex nonlinear patterns. 

The neural network component in the NARFIMA model is structured as a single hidden-layer architecture, enabling it to learn intricate nonlinear relationships between lagged values of exchange rate series, ARFIMAx residuals, and economic drivers. This single-layered framework offers a balance between computational efficiency and predictive power, restricting overfitting, and making it suitable for forecasting exchange rates in the presence of both linear and nonlinear dynamics. Due to limited data availability for macroeconomic modeling, highly computational deep learning models often fail to capture the data dynamics. On the contrary, NARFIMA utilizes an artificial neural network structure with only one hidden layer having $k$ neurons; therefore, it does not overfit. The network is designed to receive inputs consisting of $p$ historical exchange rate observations, $q$ lagged values of ARFIMAx residuals, and one lagged value for each of the $r$ macroeconomic covariates. The output of the network is a one-step-ahead forecast of the exchange rate series, which is expressed as:
$$
\hat{y}_{t+1} = f\left(y_{t}, y_{t - 1}, \ldots, y_{t - p +1}, e_{t}, e_{t - 1}, \ldots, e_{t - q +1}, X_{1, t}, X_{2, t}, \ldots, X_{r, t}\right),
$$
where $t = \max(p,q), \cdots, T$ and $f$ represents the neural network function. By learning from historical data and residuals, the neural network effectively captures both linear and nonlinear relationships between input features and exchange rate dynamics, along with long memory dependence. In the proposed settings, two variants of the network are designed based on the inclusion or exclusion of skip connections. These skip connections allow input features to directly influence the output in addition to passing through the hidden layer, influencing the learning process and impacting model performance. The presence of a skip connection ensures that both linear and nonlinear components of the data are effectively integrated. The skip connection between the input and the output layer preserves the linear dynamics of the exchange rate series while allowing the hidden layer to learn the nonlinear interactions. Additionally, the presence of skip connections enhances training stability by mitigating the vanishing gradient problem and serves as a regularizing mechanism, which reduces overfitting by allowing direct information propagation and preventing unnecessary transformations. Thus, using the single-hidden layer neural network with a skip connection, the one-step-ahead forecasts of the exchange rate series can be generated as:
\begin{align*}\label{Eq_NARFIMA}
\hat{y}_{t+1} &= \mu_0 
+ \sum_{l=1}^k \mu_l \; \sigma\Bigg( 
    \tilde{\alpha}_{l} 
    + \sum_{i=1}^p \beta_{i,l} \; y_{t-i +1} 
    + \sum_{j = 1}^q \gamma_{j,l} \; e_{t-j +1} 
    + \sum_{m = 1}^r \delta_{m,l} \; X_{m,t}  
\Bigg) \notag\\
&\quad + \sum_{u=1}^p \phi_{u} \; y_{t-u+1} 
+ \sum_{v=1}^q \eta_{v} \; e_{t-v+1} 
+ \sum_{w=1}^r \zeta_{w} \; X_{w,t},
\end{align*}
where $k$ is the number of hidden nodes, $\left\{\tilde{\alpha}_{l} ,\; \beta_{i,l}, \; \gamma_{j,l}, \; \delta_{m,l}\right\}$ are the connection weights between input and hidden layers, $\mu_l$ is the weight vector between hidden and output layers, $\left\{\phi_{u}, \eta_{v}, \zeta_{w}\right\}$ represent skip connection weights, $\mu_0$ is the bias term, and $\sigma\left(\cdot\right)$ is the nonlinear activation function. The network weights are initialized randomly and trained using the gradient descent backpropagation approach \citep{rumelhart1986learning}. In the variant of the neural network without skip connections, the mechanism follows a similar structure, but the skip connection weights $\left\{\phi_{u}, \eta_{v}, \zeta_{w}\right\}$ are set to zero, removing the skip connection between input and output layers. The above procedure generates a one-step-ahead forecast of the exchange rate series. For the multi-step ahead forecasts, we employ a recursive framework where the input layer is updated with the latest predictions at each step. Alongside point forecasts, the NARFIMA framework can be readily integrated with conformal prediction techniques to produce reliable prediction intervals (see Section \ref{uq}). 

\subsection{Choice of Parameters}

The NARFIMA model consists of primarily four tunable parameters, namely the number of lagged exchange rate observations $(p)$, the number of historical values for ARFIMAx residuals $(q)$, the number of nodes in the hidden layer $(k)$, and the network structure indicating whether a skip connection is included $(skip)$. To determine the optimal values of the parameters, we utilize a time series cross-validation strategy and select $\left(p, q, k\right)$ by minimizing the root mean square error (RMSE) on the validation set ($\mathcal{V}$) as follows:
$$
 \left(p, q, k\right) = \underset{\left(p, q, k\right)}{\operatorname{arg min}} \sqrt{\frac{1}{|\mathcal{V}|} \sum_{{t'} \in \mathcal{V}} \left(y_{t'} - \hat{y}_{t'} \right)^2},
$$
where $y_{t'}$ and $\hat{y}_{t'}$ are the ground truth and forecasts generated by NARFIMA at time $t' \in \mathcal{V}$, respectively. RMSE is selected as the optimization criterion because its quadratic form heavily penalizes larger errors. Given the severe economic consequences of large exchange rate deviations, this focus on minimizing extreme errors is appropriate. For all the datasets, we build two different versions of the NARFIMA model, one with skip connections (general case) and one without skip connections. The optimal values of other parameters are identified through temporal cross-validation. To address potential overfitting concerns arising from the model's complexity relative to available data, the NARFIMA framework incorporates several safeguards that ensure robust generalization. The neural network architecture is deliberately constrained to a single hidden layer with a limited number of neurons ($k \leq 5$), preventing excessive parameterization while maintaining sufficient nonlinear modeling capacity. Subsequently, the NARFIMA model is trained with the optimal parameters on the entire training dataset to accurately forecast the exchange rate dynamics. The integration of ARFIMAx, which provides a robust foundation for modeling linear long-term dependencies, with the neural network, which excels in learning complex, nonlinear interactions, allows the NARFIMA architecture to capture both nonlinear and long-range dependencies. This ensures a comprehensive representation of the exchange rate series dynamics by combining long-memory dependencies, macroeconomic influences, and nonlinear fluctuations. 
In the next subsection, we study the geometric ergodicity and asymptotic stationarity of the proposed NARFIMA model from a nonlinear time series perspective.

\subsection{Geometric Ergodicity and Asymptotic Stationarity}\label{AS_GE}

Stationarity and ergodicity are fundamental for statistical inference in nonlinear time-series analysis. When a process is both stationary and ergodic, a single long realization is sufficient for time averages to recover the data-generating law. Asymptotic stationarity guarantees that distributional features stabilize as time grows, even in the presence of long memory or transient dynamics. Geometric ergodicity of the Markov chain induced by the model’s state ensures exponentially fast convergence to the invariant distribution. Together, these properties justify using NARFIMA for econometric and financial forecasting under long-memory and nonlinear regimes. Building on \citealp{trapletti1999ergodicity, trapletti2000stationary, chakraborty2020unemployment}, which analyze autoregressive neural networks (ARNN) and ARIMA-ARNN models, we develop a unified framework for NARFIMA with skip connections, a strictly more general specification. We present the theory along with its economic implications and applications.

We demonstrate the asymptotic properties for the NARFIMA$(1,1,k)$ process with skip connections. However, for simplicity, we omit exogenous variables for establishing the theoretical results. The simple NARFIMA process is given by:
\begin{equation}\label{narfima_eq_simple}
 y_t=f(y_{t-1}, e_{t-1}, \Theta)+ \varepsilon_t,   
\end{equation}
where $\Theta$ denotes the weight vector, $\varepsilon_t$ is a sequence of independently and identically distributed (i.i.d.) random noise, and $f(y_{t-1}, e_{t-1}, \Theta)$ represents an autoregressive neural network with $k$ hidden units, inputs $y_{t-1}$, and the ARFIMA feedback $e_{t-1}$, as defined in Section~\ref{model_formulation}. 
The output of the simple NARFIMA$(1,1,k)$ process with an activation function $G$ is as follows: 
\begin{align*} 
    f(y_{t-1}, e_{t-1}, \Theta) &= \psi_1 y_{t-1}+\psi_2 e_{t-1}+ \beta_0 \nonumber + \sum_{i=1}^{k}\beta_i \: G\left(\mu_i + \phi_{i,1}y_{t-1}+\phi_{i,2} e_{t-1}\right) \nonumber \\ 
    &\equiv\psi_1 y_{t-1}+\psi_2 e_{t-1}+g\left(y_{t-1},e_{t-1},\beta,\phi\right),
\end{align*}
where $\psi = {(\psi_1,\psi_2)}^\top$ is a weight vector representing the skip connections, the input to hidden layer weight vector is denoted by $\phi = {\left(\phi_{1,1},\ldots,\phi_{k,1},\phi_{1,2},\ldots,\phi_{k,2},\mu_1,\ldots,\mu_k\right)}^\top$, and $\beta = {(\beta_0, \beta_1, \ldots, \beta_k)}^\top$ is the hidden to output layer weight vector. All three components, $\psi$, $\phi$, and $\beta$, are incorporated into the overall weight vector $\Theta \in \mathcal{R}^D$  of the neural network, where $D$ denotes the number of trainable weights and biases in the NARFIMA model. The function $g$ represents the nonlinear transformation performed by the hidden layer, combining $\phi$, $\beta$, $y_{t-1}$, and $e_{t-1}$ through the activation function $G$. The one-step model is given by:
$$y_t={\psi_1 y_{t-1}+\psi_2 e_{t-1}}+{\beta_0+\sum_{i=1}^k \beta_i G\left(\mu_i+\phi_{i,1} y_{t-1}+\phi_{i,2} e_{t-1}\right)}+\varepsilon_t.$$
Now, we write the NARFIMA process in the state space form as follows: 
\begin{equation}\label{eqMain}
    x_t = \Psi x_{t-1} + F(x_{t-1}) + S e_t + \sum \varepsilon_t,
\end{equation}
where $x_t = \begin{bmatrix}
y_t\\
e_t
\end{bmatrix}, \; S =\begin{bmatrix}
0\\
1
\end{bmatrix}, \; \sum =\begin{bmatrix}
1\\
0
\end{bmatrix}, \;
\Psi = \begin{bmatrix}
\psi_1 & \psi_2\\
0 & 0 
\end{bmatrix}$, and $F(x_{t-1}) = \begin{bmatrix}
g(y_{t-1},e_{t-1}, \beta,\phi)\\
0 
\end{bmatrix}$ is the nonlinear part. We say $\{x_t\}$ is a Markov chain with state space $\mathcal{X} \subseteq \mathcal{R}^2$ equipped with Borel $\sigma$-field $\mathcal{B}$ and Lebesgue measure $\lambda$. \\

In order to show the asymptotic stationarity and ergodicity of NARFIMA, we state several key assumptions. Assumption~\eqref{A1} requires the residual process $\{e_t\}$ to be weakly stationary. This ensures that the ARFIMA feedback entering the neural network evolves with a stable distribution over time, preventing the input from exhibiting drifting or explosive behavior.
Assumption~\eqref{A2} relies on the characteristics of the activation function to be used in the neural network architecture of the NARFIMA process. Assumption~\eqref{A2} is essential for the stability of the dynamical system since it ensures that the behavior of the nonlinear part in Eqn.~\eqref{eqMain} is predictable and does not exhibit erratic changes. This assumption is satisfied by popularly used activation functions, such as sigmoid and tanh, and it is considered in the choice of activation function during the implementation of the NARFIMA model. Additionally, the neural network may have skip connections where output also depends linearly on inputs like previous lag and past residual. Assumptions~\eqref {A3} and~\eqref{A5} ensure contraction in the linear part and bound the influence of past states. The innovation $\varepsilon_t$ satisfies Assumption~\eqref {A4}, which are usual restrictions on the noise process. We verify these assumptions empirically in Section \ref{verification} while implementing the NARFIMA model on BRIC exchange rate datasets. 
\begin{assumption}\label{A1}
The residual process $\{e_t\}$ is weakly stationary.
\end{assumption}
\begin{assumption}\label{A2}The activation function $G$ is bounded, nonconstant, and asymptotically constant function. This makes the neural part $F$ in Eqn.~\eqref{eqMain} bounded and Lipschitz.
\end{assumption}
\begin{assumption}\label{A3}The linear part in Eqn.~\eqref{eqMain} is controllable, that is, every point of the state space can be reached irrespective of the starting point. This condition is satisfied if $\psi_1 + \psi_2 \neq 0$. 
\end{assumption}
\begin{assumption}\label{A4}The distribution of the noise process $\{\varepsilon_t\}$ is absolutely continuous with respect to the Lebesgue measure $\lambda$ with a continuous $f_{\varepsilon}(\cdot)$ that is strictly positive on $\mathcal{R}$. The random noise process $\varepsilon_t$ is bounded with finite variance, i.e., $\mathbb{E}[\varepsilon_t^2] < \infty$.
\end{assumption}
\begin{assumption}\label{A5}The skip connection weight for the autoregressive part satisfies $|\psi_1| < 1$. For $p, q>1$, the spectral radius of the skip companion matrix $\mathcal{M}(\Psi)<1$ (the roots of the skip autoregressive polynomial outside the unit circle). 
\end{assumption}
\noindent With the assumptions established, we now proceed to prove the ergodicity and asymptotic stationarity of the NARFIMA$(1,1,k)$ process. The state space representation in Eqn.~\eqref{eqMain} makes it explicit that $y_t$ depends on both its own past and the residual process $\{e_t\}$. 
Under Assumption~\eqref{A1}, the residual process $\{e_t\}$ is weakly stationary  and thus possesses a well-defined invariant distribution. By conditioning on $e_{t-1}$ and subsequently integrating with respect to this stationary distribution, one obtains a marginalized transition kernel for $y_t$ that depends solely on $y_{t-1}$. It is with respect to this marginalized kernel that we refer to $\{y_t\}$ as a Markov chain throughout the analysis that follows. The proof proceeds in two steps. First, we establish that $\{y_t\}$ is irreducible (Lemma~\ref{lemma_irreducible}). Second, we construct a suitable Lyapunov function and verify a geometric drift condition, which, together with irreducibility, ensures the existence of a unique invariant distribution to which the process converges geometrically fast (Theorem~\ref{theorem_ergodicity}). To start with, a formal definition of irreducibility is provided below \citep{meyn2012markov}.

\begin{definition}\label{def:irreducible}
A Markov chain is called irreducible if for every $x \in \mathcal{X}$ and every $A \in \mathcal{B}(\mathcal{X})$ with $\lambda(A) > 0$, there exists $n \geq 1$ such that $\mathbb{P}^n(x, A) > 0$, where $\mathbb{P}^n(x, A)$ denotes the $n$-step transition probability from state $x$ to set $A$.
\end{definition}

\noindent If for each current state $x$, the one-step law $\mathbb{P}(x, \cdot)$ admits a density $p(x, \cdot)$ w.r.t. Lebesgue measure that is strictly positive on every nonempty open set, then the chain is irreducible. \noindent In the following lemma, we establish the irreducibility of the Markov chain associated with the NARFIMA process. The noise has a strictly positive density, which implies irreducibility. The proof is provided in Appendix \ref{proof_lemma_irreducible}. 

\begin{lemma}\label{lemma_irreducible} Suppose that Assumption~\eqref{A4} holds. Then, the process $\{y_t\}$ is an irreducible Markov chain on the state space $(\mathcal{R}, \mathcal{B})$.
\end{lemma}

\begin{remark}
A trivial example that satisfies the conditions of Lemma~\ref{lemma_irreducible} is Gaussian white noise. Although this condition is sufficient but not necessary, if the support of the noise distribution is sufficiently large, the associated Markov chain becomes irreducible.
\end{remark}

Once irreducibility is established, we proceed to demonstrate the stationarity of $\{y_t\}$. The stationarity of a Markov chain is closely related to the geometric ergodicity of the underlying process. A geometric ergodic process implies that the underlying distribution of the process converges to the unique stationary solution at a geometric rate, regardless of the initial state. Heuristically, geometric ergodicity implies that the Markov chain converges to its stationary distribution. A formal definition of geometric ergodicity and asymptotic stationarity is provided below \citep{meyn2012markov}. 

\begin{definition}
A Markov chain $\{x_t\}$ is called geometrically ergodic if there exists a probability measure $\Pi$ on a probability triple $(\mathcal{X, B}, \lambda)$ and a constant $\rho >1$ such that $\displaystyle\lim_{n\to\infty} \rho^n ||\mathbb{P}^n(x,\cdot)-\Pi(\cdot)||=0$ for each $x\in \mathcal{X}$ and $||\cdot||$ denotes the total variation norm. Then, we say the distribution of $\{x_t\}$ converges to $\Pi$ and $\{x_t\}$ is asymptotically stationary. 
\end{definition}

To establish geometric ergodicity for the NARFIMA$(1,1,k)$ process, we verify a geometric drift condition. A Lyapunov function $V(\cdot)$ satisfies the drift condition if
\begin{equation*}\label{drift}
\mathbb{E}\left[V(y_t) \mid y_{t-1} = y, e_{t-1} = e\right] \leq (1 - \delta) V(y) + B,
\end{equation*}
for some $\delta \in (0,1)$ and $B < \infty$ \citep{meyn2012markov}. Combining the geometric drift condition with irreducibility implies geometric ergodicity and thus asymptotic stationarity. The proof of the following theorem is provided in Appendix~\ref{proof_theorem_ergodicity}.

\begin{theorem}\label{theorem_ergodicity} (Geometric Ergodicity)
Suppose that Assumptions~\eqref{A2}-\eqref{A4} are satisfied. Then, Assumptions~\eqref{A1} and ~\eqref{A5} are sufficient conditions for the geometric ergodicity of the NARFIMA$(1,1,k)$ process.
\end{theorem}

\begin{remark}
Theorem \ref{theorem_ergodicity} says that, under the stated conditions (bounded activation function, contractive skip weights, stationary ARFIMA residuals, and noise process with a positive continuous density), the NARFIMA $(1,1, k)$ has a unique invariant law and converges to it at a geometric rate from any starting value. Practically, this implies well-defined long-run moments, reliable long-horizon forecasts (no explosive behavior), and valid ergodic averages for estimation and inference. The corollary below states that the process is strictly stationary if initialized at the invariant law. 
\end{remark}

\begin{corollary}\label{corollary_stationarity} (Asymptotic Stationarity of NARFIMA Process)
Let $\left\{y_t\right\}$ be the NARFIMA($1, 1, k$) process satisfying the conditions of Theorem~\ref{theorem_ergodicity}, then $\left\{y_t\right\}$ is asymptotically stationary.
\end{corollary}
\noindent Corollary~\ref{corollary_stationarity} follows directly from the proof of Theorem~\ref{theorem_ergodicity}, and it confirms that the NARFIMA process is asymptotically stationary, meaning its distribution stabilizes over time regardless of initialization. These theoretical results, stated for the $(1,1,k)$ case, extend to the general $(p,q,k)$ model under analogous conditions on the expanded parameter set. This framework allows us to use long-run statistical properties, ensures the consistency of estimators, and guarantees meaningful long-term forecasting behavior \citep{trapletti2000stationary}.

\subsubsection{Economic Implications and Practical Applications}
The established results on geometric ergodicity and asymptotic stationarity of the NARFIMA model have significant practical relevance in financial and macroeconomic time series analysis. Although the NARFIMA model captures long memory (frequently observed in exchange rates) dynamics, the process remains asymptotically stationary under appropriate constraints on the neural component and weak stationarity of the feedback residuals.
These theoretical properties justify the model’s implementation in applied forecasting and policy modeling in the following ways:
\begin{enumerate}[label=(\alph*)]
\item Central banks and financial institutions require forecasting models that are stable and generalizable. Asymptotic stationarity and geometric ergodicity guarantee that NARFIMA produces well-behaved forecasts even in the presence of long memory. This reliability is critical in volatile economic environments, where model instability or divergence would carry substantial policy consequences.
\item From a practitioner's perspective, when an irreducible NARFIMA model is estimated from observed time series data, the estimated weights are likely to be close to the true underlying parameters. However, if the irreducibility conditions are not satisfied, the model may be misspecified, and any estimated weights should be interpreted with caution.
\item The neural network component of NARFIMA is capable of capturing nonlinearity, asymmetry, and higher-order interactions. When combined with stationarity and ergodicity guarantees, this nonlinear structure becomes a reliable tool for modeling regime-dependent behaviors such as crisis dynamics and monetary policy shifts.
\end{enumerate}

\section{Experimental Analysis}\label{Section_Experimental_Evaluation}
This section first analyzes the potential drivers of monthly spot exchange rates for BRIC countries by examining historical data and several global and country-specific economic indicators. We investigate the role of GEPU, US EMV, US MPU, oil price growth rates, short-term interest rates, CPI inflation, and GPR index in forecasting exchange rates. This investigation allows us to identify the most relevant predictors to be used as exogenous variables in the proposed NARFIMA approach. Further, we evaluate the effectiveness of the NARFIMA model in forecasting the spot exchange rate of the BRIC economies by comparing its performance against state-of-the-art architectures from various paradigms. To assess its generalizability, we employed a rolling window forecasting approach with six different time horizons of lengths 1 month, 3 months, 6 months, 12 months, 24 months, and 48 months. 

\subsection{Causal Analysis}\label{Causal_anal}
To assess the causal impact of macroeconomic covariates on the exchange rates of the BRIC economies, we employ a nonlinear Granger causality test. This choice is motivated by the global characteristics of the macroeconomic variables, which predominantly exhibit nonlinear patterns, as identified in Section~\ref{Section_Data_and_Methodology}. 
The nonlinear Granger causality (GC) test evaluates the temporal predictive causality, determining whether the lagged values of one variable improve the explanatory power of another variable beyond what is provided by its past observations \citep{granger1969investigating, hiemstra1994testing}. This test specifically assesses the presence of nonlinear causal relationships between the exchange rate and macroeconomic variables. 
Table \ref{tab:nonlinear_granger_causality} presents the results of the nonlinear GC test across BRIC nations. As presented in the table, the p-value of the nonlinear GC test is below 0.05 for several global covariates, including GEPU, US EMV, US MPU, oil price growth rate, and the country-specific short-term IRD. This indicates a significant nonlinear causation between these macroeconomic drivers and the exchange rates. 
The causality analysis reveals varying degrees of association between the examined variables and spot exchange rates. Notably, some indicators, such as CPI inflation and GPR, do not demonstrate strong causal relationships with exchange rates across all the BRIC economies. Based on these findings, we refine our selection of auxiliary variables for the forecasting exercise, focusing on GEPU, US EMV, US MPU, oil price growth rate, and the country-specific short-term IRD that show the most consistent and significant associations with exchange rate movements.

\begin{table}[ht]
\centering
\caption{Nonlinear Granger causality test results assessing the influence of exogenous covariates on exchange rates in BRIC countries.}
\begin{adjustbox}{width=\textwidth}
\begin{tabular}{ccccccccc}
\hline
\multirow{3}{*}{Exogenous Variable} & \multicolumn{2}{c}{\multirow{2}{*}{Brazil exchange rates}} & \multicolumn{2}{c}{\multirow{2}{*}{Russia exchange rates}} & \multicolumn{2}{c}{\multirow{2}{*}{India exchange rates}} & \multicolumn{2}{c}{\multirow{2}{*}{China exchange rates}} \\
& & & & \\ \cline{2-9}
& p-value & Conclusion & p-value & Conclusion & p-value & Conclusion & p-value & Conclusion \\ \hline
GEPU & 0.000 & Causality & 0.000 & Causality & 0.000 & Causality &  0.000 & Causality \\
US EMV  & 0.000 & Causality & 0.000 & Causality & 0.000 & Causality &  0.000 & Causality \\
US MPU & 0.000 & Causality & 0.000 & Causality & 0.000 & Causality & 0.000 & Causality \\
GPR  & 1.000 & No Causality & 0.997 & No Causality & 1.000 & No Causality & 1.000 & No Causality\\
Oil Price Growth Rate & 0.000 & Causality & 0.000 & Causality & 0.000 & Causality & 0.000 & Causality \\
Short-term Interest Rate  & 1.000 & No Causality & 1.000 & No Causality & 1.000 & No Causality & 1.000 & No Causality \\ 
US Short-term Interest Rate & 0.999 & No Causality & 0.975 & No Causality & 0.999 & No Causality & 0.728 & No Causality  \\ 
Short-term IRD  & 0.000 & Causality & 0.000 & Causality & 0.000 & Causality & 0.000 & Causality \\
CPI Inflation  & 1.000 & No Causality & 1.000 & No Causality & 1.000 & No Causality & 1.000 & No Causality  \\
US CPI Inflation & 1.000 & No Causality & 0.996 & No Causality & 1.000 & No Causality & 1.000 & No Causality \\
CPI Inflation Differential  & 1.000 & No Causality & 1.000 & No Causality & 1.000 & No Causality & 1.000 & No Causality \\
\hline
\end{tabular}
\label{tab:nonlinear_granger_causality}
\end{adjustbox}
\end{table}

\subsection{Baseline Models}\label{Sec_Baseline}
To comprehensively assess the forecasting performance of the proposed NARFIMA model, we compare it against sixteen benchmark statistical and deep learning models, several of which can incorporate auxiliary covariates. The statistical benchmarks include the \textit{Naïve}, \textit{Autoregressive} (AR), \textit{Autoregressive Integrated Moving Average with exogenous variables} (ARIMAx), \textit{Autoregressive Fractionally Integrated Moving Average with exogenous variables} (ARFIMAx), \textit{Exponential Smoothing} (ETS), \textit{Self-Exciting Threshold Autoregressive} (SETAR), \textit{Trigonometric Box-Cox ARIMA Trend Seasonal} (TBATS), \textit{Generalized Autoregressive Conditional Heteroscedasticity} (GARCH), and \textit{Bayesian Structural Time Series with exogenous variables} (BSTSx) models. The deep learning benchmarks comprise the \textit{Autoregressive Neural Network with exogenous variables} (ARNNx), \textit{Deep learning-based Autoregressive} (DeepAR), \textit{Neural Basis Expansion Analysis for Time Series with exogenous variables} (NBeatsx), \textit{Neural Hierarchical Interpolation for Time Series with exogenous variables} (NHiTSx), \textit{Decomposition-based Linear model with exogenous variables} (DLinearx), \textit{Normalization-based Linear model with exogenous variables} (NLinearx), and \textit{Time Series Mixer with exogenous variables} (TSMixerx). Detailed descriptions of these benchmark models are provided in Appendix~\ref{Appendix_Baseline}. In our analysis, all benchmark models are implemented to generate multi-horizon exchange rate forecasts for the BRIC economies. The classical forecasting models ARIMAx, ARFIMAx, ETS, TBATS, and ARNNx are implemented using the \texttt{forecast} package in \textbf{R}, while the Naïve and AR models are implemented through the \texttt{stats} package. The SETAR, BSTSx, and GARCH models are implemented using the \texttt{tsDyn}, \texttt{bsts}, and \texttt{tseries} packages, respectively. In addition, the deep learning models DeepAR, NBeatsx, NHiTSx, DLinearx, NLinearx, and TSMixerx are implemented using the \texttt{darts} library in \textbf{Python}. Among recurrent neural network architectures, DeepAR is selected as the representative model because it employs an LSTM-based encoder-decoder framework with an autoregressive training strategy specifically designed for probabilistic multi-step forecasting \citep{salinas2020deepar}. 


\subsection{Evaluation Metrics}\label{Sec_evaluation_Metric}
To assess the performance of the proposed NARFIMA model and the baseline frameworks in forecasting the exchange rate series of the BRIC countries, we used five popularly used evaluation metrics, namely Mean Absolute Percentage Error (MAPE), Symmetric Mean Absolute Percentage Error (SMAPE), Mean Absolute Error (MAE),  Mean Absolute Scaled Error (MASE), and Root Mean Square Error (RMSE). The mathematical formulations of these metrics are provided below:
{\footnotesize
\begin{equation*}
\begin{gathered}
\text{MAPE} = \frac{1}{h} \sum_{t=1}^h \frac{|\hat{y}_t - y_t|}{|y_t|} \times 100 \%; \quad \text{SMAPE} = \frac{1}{h} \sum_{t=1}^h \frac{|\hat{y}_t - y_t|}{\left(|\hat{y}_t|+ |y_t|\right)/2} \times 100 \%; \quad\\
\text{MAE} = \frac{1}{h} \sum_{t=1}^{h} \left| y_t - \hat{y}_t \right|; \quad \text{MASE} = \frac{\sum_{t =1}^{h} |\hat{y}_t - y_t|}{\frac{h}{T-1} \sum_{t = 2}^T |y_t - y_{t-1}|}; \quad \text{RMSE} = \sqrt{\frac{1}{h}\sum_{t=1}^{h} (\hat{y}_t - y_t)^2}, 
\end{gathered}
\end{equation*}}
where $y_t$ denotes the ground truth observation at time $t$ with the corresponding forecast $\hat{y}_t$, $T$ indicates the number of training observations, and $h$ is the forecast horizon. By definition, the model with the smallest error metric value is identified as the best-performing model \citep{hyndman2018forecasting, panja2023epicasting}. 

\subsection{Implementation of NARFIMA Model} \label{Sec_setup_result}

This section presents the implementation of the proposed NARFIMA model for forecasting the exchange rates of the BRIC economies. The NARFIMA framework is implemented in \textbf{R} using a two-stage modeling strategy. In the first stage, an ARFIMAx model is fitted using the \textit{arfima} function from the \texttt{forecast} package. This component captures the long-memory dynamics of the exchange rate series while incorporating significant exogenous variables, namely GEPU, US EMV, US MPU, oil price growth rate, and the country-specific short-term IRD. The estimated ARFIMAx parameters are reported in Table~\ref{ARFIMA parameters}. A key requirement for the proposed hybrid framework is that the ARFIMAx residuals retain nonlinear information not explained by the linear long-memory component. To verify this assumption, we apply the Terasvirta and BDS tests to the residual series. Both tests reject the null hypothesis of linearity when the p-value is below 0.05 \citep{prabowo2020performance, huang2023nonlinearity}. The results, presented in Appendix~\ref{Appendix_Nonlinearity}, confirm significant nonlinear patterns in the ARFIMAx residuals. This finding suggests that the ARFIMAx model adequately captures the linear dependence structure while leaving a nonlinear component that can be further exploited by a neural network. Accordingly, in the second stage, the ARFIMAx residuals, exchange rate observations, and exogenous covariates are used as inputs to a single-hidden-layer feed-forward neural network implemented through the \textit{nnet} function from the \texttt{nnet} package in \textbf{R}. Specifically, the network receives $p$ lagged values of the exchange rate series, $q$ lagged ARFIMAx residuals, and one lagged value of each exogenous variable. These inputs are processed through $k$ hidden neurons to generate a one-step-ahead forecast. To improve stability and reduce sensitivity to random initialization, each NARFIMA specification is estimated as an ensemble of 1000 neural networks, with the final forecast obtained by averaging the ensemble outputs. Multi-step-ahead forecasts are produced recursively. To select the model hyperparameters a time series cross-validation approach is adopted. The hyperparameters $(p,q,k)$ are searched over the range 1--5, and the optimal configuration is chosen by minimizing the RMSE. In addition, two network architectures are considered: one with direct input-to-output connections ($skip=\text{TRUE}$) and another without such connections ($skip=\text{FALSE}$). The resulting optimal NARFIMA$(p,q,k,skip)$ specifications for each BRIC country and forecast horizon are summarized in Table~\ref{NARFIMA parameters}. Alternative loss functions were also examined and produced broadly similar optimal configurations, with only minor differences in a small number of cases, indicating that the selected model architecture is robust to the choice of error metric.

\begin{table}[!htbp]
\tiny
\caption{ARFIMAx parameters \( (\tilde{p}, d, \tilde{q}) \) for BRIC nations across 1-month-ahead $(h = 1)$, 3-month-ahead $(h = 3)$, 6-month-ahead $(h = 6)$, 12-month-ahead $(h = 12)$, 24-month-ahead $(h = 24)$, and 48-month-ahead $(h = 48)$ rolling window forecasts of exchange rate.}

\begin{adjustbox}{width=\textwidth}
   \centering
    \begin{tabular}{ccccccc} \hline
    Country & $(\tilde{p}, d, \tilde{q})_{h=1}$ & $(\tilde{p}, d, \tilde{q})_{h=3}$ & $(\tilde{p}, d, \tilde{q})_{h=6}$ & $(\tilde{p}, d, \tilde{q})_{h=12}$ &$(\tilde{p}, d, \tilde{q})_{h=24}$ & $(\tilde{p}, d, \tilde{q})_{h=48}$\\ \hline
    Brazil & (0,0.493,5) & (0,0.492,5) & (0,0.492,5) & (0,0.492,5) & (0,0.493,5) & (0,0.490,5) \\ \hline
    
    Russia  & (0,0.490,5) & (0,0.490,5) & (0,0.489,5) & (0,0.494,2) & (0,0.489,5) & (0,0.492,3) \\ \hline
    
    India  & (0,0.494,5) & (0,0.493,5) & (0,0.493,5) & (0,0.493,5) & (0,0.492,5) & (0,0.491,5) \\ \hline

    China  & (0,0.496,4) & (0,0.496,4) & (0,0.496,4) & (0,0.496,4) & (0,0.497,4) & (0,0.499,1) \\ \hline
    \end{tabular}
    \label{ARFIMA parameters}
    \end{adjustbox}
\end{table}


\begin{table}[!htbp]
\caption{Optimal NARFIMA parameters \( (p, q, k, skip) \) for BRIC nations across 1-month-ahead $(h = 1)$, 3-month-ahead $(h = 3)$, 6-month-ahead $(h = 6)$, 12-month-ahead $(h = 12)$, 24-month-ahead $(h = 24)$, and 48-month-ahead $(h = 48)$ rolling window forecasts of exchange rate.}

\begin{adjustbox}{width=\textwidth}
   \centering
    \begin{tabular}{ccccccc} \hline
    Country & $(p,q,k,skip)_{h=1}$ & $(p,q,k,skip)_{h=3}$ & $(p,q,k,skip)_{h=6}$ & $(p,q,k,skip)_{h=12}$ &$(p,q,k,skip)_{h=24}$ & $(p,q,k,skip)_{h=48}$\\ \hline
    Brazil & (5,4,4,\text{F}) & (2,5,2,\text{F}) & (1,2,1,\text{F}) & (1,1,1,\text{F}) & (2,5,1,\text{F}) & (4,2,1,\text{T}) \\ \hline
    
    Russia  & (1,2,2,\text{T}) & (1,1,1,\text{F}) & (1,1,1,\text{F}) & (5,2,5,\text{T}) & (1,1,5,\text{T}) & (3,2,1,\text{F}) \\ \hline
    
    India  & (2,1,1,\text{T}) & (4,3,4,\text{T}) & (4,2,1,\text{T}) & (1,3,4,\text{T}) & (5,4,1,\text{T}) & (2,4,4,\text{T}) \\ \hline

    China  & (5,4,1,\text{T}) & (1,2,1,\text{T}) & (1,1,2,\text{F}) & (5,4,1,\text{T}) & (1,2,4,\text{T}) & (4,1,2,\text{T}) \\ \hline
    \end{tabular}
    \label{NARFIMA parameters}
    \end{adjustbox}
\end{table}

\subsection{Validation of Theoretical Results on Empirical Data}\label{verification}
We empirically examine the validity of the theoretical assumptions underlying the NARFIMA framework. Assumption~\eqref{A1} requires the ARFIMAx residuals to be stationary. Unit root tests performed on the residual series reject the null hypothesis of a unit root for all BRIC economies, providing evidence of weak stationarity. Therefore, Assumption~\eqref{A1} is supported empirically. Detailed results are presented in Appendix~\ref{Appendix_Nonlinearity}. 
Assumption~\eqref{A2} is ensured by the model design, since the hidden layer employs a logistic (sigmoid) activation function, defined as $
\sigma(x)=\frac{1}{1+e^{-x}}$. The function is (i) bounded in $(0,1)$; (ii) $\lim _{x \rightarrow-\infty} \sigma(x)=0, \lim _{x \rightarrow+\infty} \sigma(x)=1$ (asymptotically constant); (iii) $\sigma^{\prime}(x)=\sigma(x)(1-\sigma(x)) \leq \frac{1}{4}$ implies $1 / 4$-Lipschitz and $\sigma_\alpha(x)=\sigma(\alpha x)$ is $\alpha / 4$-Lipschitz. These properties ensure that Assumption~\eqref{A2} is satisfied in the NARFIMA implementation. We further evaluate Assumptions~\eqref{A3} and~\eqref{A5} using BRIC exchange rate data, extending the theoretical framework to the NARFIMA$(p,q,k, skip)$ process. In this setting, Assumption~\eqref{A3} generalizes to the condition $\sum_{i=1}^{p} \psi_{1,i} + \sum_{j=1}^{q} \psi_{2,j} \neq 0,$ where $\psi_{1,i}$ and $\psi_{2,j}$ denote the autoregressive and residual skip weights, respectively, while Assumption~\eqref{A5} requires $\Big|\sum_{i=1}^{p} \psi_{1,i}\Big| < 1$. Table~\ref{tab:Assumptions_Validation} reports the results for the longer forecast horizons of 12, 24, and 48 months, for which the optimal parameter configurations in Table~\ref{NARFIMA parameters} had skip connections ($skip = \text{TRUE}$). A bootstrap resampling procedure involving 1000 independent iterations was conducted to account for the stochastic nature of neural network weight initialization and ensure the stability of the empirical results. Each iteration utilized an internal ensemble of 100 model fits to stabilize the base estimates. We report the median skip connection weights and the corresponding values for the conditions specified in Assumptions~\eqref{A3} and~\eqref{A5}. 
This median-based approach mitigates the influence of extreme estimates across bootstrap iterations, reflecting the central tendency of the estimated weights. Empirically, both assumptions were satisfied across all countries, confirming the validity of the theoretical framework in practice.











\begin{table}[!htbp]
\caption{Empirical validation of Assumptions~\eqref{A3} and~\eqref{A5} for the NARFIMA model. Reported values are median weight estimates and the associated assumption conditions, obtained from $1000$ bootstrap iterations ($100$ internal ensemble fits per iteration). Long horizons with $skip = \text{FALSE}$ are omitted, as skip connections were removed.}
\centering
\begin{adjustbox}{width=\textwidth}
\begin{tabular}{ccccccc}
\hline
Country & Horizon & NARFIMA$(p,q,k,skip)$ & AR Skip Weights & Residual Skip Weights & Assumption~\eqref{A3} & Assumption~\eqref{A5} \\ \hline

Brazil & 48 & (4,2,1,T) & (-0.165,0.958,-0.195,0.286)   & (-0.175,0.028) & 0.737 & 0.884\\\hline

Russia & 12 & (5,2,5,T) & (0.14,0.089,0.375,0.181,0.083) & (0.021,0.035)  & 0.923 & 0.867\\

  & 24 & (1,1,5,T) & (0.114) & (0.591)   & 0.704 & 0.114\\\hline

India & 12 & (1,3,4,T) & (0.077) & (0.958,0.058,-0.018)   & 1.075 & 0.077 \\

  & 24 & (5,4,1,T) & (0.321,0.175,0.173,0.172,0.1) & (0.12,-0.048,-0.109,-0.068) & 0.834 & 0.939 \\

  & 48 & (2,4,4,T) & (0.087,0.427) & (0.384,0.116,0.009,0.139)  & 1.162 & 0.514\\ \hline

  China & 12 & (5,4,1,T) & (0.313,0.632,-0.002,0.13,-0.287) & (0.408,0.003,0.003,0.001) & 1.200 & 0.785\\

  & 24 & (1,2,4,T) & (0.033)  & (0.976,0.021)  & 1.031 & 0.033\\

  & 48 & (4,1,2,T) & (0.063,0.868,-0.294,0.34) & (-0.251)  & 0.727 & 0.978 \\ \hline
\end{tabular}
\label{tab:Assumptions_Validation}
\end{adjustbox}
\end{table}

\subsection{Baseline Comparisons}

The forecasting performance of NARFIMA and the competing baseline models for Brazil, Russia, India, and China is reported in Tables \ref{tab:performance_metrics_braz}, \ref{tab:performance_metrics_rus}, \ref{tab:performance_metrics_ind}, and \ref{tab:performance_metrics_chn}, respectively. Overall, the results indicate that the proposed NARFIMA framework consistently achieves superior predictive accuracy across most forecasting horizons and evaluation metrics, particularly when compared with both classical statistical models and deep learning architectures. For short-term forecasting (1-month-ahead), NARFIMA clearly outperforms all competing models across all BRIC economies, demonstrating its strong capability in capturing immediate dynamics. At the 3-month horizon, it continues to perform best for Brazil and China, while ARIMAx shows competitive performance for Russia and India. In the 6-month-ahead setting, NARFIMA further improves forecast accuracy relative to its individual components (ARFIMAx and ARNNx), and yields the best results for India and China, whereas ARIMAx and ETS perform comparably for Brazil and Russia among the baselines. For medium-term horizons (12 months), NARFIMA remains the leading model for Brazil and Russia, while Naïve and ARIMAx exhibit relatively stronger performance for India and China, respectively. At the 24-month horizon, NARFIMA continues to dominate in most cases, except for India where NLinearx produces slightly better forecasts. Finally, for the longest horizon (48 months), the proposed model consistently achieves the best performance across all BRIC countries, reinforcing its ability to capture both long-range dependencies and nonlinear dynamics in exchange rate behavior.

From a country-specific perspective, NARFIMA delivers the most accurate forecasts in five out of six horizons for Brazil and China. For Russia, it outperforms baseline models in four out of six cases while remaining competitive with ARIMAx and ETS models in the other two horizons. However, for India's exchange rate series, the NARFIMA model performs best in only three forecasting horizons. This is primarily attributed to the linearity of India's exchange rate data, which is better modeled with ARIMAx, Naïve, and NLinearx frameworks. These empirical results confirm that the NARFIMA approach can effectively model volatility, non-stationarity, and nonlinearity in time series data. However, its forecasting performance declines for inherently linear series, where traditional statistical models perform better. Nevertheless, as most macroeconomic variables exhibit nonlinear characteristics \citep{franses2000non}, NARFIMA remains well-suited for forecasting complex financial time series. On the other hand, deep learning models yield relatively inaccurate exchange rate forecasts compared to both statistical baselines and the NARFIMA model. This is likely due to the low sample size of the dataset, which pose significant challenges in accurately training deep learning architectures. While some baseline models, including ARIMAx and ETS, achieve comparable short-term and semi-long-term performance, their accuracy deteriorates significantly over longer horizons. In contrast, NARFIMA consistently performs well across all time horizons, demonstrating its robustness and generalizability. Additionally, the integration of ARFIMAx feedback residuals with the target series and exogenous variables in NARFIMA proves to be a highly effective technique for capturing long-term dependencies, especially for non-stationary and nonlinear exchange rate series.

\begin{table*}[!htbp]
\caption{Evaluation of the proposed NARFIMA model's performance relative to baseline forecasters across all forecast horizons for Brazil (\textbf{\underline{best}} and \textbf{\textit{second-best}} results are highlighted). The MASE metric is not defined for a forecast horizon of length one.}
\centering
\begin{adjustbox}{width=1\textwidth, height=3.5cm}
\begin{tabular}{ccccccccccccccccccc}
\hline
Horizon & Metrics & Naïve & AR & ARIMAx & ARNNx & ARFIMAx & ETS & SETAR & TBATS & GARCH & BSTSx & DeepAR & NBeatsx & NHiTSx & DLinearx & NLinearx & TSMixerx & {\color{blue}NARFIMA} \\\hline
1 & MAPE & 2.318 & 2.855 & 5.613 & 5.446 & 2.270 & 1.949 & 2.084 & 2.351 & 2.350 & 5.965 & \textbf{\textit{1.812}} & 8.493 & 9.267 & 11.001 & 16.562 & 22.367 & \textbf{\underline{0.041}} \\
& SMAPE & 2.345 & 2.896 & 5.775 & 5.599 & 2.296 & 1.969 & 2.106 & 2.379 & 2.378 & 6.148 & \textbf{\textit{1.829}} & 8.869 & 8.857 & 10.428 & 15.296 & 25.183 & \textbf{\underline{0.042}} \\
& MAE & 0.117 & 0.144 & 0.284 & 0.275 & 0.115 & 0.099 & 0.105 & 0.119 & 0.119 & 0.302 & \textbf{\textit{0.092}} & 0.429 & 0.469 & 0.556 & 0.837 & 1.131 & \textbf{\underline{0.002}} \\
& RMSE & 0.117 & 0.144 & 0.284 & 0.275 & 0.115 & 0.099 & 0.105 & 0.119 & 0.119 & 0.302 & \textbf{\textit{0.092}} & 0.429 & 0.469 & 0.556 & 0.837 & 1.131 & \textbf{\underline{0.002}} \\
\hline

3 & MAPE & 3.301 & 4.688 & 6.368 & 9.182 & 4.995 & 5.137 & 2.603 & 3.254 & 3.304 & 6.789 & \textbf{\textit{0.761}} & 2.512 & 1.284 & 17.789 & 17.603 & 13.737 & \textbf{\underline{0.541}} \\
& SMAPE & 3.365 & 4.818 & 6.585 & 9.698 & 5.142 & 5.291 & 2.642 & 3.316 & 3.368 & 7.033 & \textbf{\textit{0.761}} & 2.478 & 1.276 & 16.324 & 16.170 & 14.757 & \textbf{\underline{0.541}} \\
& MAE & 0.165 & 0.234 & 0.317 & 0.458 & 0.249 & 0.256 & 0.130 & 0.162 & 0.165 & 0.338 & \textbf{\textit{0.038}} & 0.125 & 0.063 & 0.882 & 0.875 & 0.682 & \textbf{\underline{0.027}} \\
& MASE & 2.130 & 3.025 & 4.098 & 5.924 & 3.222 & 3.313 & 1.679 & 2.100 & 2.132 & 4.367 & \textbf{\textit{0.490}} & 1.612 & 0.820 & 11.409 & 11.310 & 8.812 & \textbf{\underline{0.348}} \\
& RMSE & 0.177 & 0.251 & 0.324 & 0.494 & 0.267 & 0.274 & 0.139 & 0.175 & 0.178 & 0.342 & \textbf{\textit{0.040}} & 0.153 & 0.068 & 0.886 & 0.879 & 0.683 & \textbf{\underline{0.031}} \\
\hline

6 & MAPE & 2.280 & 2.263 & \textbf{\underline{1.265}} & 7.136 & 4.903 & 2.198 & 2.503 & 1.444 & 2.251 & 2.239 & 3.841 & 16.281 & 4.841 & 15.575 & 10.943 & 14.721 & \textbf{\textit{1.370}} \\
& SMAPE & 2.245 & 2.300 & \textbf{\underline{1.271}} & 7.536 & 5.075 & 2.235 & 2.458 & 1.438 & 2.217 & 2.269 & 3.754 & 15.635 & 4.690 & 14.377 & 10.213 & 16.011 & \textbf{\textit{1.368}} \\
& MAE & 0.111 & 0.112 & \textbf{\underline{0.063}} & 0.354 & 0.243 & 0.109 & 0.122 & 0.071 & 0.110 & 0.111 & 0.188 & 0.800 & 0.238 & 0.772 & 0.540 & 0.725 & \textbf{\textit{0.067}} \\
& MASE & 1.273 & 1.288 & \textbf{\underline{0.718}} & 4.055 & 2.785 & 1.252 & 1.396 & 0.812 & 1.257 & 1.272 & 2.150 & 9.169 & 2.723 & 8.849 & 6.195 & 8.307 & \textbf{\textit{0.772}} \\
& RMSE & 0.130 & 0.146 & \textbf{\underline{0.075}} & 0.432 & 0.286 & 0.142 & 0.148 & 0.085 & 0.128 & 0.122 & 0.206 & 0.929 & 0.275 & 0.891 & 0.628 & 0.756 & \textbf{\textit{0.082}} \\
\hline

12 & MAPE & 4.069 & \textbf{\textit{2.328}} & 3.853 & 4.050 & 5.784 & 5.259 & 4.978 & 3.482 & 3.939 & 3.374 & 3.366 & 25.537 & 6.992 & 14.187 & 15.544 & 26.956 & \textbf{\underline{1.548}} \\
& SMAPE & 3.941 & \textbf{\textit{2.373}} & 3.822 & 4.163 & 5.991 & 5.064 & 4.803 & 3.386 & 3.818 & 3.446 & 3.441 & 23.427 & 7.150 & 12.946 & 13.967 & 31.741 & \textbf{\underline{1.562}} \\
& MAE & 0.201 & \textbf{\textit{0.118}} & 0.194 & 0.206 & 0.291 & 0.260 & 0.246 & 0.172 & 0.194 & 0.173 & 0.173 & 1.284 & 0.356 & 0.705 & 0.771 & 1.373 & \textbf{\underline{0.079}} \\
& MASE & 2.837 & \textbf{\textit{1.669}} & 2.737 & 2.907 & 4.114 & 3.677 & 3.480 & 2.426 & 2.746 & 2.450 & 2.441 & 18.139 & 5.031 & 9.963 & 10.893 & 19.407 & \textbf{\underline{1.118}} \\
& RMSE & 0.252 & \textbf{\textit{0.150}} & 0.207 & 0.238 & 0.317 & 0.313 & 0.296 & 0.220 & 0.245 & 0.212 & 0.199 & 1.477 & 0.402 & 0.873 & 0.950 & 1.453 & \textbf{\underline{0.096}} \\
\hline

24 & MAPE & 8.246 & \textbf{\textit{7.324}} & 10.068 & 19.586 & 10.250 & 18.685 & 13.438 & 10.277 & 7.938 & 9.671 & 15.602 & 20.290 & 18.297 & 13.855 & 7.850 & 26.467 & \textbf{\underline{3.341}} \\
& SMAPE & 7.836 & 7.632 & 9.477 & 22.621 & 10.989 & 16.875 & 12.395 & 9.673 & \textbf{\textit{7.555}} & 9.131 & 16.996 & 19.500 & 18.455 & 14.425 & 8.487 & 32.126 & \textbf{\underline{3.361}} \\
& MAE & 0.414 & \textbf{\textit{0.374}} & 0.507 & 1.002 & 0.524 & 0.944 & 0.676 & 0.517 & 0.398 & 0.487 & 0.805 & 1.029 & 0.939 & 0.729 & 0.417 & 1.382 & \textbf{\underline{0.173}} \\
& MASE & 3.480 & \textbf{\textit{3.140}} & 4.261 & 8.422 & 4.405 & 7.938 & 5.683 & 4.347 & 3.349 & 4.091 & 6.766 & 8.653 & 7.897 & 6.129 & 3.502 & 11.619 & \textbf{\underline{1.455}} \\
& RMSE & 0.464 & \textbf{\textit{0.417}} & 0.566 & 1.160 & 0.601 & 1.007 & 0.754 & 0.566 & 0.447 & 0.540 & 0.827 & 1.276 & 1.079 & 0.905 & 0.585 & 1.594 & \textbf{\underline{0.204}} \\
\hline

48 & MAPE & 20.214 & 36.403 & 17.081 & 45.771 & 34.701 & 18.714 & 28.760 & 20.465 & 20.309 & \textbf{\textit{15.632}} & 36.783 & 43.228 & 32.738 & 25.167 & 26.632 & 31.646 & \textbf{\underline{11.888}} \\
& SMAPE & 22.753 & 45.436 & 18.914 & 61.606 & 42.697 & 20.895 & 34.036 & 23.063 & 22.871 & \textbf{\textit{17.239}} & 45.540 & 34.096 & 27.344 & 22.149 & 23.322 & 38.054 & \textbf{\underline{12.856}} \\
& MAE & 1.064 & 1.901 & 0.900 & 2.382 & 1.815 & 0.986 & 1.507 & 1.077 & 1.069 & \textbf{\textit{0.827}} & 1.919 & 2.198 & 1.662 & 1.279 & 1.352 & 1.641 & \textbf{\underline{0.630}} \\
& MASE & 6.700 & 11.969 & 5.669 & 15.000 & 11.431 & 6.210 & 9.494 & 6.780 & 6.730 & \textbf{\textit{5.210}} & 12.083 & 13.841 & 10.470 & 8.055 & 8.512 & 10.332 & \textbf{\underline{3.968}} \\
& RMSE & 1.126 & 1.987 & 0.968 & 2.508 & 1.891 & 1.049 & 1.574 & 1.138 & 1.131 & \textbf{\textit{0.912}} & 1.974 & 2.475 & 1.833 & 1.322 & 1.383 & 1.700 & \textbf{\underline{0.715}} \\
\hline
\end{tabular}
\label{tab:performance_metrics_braz}
\end{adjustbox}
\end{table*}

\begin{table*}[!htbp]
\caption{Evaluation of the proposed NARFIMA model's performance relative to baseline forecasters across all forecast horizons for Russia (\textbf{\underline{`best'}} and \textbf{\textit{`second-best'}} results are highlighted). The MASE metric is not defined for a forecast horizon of length one.}
\centering
\begin{adjustbox}{width=1\textwidth, height=3.5cm}
\begin{tabular}{ccccccccccccccccccc}
\hline
Horizon & Metrics & Naïve & AR & ARIMAx & ARNNx & ARFIMAx & ETS & SETAR & TBATS & GARCH & BSTSx & DeepAR & NBeatsx & NHiTSx & DLinearx & NLinearx & TSMixerx & {\color{blue}NARFIMA} \\\hline
1 & MAPE & 4.473 & 3.028 & 5.213 & 12.051 & 2.891 & 6.237 & 4.548 & 4.604 & \textbf{\textit{1.193}} & 5.164 & 74.878 & 25.264 & 18.440 & 3.997 & 10.679 & 40.493 & \textbf{\underline{0.043}} \\
& SMAPE & 4.376 & 2.983 & 5.080 & 12.824 & 2.849 & 6.048 & 4.447 & 4.500 & \textbf{\textit{1.186}} & 5.034 & 119.689 & 28.917 & 20.312 & 4.079 & 11.281 & 50.772 & \textbf{\underline{0.043}} \\
& MAE & 4.171 & 2.824 & 4.860 & 11.237 & 2.695 & 5.816 & 4.241 & 4.293 & \textbf{\textit{1.113}} & 4.815 & 69.819 & 23.557 & 17.194 & 3.727 & 9.958 & 37.757 & \textbf{\underline{0.040}} \\
& RMSE & 4.171 & 2.824 & 4.860 & 11.237 & 2.695 & 5.816 & 4.241 & 4.293 & \textbf{\textit{1.113}} & 4.815 & 69.819 & 23.557 & 17.194 & 3.727 & 9.958 & 37.757 & \textbf{\underline{0.040}} \\
\hline

3 & MAPE & 4.732 & 7.064 & \textbf{\underline{2.576}} & 27.106 & 6.487 & \textbf{\textit{2.830}} & 9.698 & 4.224 & 8.972 & 2.932 & 76.133 & 25.671 & 23.114 & 6.731 & 11.367 & 53.993 & 3.181 \\
& SMAPE & 4.863 & 7.332 & \textbf{\underline{2.627}} & 31.593 & 6.716 & \textbf{\textit{2.784}} & 10.221 & 4.331 & 9.409 & 2.996 & 122.931 & 29.515 & 26.169 & 7.027 & 12.092 & 73.996 & 3.258 \\
& MAE & 4.551 & 6.767 & \textbf{\underline{2.493}} & 25.858 & 6.197 & \textbf{\textit{2.668}} & 9.260 & 4.065 & 8.599 & 2.835 & 72.736 & 24.573 & 22.118 & 6.481 & 10.897 & 51.599 & 3.076 \\
& MASE & 1.609 & 2.392 & \textbf{\underline{0.881}} & 9.141 & 2.191 & \textbf{\textit{0.943}} & 3.273 & 1.437 & 3.040 & 1.002 & 25.713 & 8.687 & 7.819 & 2.291 & 3.852 & 18.240 & 1.088 \\
& RMSE & 4.867 & 6.901 & \textbf{\underline{3.059}} & 26.369 & 6.354 & \textbf{\textit{3.284}} & 9.506 & 4.416 & 8.768 & 3.421 & 72.757 & 24.779 & 22.255 & 7.260 & 11.205 & 51.651 & 3.762 \\
\hline

6 & MAPE & 11.051 & 21.548 & 11.707 & 38.045 & 16.097 & \textbf{\underline{5.501}} & 18.408 & 8.867 & 14.591 & 11.098 & 74.090 & 16.626 & 11.537 & 9.177 & 16.571 & 59.835 & \textbf{\textit{8.797}} \\
& SMAPE & 11.893 & 24.438 & 12.636 & 47.493 & 18.004 & \textbf{\underline{5.683}} & 20.701 & 9.389 & 15.932 & 11.908 & 117.747 & 18.149 & 12.311 & 9.696 & 18.339 & 85.465 & \textbf{\textit{9.306}} \\
& MAE & 10.372 & 19.947 & 10.971 & 34.973 & 15.123 & \textbf{\underline{5.130}} & 17.186 & 8.311 & 13.577 & 10.378 & 67.432 & 15.134 & 10.656 & \textbf{\textit{8.121}} & 14.794 & 54.391 & 8.241 \\
& MASE & 2.481 & 4.772 & 2.625 & 8.367 & 3.618 & \textbf{\underline{1.227}} & 4.112 & 1.988 & 3.248 & 2.483 & 16.133 & 3.621 & 2.549 & \textbf{\textit{1.943}} & 3.539 & 13.013 & 1.972 \\
& RMSE & 11.817 & 21.085 & 12.385 & 35.989 & 17.433 & \textbf{\underline{5.806}} & 18.987 & 9.421 & 14.724 & 11.580 & 67.671 & 15.257 & 11.209 & \textbf{\textit{8.970}} & 15.700 & 54.556 & 9.297 \\
\hline

12 & MAPE & 23.259 & 25.521 & 25.434 & 22.431 & 22.705 & 14.957 & 19.294 & 22.497 & 24.575 & 21.056 & 72.813 & 19.963 & 22.332 & 9.857 & \textbf{\textit{9.176}} & 44.246 & \textbf{\underline{7.201}} \\
& SMAPE & 27.127 & 30.280 & 29.862 & 25.967 & 26.535 & 16.540 & 21.942 & 26.105 & 28.858 & 24.238 & 114.790 & 17.722 & 19.852 & 10.655 & \textbf{\textit{8.595}} & 56.883 & \textbf{\underline{7.539}} \\
& MAE & 20.152 & 22.122 & 21.827 & 19.391 & 19.776 & 13.055 & 16.780 & 19.502 & 21.236 & 18.294 & 59.849 & 15.490 & 17.499 & 8.265 & \textbf{\textit{7.194}} & 36.201 & \textbf{\underline{6.202}} \\
& MASE & 5.278 & 5.794 & 5.716 & 5.078 & 5.179 & 3.419 & 4.395 & 5.108 & 5.562 & 4.791 & 15.674 & 4.057 & 4.583 & 2.165 & \textbf{\textit{1.884}} & 9.481 & \textbf{\underline{1.624}} \\
& RMSE & 22.908 & 25.172 & 24.205 & 21.911 & 23.012 & 15.353 & 19.374 & 22.228 & 23.954 & 21.016 & 60.862 & 16.750 & 17.939 & 10.523 & \textbf{\textit{8.553}} & 36.650 & \textbf{\underline{7.613}} \\
\hline

24 & MAPE & 14.531 &  17.836 & \textbf{\textit{13.614}} & 15.995 & 18.086 & 14.531 & 14.509 & 14.779 & 14.480 & 13.908 & 69.848 & 21.984 & 16.352 & 36.526 & 39.446 & 19.307 & \textbf{\underline{7.913}} \\
& SMAPE & 14.905 & 20.107 & \textbf{\textit{13.064}} & 16.787 & 20.722 & 14.905 & 14.838 & 15.384 & 14.642 & 13.860 & 107.824 & 19.221 & 14.614 & 29.079 & 30.997 & 22.362 & \textbf{\underline{7.722}} \\
& MAE & 11.044 & 14.493 & \textbf{\textit{9.652}} & 12.329 & 14.903 & 11.044 & 10.996 & 11.383 & 10.856 & 10.276 & 53.031 & 15.592 & 11.020 & 24.952 & 27.328 & 15.234 & \textbf{\underline{5.347}} \\
& MASE & 2.441 & 3.204 & \textbf{\textit{2.134}} & 2.725 & 3.295 & 2.441 & 2.431 & 2.516 & 2.400 & 2.272 & 11.723 & 3.447 & 2.436 & 5.516 & 6.041 & 3.368 & \textbf{\underline{1.182}} \\
& RMSE & 13.305 & 18.504 & \textbf{\textit{11.396}} & 14.174 & 19.165 & 13.305 & 13.311 & 13.730 & 13.053 & 12.064 & 54.491 & 17.803 & 13.578 & 28.648 & 31.076 & 18.372 & \textbf{\underline{7.488}} \\
\hline

48 & MAPE & 14.478 & 27.687 & 11.599 & 30.146 & 27.016 & 14.478 & 13.559 & 14.462 & 14.416 & \textbf{\textit{11.321}} & 73.245 & 66.489 & 44.946 & 29.673 & 40.079 & 29.388 & \textbf{\underline{9.371}} \\
& SMAPE & 15.866 & 33.438 & 12.120 & 38.900 & 32.244 & 15.866 & 14.684 & 15.846 & 15.805 & \textbf{\textit{11.600}} & 115.810 & 46.700 & 35.699 & 23.658 & 30.969 & 35.623 & \textbf{\underline{9.383}} \\
& MAE & 11.324 & 21.460 & 8.793 & 23.286 & 20.931 & 11.324 & 10.553 & 11.311 & 11.286 & \textbf{\textit{8.434}} & 54.370 & 47.375 & 31.682 & 20.691 & 28.248 & 22.386 & \textbf{\underline{6.904}} \\
& MASE & 3.268 & 6.193 & 2.537 & 6.720 & 6.040 & 3.268 & 3.045 & 3.264 & 3.257 & \textbf{\textit{2.434}} & 15.690 & 13.671 & 9.142 & 5.971 & 8.152 & 6.460 & \textbf{\underline{1.992}} \\
& RMSE & 13.730 & 24.704 & 10.520 & 29.103 & 23.667 & 13.730 & 12.836 & 13.715 & 13.747 & \textbf{\textit{10.021}} & 55.195 & 53.181 & 33.387 & 26.450 & 33.566 & 24.753 & \textbf{\underline{9.709}} \\
\hline
\end{tabular}
\label{tab:performance_metrics_rus}
\end{adjustbox}
\end{table*}

\begin{table*}[!htbp]
\caption{Evaluation of the proposed NARFIMA model's performance relative to baseline forecasters across all forecast horizons for India (\textbf{\underline{`best'}} and \textbf{\textit{`second-best'}} results are highlighted). The MASE metric is not defined for a forecast horizon of length one.}
\centering
\begin{adjustbox}{width=1\textwidth, height=3.5cm}
\begin{tabular}{ccccccccccccccccccc}
\hline
Horizon & Metrics & Naïve & AR & ARIMAx & ARNNx & ARFIMAx & ETS & SETAR & TBATS & GARCH & BSTSx & DeepAR & NBeatsx & NHiTSx & DLinearx & NLinearx & TSMixerx & {\color{blue}NARFIMA} \\ \hline
1 & MAPE & 0.176 & 0.614 & 0.773 & 3.866 & 0.774 & 0.022 & \textbf{\textit{0.013}} & 0.155 & 0.394 & 0.987 & 69.916 & 1.964 & 0.577 & 4.459 & 3.082 & 18.362 & \textbf{\underline{0.006}} \\
& SMAPE & 0.176 & 0.616 & 0.770 & 3.942 & 0.777 & 0.022 & \textbf{\textit{0.013}} & 0.156 & 0.395 & 0.992 & 107.494 & 1.945 & 0.578 & 4.561 & 3.130 & 20.218 & \textbf{\underline{0.006}} \\

& MAE & 0.146 & 0.511 & 0.643 & 3.217 & 0.644 & 0.018 & \textbf{\textit{0.011}} & 0.129 & 0.328 & 0.821 & 58.182 & 1.634 & 0.480 & 3.711 & 2.565 & 15.280 & \textbf{\underline{0.005}} \\
& RMSE & 0.146 & 0.511 & 0.643 & 3.217 & 0.644 & 0.018 & \textbf{\textit{0.011}} & 0.129 & 0.328 & 0.821 & 58.182 & 1.634 & 0.480 & 3.711 & 2.565 & 15.280 & \textbf{\underline{0.005}} \\
\hline

3 & MAPE & 1.055 & 1.899 & \textbf{\underline{0.133}} & 7.131 & 3.087 & 1.063 & 0.830 & 1.060 & 1.331 & \textbf{\textit{0.428}} & 70.055 & 2.394 & 5.507 & 4.102 & 7.598 & 18.409 & 0.546 \\
& SMAPE & 1.061 & 1.918 & \textbf{\underline{0.133}} & 7.421 & 3.142 & 1.069 & 0.834 & 1.066 & 1.340 & \textbf{\textit{0.429}} & 107.823 & 2.360 & 5.337 & 4.191 & 7.912 & 20.279 & 0.548 \\
& MAE & 0.876 & 1.577 & \textbf{\underline{0.110}} & 5.925 & 2.565 & 0.883 & 0.690 & 0.880 & 1.106 & \textbf{\textit{0.355}} & 58.169 & 1.986 & 4.576 & 3.407 & 6.311 & 15.286 & 0.454 \\
& MASE & 4.338 & 7.808 & \textbf{\underline{0.547}} & 29.330 & 12.697 & 4.371 & 3.414 & 4.358 & 5.474 & \textbf{\textit{1.758}} & 287.964 & 9.833 & 22.653 & 16.867 & 31.243 & 75.674 & 2.246 \\
 & RMSE & 0.892 & 1.640 & \textbf{\underline{0.121}} & 6.199 & 2.745 & 0.899 & 0.695 & 0.896 & 1.135 & \textbf{\textit{0.355}} & 58.169 & 2.178 & 4.923 & 3.473 & 6.449 & 15.298 & 0.477 \\
\hline

6 & MAPE & 0.798 & 2.324 & 0.578 & 8.584 & 4.733 & 0.892 & 0.436 & 0.828 & 1.235 & \textbf{\textit{0.329}} & 70.874 & 9.691 & 3.016 & 3.224 & 4.742 & 18.645 & \textbf{\underline{0.176}} \\
& SMAPE & 0.802 & 2.359 & 0.575 & 9.024 & 4.886 & 0.898 & 0.437 & 0.833 & 1.245 & \textbf{\textit{0.328}} & 109.777 & 9.223 & 2.966 & 3.306 & 4.860 & 20.565 & \textbf{\underline{0.176}} \\
& MAE & 0.661 & 1.925 & 0.476 & 7.104 & 3.921 & 0.740 & 0.361 & 0.686 & 1.023 & \textbf{\textit{0.271}} & 58.564 & 8.016 & 2.493 & 2.654 & 3.916 & 15.408 & \textbf{\underline{0.145}} \\
& MASE & 2.753 & 8.016 & 1.980 & 29.578 & 16.326 & 3.079 & 1.503 & 2.857 & 4.260 & \textbf{\textit{1.130}} & 243.834 & 33.373 & 10.382 & 11.052 & 16.305 & 64.151 & \textbf{\underline{0.603}} \\
& RMSE & 0.784 & 2.170 & 0.593 & 7.571 & 4.517 & 0.864 & 0.428 & 0.806 & 1.186 & \textbf{\textit{0.314}} & 58.565 & 8.216 & 2.645 & 3.294 & 3.964 & 15.418 & \textbf{\underline{0.236}} \\
\hline

12 & MAPE & \textbf{\underline{0.447}} & 3.333 & \textbf{\textit{0.535}} & 9.260 & 6.401 & 2.511 & 1.544 & 0.712 & 1.470 & 1.559 & 72.896 & 14.840 & 4.956 & 2.240 & 4.377 & 19.576 & 1.296 \\
& SMAPE & \textbf{\underline{0.448}} & 3.411 & \textbf{\textit{0.533}} & 9.762 & 6.705 & 2.477 & 1.531 & 0.709 & 1.485 & 1.546 & 114.703 & 13.786 & 4.774 & 2.226 & 4.496 & 21.721 & 1.296 \\
& MAE & \textbf{\underline{0.369}} & 2.754 & \textbf{\textit{0.439}} & 7.637 & 5.286 & 2.068 & 1.272 & 0.585 & 1.215 & 1.285 & 60.049 & 12.228 & 4.093 & 1.846 & 3.599 & 16.122 & 1.068 \\
& MASE & \textbf{\underline{0.894}} & 6.663 & \textbf{\textit{1.063}} & 18.481 & 12.790 & 5.004 & 3.078 & 1.415 & 2.939 & 3.110 & 145.311 & 29.590 & 9.906 & 4.467 & 8.708 & 39.014 & 2.584 \\
& RMSE & \textbf{\underline{0.471}} & 3.219 & \textbf{\textit{0.533}} & 8.040 & 6.295 & 2.142 & 1.336 & 0.677 & 1.434 & 1.360 & 60.051 & 12.432 & 5.115 & 2.231 & 3.946 & 16.184 & 1.160 \\
\hline

24 & MAPE & 6.176  & 10.081 & 4.962 & 10.541 & 12.650 & 5.159 & 5.831 & 5.729 & 8.251 & \textbf{\textit{4.212}} & 72.293 & 17.215 & 10.885 & 5.669 & \textbf{\underline{3.275}} & 19.947 & 4.555 \\
& SMAPE & 6.440 & 10.795 & 5.125 & 11.224 & 13.770 & 5.344 & 6.065 & 5.955 & 8.724 & \textbf{\textit{4.329}} & 113.241 & 15.797 & 10.267 & 5.448 & \textbf{\underline{3.226}} & 22.176 & 4.671 \\
& MAE & 5.041 & 8.222 & 4.043 & 8.539 & 10.308 & 4.213 & 4.759 & 4.676 & 6.731 & \textbf{\textit{3.432}} & 57.782 & 13.743 & 8.648 & 4.553 & \textbf{\underline{2.629}} & 15.981 & 3.637 \\
& MASE & 7.652 & 12.482 & 6.138 & 12.963 & 15.648 & 6.395 & 7.224 & 7.098 & 10.217 & \textbf{\textit{5.210}} & 87.715 & 20.862 & 13.128 & 6.912 & \textbf{\underline{3.992}} & 24.260 & 5.520 \\
& RMSE & 5.827 & 9.454 & 4.610 & 9.221 & 11.732 & 4.915 & 5.505 & 5.423 & 7.745 & 3.932 & 57.863 & 14.059 & 9.071 & 5.475 & \textbf{\underline{3.129}} & 16.093 & \textbf{\textit{3.801}} \\
\hline

48 & MAPE & 7.308 & 16.106 & 4.749 & 18.908 & 17.576 & 6.775 & 8.633 & 7.311 & 11.368 & 3.943 & 72.878 & 9.460 & 26.766 & \textbf{\textit{3.282}} & 3.610 & 31.710 & \textbf{\underline{2.872}} \\
& SMAPE & 7.704 & 17.952 & 4.926 & 21.937 & 19.594 & 7.130 & 9.165 & 7.707 & 12.302 & 4.070 & 114.694 & 8.930 & 23.570 & \textbf{\textit{3.315}} & 3.582 & 37.715 & \textbf{\underline{2.873}} \\
& MAE & 5.791 & 12.674 & 3.772 & 14.980 & 13.756 & 5.381 & 6.823 & 5.793 & 8.976 & 3.137 & 56.022 & 7.166 & 20.475 & \textbf{\textit{2.504}} & 2.764 & 24.400 & \textbf{\underline{2.210}} \\
& MASE & 8.409 & 18.404 & 5.477 & 21.752 & 19.975 & 7.813 & 9.907 & 8.412 & 13.034 & 4.555 & 81.348 & 10.405 & 29.731 & \textbf{\textit{3.635}} & 4.014 & 35.431 & \textbf{\underline{3.208}} \\
& RMSE & 6.973 & 14.468 & 4.702 & 18.215 & 15.045 & 6.614 & 8.021 & 6.975 & 10.505 & 4.003 & 56.157 & 8.091 & 20.573 & \textbf{\textit{3.026}} & 3.034 & 24.523 & \textbf{\underline{2.616}} \\
\hline
\end{tabular}
\label{tab:performance_metrics_ind}
\end{adjustbox}
\end{table*}

\begin{table*}[!htbp]
\caption{Evaluation of the proposed NARFIMA model's performance relative to baseline forecasters across all forecast horizons for China (\textbf{\underline{`best'}} and \textbf{\textit{`second-best'}} results are highlighted). The MASE metric is not defined for a forecast horizon of length one.}
\centering
\begin{adjustbox}{width=1\textwidth, height=3.5cm}
\begin{tabular}{ccccccccccccccccccc}
\hline
Horizon & Metrics & Naïve & AR & ARIMAx & ARNNx & ARFIMAx & ETS & SETAR & TBATS & GARCH & BSTSx & DeepAR & NBeatsx & NHiTSx & DLinearx & NLinearx & TSMixerx & {\color{blue}NARFIMA} \\\hline
1  & MAPE  & 0.126 &  \textbf{\textit{0.018}} & 2.588 & 0.487 & 0.048 & 0.381 & 0.149 & 0.262 & 0.349 & 0.202 & 0.874 & 66.993 & 45.806 & 15.326 & 11.975 & 6.737 & \textbf{\underline{0.005}} \\ 
& SMAPE & 0.126 & \textbf{\textit{0.018}} & 2.622 & 0.486 & 0.048 & 0.380 & 0.149 & 0.262 & 0.349 & 0.202 & 0.878 & 50.183 & 37.270 & 14.235 & 11.299 & 6.972 & \textbf{\underline{0.005}} \\
& MAE & 0.009 & \textbf{\textit{0.001}} & 0.189 & 0.036 & 0.003 & 0.028 & 0.011 & 0.019 & 0.025 & 0.015 & 0.064 & 4.895 & 3.347 & 1.120 & 0.875 & 0.492 & $\mathbf{\underline{3.8 \times 10^{-4}}}$ \\ 
& RMSE   & 0.009 & \textbf{\textit{0.001}} & 0.018 & 0.028 & 0.036 & 0.028 & 0.011 & 0.019 & 0.025 & 0.015 & 0.064 & 4.895 & 3.347 & 1.120 & 0.875 & 0.492 & $\mathbf{\underline{3.8 \times 10^{-4}}}$ \\ 
\hline

3  & MAPE   & 1.347 & 1.249 & 3.645 & 0.404 & 3.067 & \textbf{\textit{0.310}} & 4.195 & 1.727 & 1.551 & 1.552 & 0.616 & 51.420 & 54.301 & 19.451 & 12.578 & 6.096 & \textbf{\underline{0.142}} \\ 
& SMAPE & 1.357 & 1.257 & 3.713 & 0.403 & 3.125 & \textbf{\textit{0.310}} & 4.300 & 1.743 & 1.563 & 1.565 & 0.618 & 40.470 & 42.692 & 17.725 & 11.834 & 6.288 & \textbf{\underline{0.142}} \\
 & MAE    & 0.098 & 0.091 & 0.266 & 0.030 & 0.224 & \textbf{\textit{0.023}} & 0.306 & 0.126 & 0.113 & 0.113 & 0.045 & 3.746 & 3.955 & 1.417 & 0.916 & 0.444 & \textbf{\underline{0.010}} \\ 
& MASE   & 3.358 & 3.113 & 9.080 & 1.009 & 7.651 & \textbf{\textit{0.771}} & 10.462 & 4.305 & 3.865 & 3.868 & 1.536 & 128.061 & 135.213 & 48.435 & 31.323 & 15.184 & \textbf{\underline{0.354}} \\ 
& RMSE   & 0.102 & 0.094 & 0.267 & 0.044 & 0.245 & \textbf{\textit{0.024}} & 0.330 & 0.130 & 0.116 & 0.114 & 0.051 & 3.868 & 3.958 & 1.417 & 0.916 & 0.445 & \textbf{\underline{0.011}} \\ 
\hline

6  & MAPE   & 4.287 & 4.177 & 6.301 & 3.028 & 5.704 & 4.105 & 4.066 & 4.631 & 4.258 & 3.842 & \textbf{\textit{1.648}} & 72.662 & 53.540 & 18.032 & 13.035 & 7.543 & \textbf{\underline{0.445}} \\ 
& SMAPE & 4.393 & 4.277 & 6.515 & 3.078 & 5.885 & 4.202 & 4.160 & 4.753 & 4.362 & 3.929 & \textbf{\textit{1.662}} & 53.191 & 42.181 & 16.533 & 12.229 & 7.852 & \textbf{\underline{0.445}} \\
& MAE    & 0.310 & 0.302 & 0.455 & 0.219 & 0.412 & 0.297 & 0.294 & 0.335 & 0.308 & 0.278 & \textbf{\textit{0.119}} & 5.226 & 3.849 & 1.296 & 0.937 & 0.544 & \textbf{\underline{0.032}} \\ 
& MASE   & 4.821 & 4.697 & 7.071 & 3.399 & 6.407 & 4.616 & 4.571 & 5.206 & 4.789 & 4.323 & \textbf{\textit{1.852}} & 81.232 & 59.819 & 20.150 & 14.560 & 8.462 & \textbf{\underline{0.495}} \\ 
 & RMSE   & 0.329 & 0.320 & 0.466 & 0.227 & 0.429 & 0.315 & 0.310 & 0.353 & 0.327 & 0.299 & \textbf{\textit{0.127}} & 5.250 & 3.859 & 1.299 & 0.941 & 0.557 & \textbf{\underline{0.039}} \\ 
\hline

12 & MAPE   & 2.525 & 2.849 & \textbf{\underline{2.104}} & 10.482 & 3.104 & 7.759 & 3.991 & 6.032 & \textbf{\textit{2.417}} & 5.424 & 3.393 & 65.403 & 62.941 & 37.394 & 21.503 & 6.537 & 2.561 \\ 
& SMAPE & 2.480 & 2.786 & \textbf{\underline{2.100}} & 9.916 & 3.107 & 7.440 & 3.872 & 5.828 & \textbf{\textit{2.386}} & 5.253 & 3.305 & 49.002 & 47.794 & 31.496 & 19.393 & 6.798 & 2.515 \\
& MAE    & 0.175 & 0.197 & \textbf{\underline{0.148}} & 0.739 & 0.220 & 0.545 & 0.277 & 0.422 & \textbf{\textit{0.168}} & 0.379 & 0.235 & 4.612 & 4.437 & 2.639 & 1.514 & 0.466 & 0.178 \\ 
& MASE   & 2.108 & 2.374 & \textbf{\underline{1.783}} & 8.891 & 2.642 & 6.553 & 3.328 & 5.080 & \textbf{\textit{2.026}} & 4.560 & 2.828 & 55.495 & 53.396 & 31.758 & 18.225 & 5.613 & 2.142 \\ 
& RMSE   & 0.221 & 0.250 & \textbf{\underline{0.158}} & 0.771 & 0.243 & 0.570 & 0.341 & 0.452 & \textbf{\textit{0.200}} & 0.413 & 0.294 & 4.678 & 4.451 & 2.641 & 1.521 & 0.508 & 0.212 \\ 
\hline

24 & MAPE   & 6.318 & 5.429 & 6.708 & 4.734 & \textbf{\textit{3.510}} & 7.290 & 5.386 & 6.909 & 6.226 & 8.356 & 7.349 & 40.017 & 53.031 & 23.797 & 23.402 & 5.306 & \textbf{\underline{1.782}} \\ 
 & SMAPE & 6.608 & 5.639 & 7.042 & 4.896 & \textbf{\textit{3.589}} & 7.686 & 5.597 & 7.264 & 5.931 & 8.864 & 6.971 & 33.087 & 41.582 & 21.217 & 20.921 & 5.467 & \textbf{\underline{1.780}} \\
 & MAE    & 0.445 & 0.382 & 0.473 & 0.333 & \textbf{\textit{0.246}} & 0.514 & 0.379 & 0.487 & 0.410 & 0.588 & 0.486 & 2.727 & 3.596 & 1.628 & 1.598 & 0.370 & \textbf{\underline{0.122}} \\ 
& MASE   & 5.613 & 4.820 & 5.963 & 4.200 & \textbf{\textit{3.105}} & 6.481 & 4.786 & 6.142 & 5.175 & 7.415 & 6.133 & 34.410 & 45.376 & 20.547 & 20.171 & 4.675 & \textbf{\underline{1.535}} \\ 
& RMSE   & 0.531 & 0.454 & 0.566 & 0.397 & \textbf{\textit{0.300}} & 0.615 & 0.454 & 0.582 & 0.513 & 0.693 & 0.583 & 2.797 & 3.656 & 1.651 & 1.610 & 0.427 & \textbf{\underline{0.150}} \\ 
\hline

48 & MAPE   & 5.337 & 5.529 & 4.571 & 14.801 & 4.306 & 5.800 & \textbf{\textit{3.698}} & 5.903 & 5.339 & 4.754 & 8.877 & 89.715 & 119.670 & 68.479 & 72.332 & 5.688 & \textbf{\underline{2.476}} \\ 
& SMAPE & 5.130 & 5.301 & 4.427 & 13.639 & 4.184 & 5.555 & \textbf{\textit{3.710}} & 5.649 & 5.132 & 4.615 & 8.399 & 61.242 & 74.691 & 50.910 & 53.004 & 5.944 & \textbf{\underline{2.480}} \\
& MAE    & 0.351 & 0.363 & 0.301 & 0.990 & 0.284 & 0.381 & \textbf{\textit{0.251}} & 0.388 & 0.351 & 0.314 & 0.588 & 6.067 & 8.084 & 4.627 & 4.888 & 0.398 & \textbf{\underline{0.169}} \\ 
& MASE   & 5.420 & 5.609 & 4.652 & 15.309 & 4.390 & 5.889 & \textbf{\textit{3.883}} & 5.993 & 5.423 & 4.856 & 9.086 & 93.786 & 124.956 & 71.519 & 75.559 & 6.147 & \textbf{\underline{2.610}} \\ 
& RMSE   & 0.432 & 0.450 & 0.375 & 1.056 & 0.349 & 0.468 & \textbf{\textit{0.298}} & 0.476 & 0.433 & 0.384 & 0.659 & 6.217 & 8.100 & 4.638 & 4.902 & 0.494 & \textbf{\underline{0.199}} \\   
\hline
\end{tabular}
\label{tab:performance_metrics_chn}
\end{adjustbox}
\end{table*}

\subsection{Robustness and Statistical Significance Tests}\label{Sec_stat_signif}
\begin{figure}[ht]
 \centering
  \includegraphics[width=0.85\textwidth, height = 11.5cm]{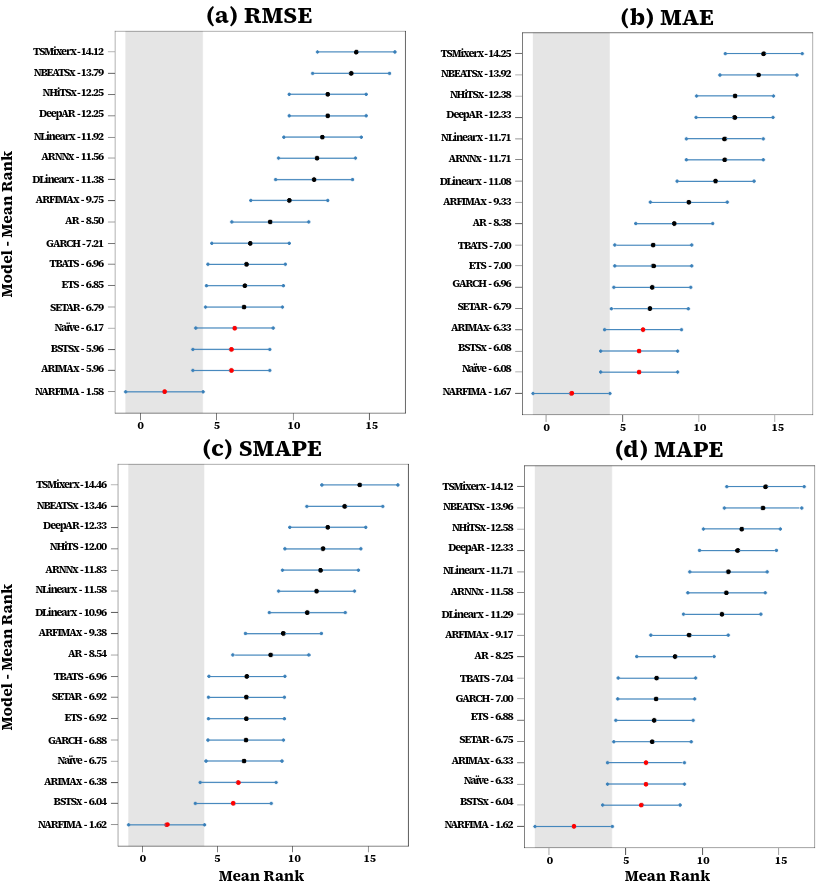}
  \caption{Multiple comparisons with the best (MCB) plot for BRIC nations based on (a) RMSE, (b) MAE, (c) SMAPE, and (d) MAPE metrics. In the plots, `$\text{NARFIMA-}1.58$' indicates that the average rank of the NARFIMA model is $1.58$, based on the RMSE metric. A similar interpretation holds across different models and metrics.}
\label{fig:MCB_test}
\end{figure}

In this section, we assess the robustness of our empirical results by evaluating the performance of different forecasting models based on differences in measurement errors, using multiple comparisons with the best (MCB) test. The model-agnostic MCB test is a nonparametric method that ranks each forecaster based on its performance across multiple datasets \citep{koning2005m3}. It then identifies the model with the lowest average rank as the best-performing framework and considers the critical distance of this model as the reference value for comparison. Fig.~\ref{fig:MCB_test} presents the MCB test results across different forecasting tasks based on RMSE, MAPE, SMAPE, and MAE metrics. The figure shows that the proposed NARFIMA model achieves the lowest average ranks across all metrics, making it the `best' performing model, followed by the ARIMAx, BSTSx, and Naïve frameworks in terms of the RMSE metric. The critical distance of the NARFIMA model (shaded area), serves as the reference value of the MCB test. Since the critical distance values of all baseline forecasters, except ARIMAx, BSTSx, and Naïve, lie well beyond this reference, their performance differs significantly from that of the best-performing NARFIMA model. Overall, the MCB test highlights the statistical significance of the performance differences, demonstrating the superiority of the NARFIMA model across various datasets and forecast horizons. 

One limitation of the MCB analysis is that it does not test for statistical equivalence among the top-ranked models. Therefore, 
we also employ the Model Confidence Set (MCS) procedure. The MCS framework constructs a subset of models that, with a given level of confidence, contains the best forecasting model among the initial set of competitors. The procedure sequentially eliminates models whose forecast accuracy is significantly inferior relative to the remaining alternatives and retains only those that are statistically indistinguishable from the best performer \citep{hansen2011model}. We implement the MCS procedure using squared loss and conduct the tests at three significance levels, specifically 0.05, 0.10, and 0.15. A detailed description of the MCS methodology and implementation is provided in Appendix~\ref{Appendix_MCS}. 
The results, summarized in Table~\ref{tab:MCS_transposed_final}, show that across all BRIC economies, NARFIMA remains in the optimal model set at every significance level. In contrast, the majority of baseline models are eliminated at one or more thresholds. For Brazil and China, every baseline model is eliminated at the 0.05 level. For Russia, all models are eliminated except for ARIMAx at the 0.05 level and BSTSx at the 0.10 level. In the case of India, BSTSx is retained across all three significance thresholds, while NLinearx remains in the confidence set until exclusion at the 0.15 level. No baseline model, however, retains inclusion across all three significance thresholds for all countries. This pattern of elimination, even among the strongest competing models, reinforces the robustness of NARFIMA's predictive superiority and complements the findings of the MCB analysis.

Motivated by these findings, we utilize Murphy diagrams to further evaluate the forecast dominance of the proposed NARFIMA model. While the MCB analysis provides a global ranking based on average performance and the MCS procedure assesses statistical equivalence, the Murphy diagram offers a complementary perspective by examining model utility across a comprehensive class of scoring functions \citep{ehm2016quanttiles}. This allows us to determine whether NARFIMA’s superiority is systematic or sensitive to the choice of loss function. For this analysis, we restrict the benchmarking to the top-performing baselines according to the MCB test, ARIMAx, and BSTSx. Although Naïve also ranked strongly in the MCB analysis, it is excluded because the MCS procedure eliminates it at the 0.05 level across all countries, indicating that its apparent performance lacks statistical robustness. ARIMAx and BSTSx, in contrast, are not uniformly eliminated and thus represent the only baseline competitors that warrant further comparison against the NARFIMA framework.

The Murphy diagram is constructed by computing the scoring function
\begin{equation}
\label{Eq:Theta}
\tilde{s}(\hat{y}_t, y_t) = 
    \begin{cases}
    |y_t - \theta| &  \min(\hat{y}_t, y_t) \leq \theta < \max(\hat{y}_t, y_t), \\
0 & \text{otherwise},
    \end{cases}
\end{equation}
where the parameter $\theta \in \mathcal{R}$ controls the shape of the loss function, $y_t$ represents the actual observation, and $\hat{y}_t$ is the point forecast generated by a model at time $t$. To compare the performance of two forecasting frameworks, this distribution-free method computes the average scores for each model as $\mathcal{S}_j(\theta) = 1/h \sum_{t = 1}^h \tilde{s}(\hat{y}_{j,t}, y_t)$, where $\hat{y}_{j,t}$ denotes the forecast generated by the $j^{th}$ model corresponding to ground truth observation $y_t$ and $h$ is the forecast horizon. The Murphy diagram then plots the extremal scores ($\mathcal{S}_j(\theta)$) for different models across a range of $\theta$ values. The parameter $\theta$, plays a crucial role in evaluating forecast accuracy, as different values emphasize different aspects of prediction error. Smaller values of $\theta$ make the scoring function more sensitive to underpredictions, penalizing models that consistently underestimate actual values, whereas higher values of $\theta$ penalize overpredictions more heavily. This adaptability makes the Murphy diagram a robust tool for comprehensive model evaluation, as it provides insights into how forecasting errors are distributed and whether a model consistently outperforms its competitors under different scoring functions.

\begin{figure}
    \centering
    \includegraphics[width=0.85\textwidth]{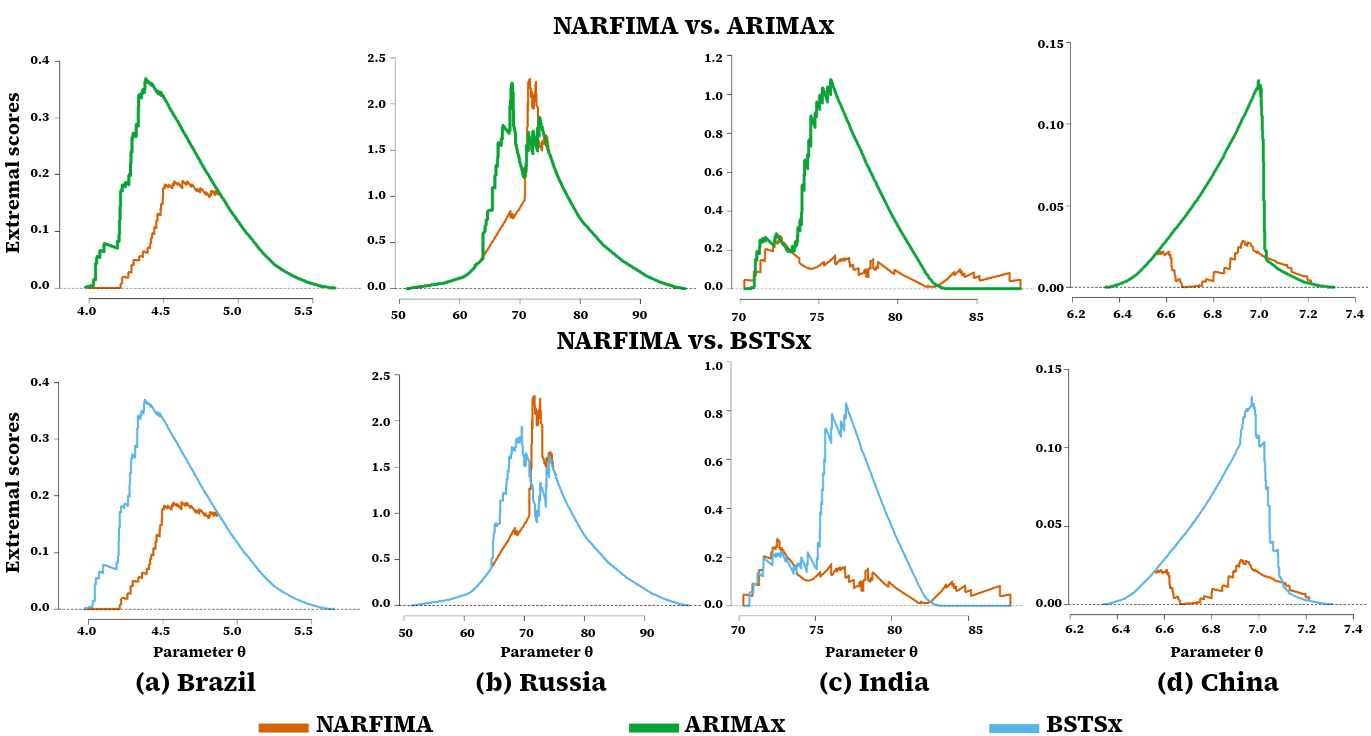} 
    \caption{Murphy diagrams of NARFIMA with baselines (ARIMAx (top) and BSTSx (bottom)) for the 48-month ahead exchange rate forecasting of (a) Brazil, (b) Russia, (c) India, and (d) China. The parameter $\theta$ represents the shape parameter as defined in Eqn.~\eqref{Eq:Theta}. Lower scores indicate better performance.}
    \label{fig:MurphyDiagram}
\end{figure}

To empirically validate the effectiveness of NARFIMA against benchmark models, we utilized the \texttt{murphydiagram} package in \textbf{R} for generating the Murphy diagrams. 
Fig.~\ref{fig:MurphyDiagram} presents the Murphy diagrams comparing NARFIMA with the ARIMAx and BSTSx frameworks, for each BRIC nation over a 48-month-ahead forecast horizon. The diagram plots the extremal scores for competing models, where a lower score indicates better model performance. The results reveal distinct patterns across different economies. In the case of Brazil, NARFIMA consistently outperforms both ARIMAx and BSTSx across all scoring functions, establishing its superiority in exchange rate forecasting. For Russia, the model provides similar or more accurate exchange rate forecasts when $\theta$ lies below 72 or above 75, suggesting that its dominance depends on specific error considerations. In India and China, NARFIMA remains competitive; however, sometimes ARIMAx and BSTSx outperform it. Overall, the combined evidence from multiple statistical assessments establishes NARFIMA as an accurate and reliable forecasting model across diverse economic conditions. Its superiority holds under formal statistical comparisons, alternative evaluation criteria, and across all BRIC economies and forecast horizons. 


\section{Uncertainty Quantification and Sensitivity Analysis}\label{uq}

\subsection{Conformal Prediction}\label{Sec_conf_pred}

\begin{figure}[ht]
 \centering
 \includegraphics[width=0.85\textwidth]{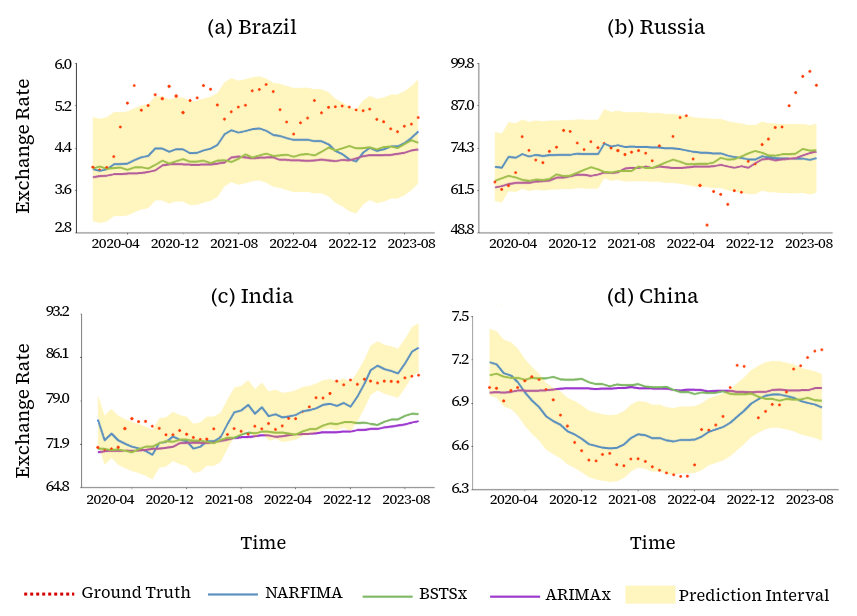}
 \caption{Visualization of the ground truth exchange rate observations (red dots) along with point forecasts from NARFIMA (blue line), BSTSx (green line), and ARIMAx (violet line), along with the conformal prediction interval for NARFIMA (yellow shaded region). Forecasts are for a 48-month-ahead horizon for (a) Brazil, (b) Russia, (c) India, and (d) China.}
 \label{fig:CP_48}
\end{figure}

Alongside the point forecasts of the exchange rate dynamics, we quantify the uncertainty associated with the NARFIMA model predictions using conformal prediction intervals. The distribution-free conformal prediction approach converts uncertainty scores to prediction intervals that contain the true outcome \citep{vovk2005algorithmic}. This model-agnostic method offers several advantages over simulation-based prediction intervals, including computational efficiency, fewer assumptions about the underlying data distribution, and guarantees coverage. In the context of time series setup, the conformal prediction framework utilizes the sequential ordering of the data to generate the prediction interval. Given the training set $\{\left(y_t, \tilde{x}_t\right)_{t=1}^T\}$, where $y_t$ represents the target series and $\tilde{x}_t$ indicates the set of features including lagged values of $y_t$, ARFIMAx residuals, and the exogenous variables, we apply the NARFIMA framework and an uncertainty model ($\widehat{\Psi}$) on $\tilde{x}_t$ to generate a measure of uncertainty. The conformal score $\tilde{S}_{t}$ is then calculated as:
$\tilde{S}_{t} = \frac{\left|y_{t} - \operatorname{NARFIMA}\left(\tilde{x}_{t} \right)\right|}{\widehat{\Psi}(\tilde{x}_{t})}.$
Due to the temporal dependencies in the series, the conformal quantiles, computed using a weighted conformal method with a fixed window of size $\tau$ defined as $\omega_{\tilde{t}} = \mathbb{1}\{ \tilde{t} \geq t - \tau\} \;\; \forall \; \tilde{t} < t$, are as follows:
\begin{equation*}
{\operatorname{CQ}}_{t} = inf\left\{ \tilde{q} : \ \frac{1}{\min\left( \tau,\tilde{t} - 1 \right) + 1}\sum_{\tilde{t} = 1}^{t - 1}{\tilde{S}_{\tilde{t}}\;\omega_{\tilde{t}}  \geq 1 - \alpha}\right\}.
\end{equation*}
Using these weight-adjusted quantiles, the conformal prediction interval at time step $t$ with $100(1-\alpha)$\% confidence is given by 
$ \left[\operatorname{NARFIMA}\left(\tilde{x}_{t} \right) \pm {\operatorname{CQ}}_{t} \; \widehat{\Psi}\left(\tilde{x}_{t} \right)\right].$ Fig.~\ref{fig:CP_48} represents the 80\% conformal prediction intervals of the NARFIMA framework for the long-term 48-month-ahead forecast horizon. The figure depicts the point forecasts of the exchange rate series generated by the NARFIMA model and the two best-performing benchmarks, namely ARIMAx and BSTSx, as identified in Section~\ref{Sec_stat_signif}. From the plots, it is evident that the proposed architecture can better capture the fluctuations in the exchange rate dynamics of the BRIC economies compared to the ARIMAx and BSTSx models. Furthermore, the width of conformal prediction intervals for the NARFIMA model varies across different countries, although it can still consistently capture most of the variations in the exchange rate datasets. However, for Brazil and Russia, both point forecasts and prediction intervals of NARFIMA and other leading models occasionally fail to capture the inherent dynamics of the exchange rate series, primarily due to abrupt macroeconomic fluctuations triggered by the COVID-19 pandemic and geopolitical conflicts, respectively. The overall analysis thus offers insights into uncertainties associated with the exchange rate forecasts for the BRIC economies.

\subsection{Sensitivity Analysis of Residual Selection}\label{Sec_res}
\begin{figure}[ht]
    \centering
    \includegraphics[width=0.85\textwidth]{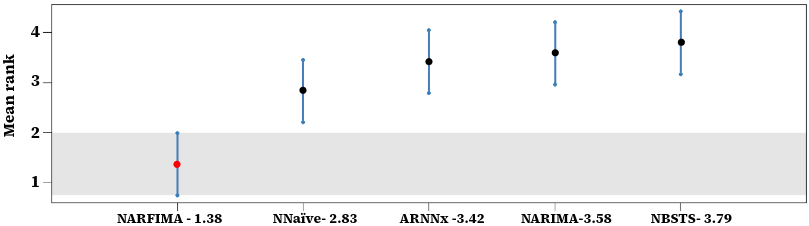}
    \caption{MCB plot comparing the performance of NARFIMA and its variants based on RMSE metrics. In the plot, `$\text{NARFIMA-} 1.38$' indicates that the average rank of NARFIMA is $1.38$, similar for other models.}
    \label{fig:MCB_Error}
\end{figure}

This section examines the sensitivity of the NARFIMA model to residual selection. The choice of feedback residuals plays a crucial role in designing the overall forecasting framework, as it directly influences how well long-term dependencies are captured. In the proposed NARFIMA architecture, long-memory modeling is incorporated into the neural network by utilizing the residuals of the ARFIMAx model, along with exchange rate series and macroeconomic covariates. To empirically validate the importance of ARFIMAx residuals, we conduct an iterative forecast evaluation in two ways. First, we replace them with residuals from ARIMAx, BSTSx, and Naïve, the three best-performing benchmark models identified in the MCB plot, while keeping the neural network structure unchanged. This results in three NARFIMA variants, namely Neural ARIMAx (NARIMA), Neural BSTSx (NBSTS), and Neural Naïve (NNaïve), respectively. Second, we examine a residual-free variant of NARFIMA, identical to ARNNx, where the neural network is trained solely on exchange rate series and macroeconomic covariates, removing residual feedback altogether. This empirical evaluation aims to assess how residual selection affects overall forecast accuracy. The MCB plot in Fig.~\ref{fig:MCB_Error} demonstrates the statistical significance of the performance improvement among NARFIMA and its variants in terms of the RMSE metric. As evident from the plot, NARFIMA achieves the lowest average rank across all BRIC economies and forecast horizons, further validating the superiority of our proposal over its variants. These findings confirm that the forecasting performance of NARFIMA relies on both the neural network and ARFIMAx residuals, which optimizes its modeling capabilities.

\section{Policy Implications}\label{Sec_Policy_Implications}

{Accurate forecasting of spot exchange rates is of immense importance for central banks, especially within emerging market economies such as the BRIC. Exchange rate movements substantially influence macroeconomic stability through multiple channels, including trade competitiveness, inflation dynamics, capital flows, and effectiveness of monetary policy. Given the varied exchange rate regimes among the BRIC economies, from managed floats to more flexible arrangements, precise forecasting is essential for mitigating external shocks, guiding foreign exchange interventions, and enhancing overall economic resilience. The recent acceleration of de-dollarization efforts in the BRIC nations, which aim to reduce dependence on the USD in trade, reserves, and settlement, has further intensified the complexity of exchange rate dynamics by introducing new drivers alongside traditional macroeconomic determinants. This shift toward greater currency diversification in trade and reserves increases the need for models that can simultaneously capture long-term dollar spillover effects and emerging local currency influences. In particular, BRIC countries deepen their integration within global trade and financial networks while simultaneously seeking greater monetary sovereignty. From the central banks' standpoint, improved accuracy in exchange rate forecasts enables enhanced management of foreign exchange reserves, smoother market interventions, and effective transmission of monetary policy. Reliable forecasts also equip policymakers to proactively address capital flow volatility, assess external vulnerabilities accurately, and adjust interest rates strategically to maintain economic stability. 

Given these insights, BRIC central banks must adapt their policy frameworks and exchange rate forecasting models to reflect de-dollarization dynamics. As BRIC economies increase trade settlement in local currencies, forecasting models must account for shifting demand patterns for BRIC currencies. The persistence of partial dollarization requires the models to incorporate both dollar-based and local currency-based trade flows, ensuring an accurate representation of exchange rate drivers. While the BRIC nations are reducing their reliance on dollar reserves, empirical data suggest that the USD still dominates global reserves. This means that exchange rate forecasting models must still incorporate US monetary policy spillovers, even as BRIC central banks increase holdings in gold, yuan, and other non-dollar assets. This aligns with empirical evidence underscoring the pivotal role of uncertainty measures in exchange rate forecasting, particularly within emerging markets pursuing de-dollarization. A recent study by \citealp{abid2020economic} confirms that heightened EPU induces short-run and long-run depreciation pressures on emerging market currencies, suggesting that integrating EPU into forecasting models becomes even more critical during transitions away from dollar dependence. Further, \citealp{christou2018role} highlights the predictive power of EPU on exchange rate volatility under extreme market conditions, underscoring its relevance in forecasting strategies during periods of monetary transition. The significant impact of US MPU on global currency volatility, as highlighted by \citealp{mueller2017exchange}, reinforces the necessity for the BRIC countries to incorporate MPU into predictive frameworks, particularly as they attempt to reduce vulnerability to Fed policy shifts through currency diversification. 

In this context, the proposed NARFIMA model provides a robust framework capable of dynamically capturing the complex influences of uncertainty measures, country-specific short-term IRD, and oil shocks on exchange rates. The model's capacity to incorporate these key causal drivers is particularly valuable as the BRIC economies navigate the transition toward de-dollarization, where currency valuations become increasingly sensitive to both global uncertainties and structural shifts in international monetary arrangements. Additionally, our proposed NARFIMA model can address nonlinearity, long memory, and structural breaks inherent in de-dollarization processes. Thus, it equips central banks with a powerful tool to anticipate exchange rate movements in an environment characterized by evolving currency preferences and settlement mechanisms. 
Finally, the study underscores the need for policy coordination among BRIC nations in the face of shared vulnerabilities to oil shocks and global uncertainty. 
}

\section{Conclusion and Discussion}\label{Section_Conclusion}
{This paper introduces a Neural ARFIMA model to accurately forecast the spot exchange rate dynamics in BRIC countries by taking into account various uncertainty measures, oil shocks, and country-specific short-term IRD. Our proposed approach effectively combines the memory properties of fractionally integrated processes with the flexible nonlinear mapping capabilities of simple neural networks, creating a powerful methodological solution for capturing both long-range dependencies and complex nonlinear patterns in exchange rate data. Our empirical results across BRIC economy exchange rates demonstrate that the proposed NARFIMA model consistently outperforms traditional statistical and state-of-the-art deep learning approaches across various forecasting horizons. The conditions for geometric ergodicity and asymptotic stationarity of the NARFIMA process are established by analyzing the asymptotic behavior of the associated Markov chain. Under appropriate parameter constraints on the skip connections and activation function, the NARFIMA process converges to a unique stationary distribution at a geometric rate, regardless of initial conditions. 
These properties provide strong guarantees on the model's long-run behavior and practical reliability,
validating its application to exchange rate forecasting problems where convergence and stability are essential properties. The nonlinear GC test uncovered the complex interplay between key macroeconomic drivers and exchange rate series. 
By focusing on spot exchange rates rather than real effective exchange rates, our framework targets the price directly faced by market participants in trade, investment, and policy operations, ensuring that the forecasts are immediately relevant for day-to-day decision-making in foreign exchange management. 

Our results contrast with the skepticism expressed in earlier influential studies (e.g., \citealp{diebold1990nonparametric}), which found little evidence of predictive gains from nonlinear modeling of exchange rates. By embedding long memory (also see \citealp{diebold1989long}) within a neural architecture and incorporating uncertainty measures and macroeconomic drivers, we find consistent forecasting improvements across horizons for BRIC nations. This suggests a more optimistic conclusion: with appropriately structured models, the nonlinear and persistent features of exchange rate dynamics can indeed be harnessed for improved predictability, offering new insights for both researchers and policymakers. The NARFIMA model, therefore, offers significant practical value for central banks and financial institutions engaged in monetary policy formulation and foreign exchange operations, particularly during periods of heightened uncertainty. While our current implementation focuses on the BRIC economies and incorporates several key uncertainty measures, future research could extend the NARFIMA framework to include additional factors such as climate risk indicators and social media-based uncertainty measures derived from platforms like Twitter (X). 
For future research, NARFIMA can be adapted to a multivariate setting to forecast other financial indicators characterized by long memory and nonlinear dynamics, including interest rates, commodity prices, and equity market volatility. 
Although the NARFIMA model is primarily developed for long memory and nonlinear time series problems arising in BRIC exchange rate series, it can also be useful for similar complex temporal data problems of epidemics, business, and climatology.}

\section*{Acknowledgement}
The authors would like to acknowledge Professor Francis X. Diebold for providing insightful suggestions on the first version of this manuscript, which significantly improved the paper.

\section*{Competing interests}
There are no competing interests to be declared.

\section*{Data and Code Availability Statement}
The spot exchange rate, along with all macroeconomic indicators used in this analysis, is obtained from the Federal Reserve Economic Data (FRED) repository: \url{https://fred.stlouisfed.org}. Data for all uncertainty indicators are sourced from the Economic Policy Uncertainty website: \url{https://www.policyuncertainty.com}. The code and data necessary to reproduce the results of this study are available at: \url{https://github.com/mad-stat/NARFIMA}. We developed an \textbf{R} package, \texttt{narfima}, that implements our approach.

\bibliography{refs}

@article{chakraborty2020unemployment,
journal={Computational Economics},
author={Tanujit Chakraborty and Ashis Kumar Chakraborty and Munmun Biswas and Sayak Banerjee and Shramana Bhattacharya},
title={Unemployment Rate Forecasting: A Hybrid Approach},
year={2021},
pages={183-201},
volume={57},
publisher={Springer}
}

@book{meyn2012markov,
  title={Markov chains and stochastic stability},
  author={Meyn, Sean P and Tweedie, Richard L},
  year={2012},
  publisher={Springer Science \& Business Media}
}

@article{diebold1990nonparametric,
  title={Nonparametric exchange rate prediction?},
  author={Diebold, Francis X and Nason, James A},
  journal={Journal of International Economics},
  volume={28},
  number={3-4},
  pages={315--332},
  year={1990},
  publisher={Elsevier}
}

@article{diebold1989long,
  title={Long memory and persistence in aggregate output},
  author={Diebold, Francis X and Rudebusch, Glenn D},
  journal={Journal of Monetary Economics},
  volume={24},
  number={2},
  pages={189--209},
  year={1989},
  publisher={Elsevier}
}

@techreport{bernanke1994financial,
 title = "The Financial Accelerator and the Flight to Quality",
 author = "Bernanke, Ben and Gertler, Mark and Gilchrist, Simon",
 institution = "National Bureau of Economic Research",
 type = "Working Paper",
 series = "Working Paper Series",
 number = "4789",
 year = "1994"
}

@techreport{calvo2005sudden,
 title = "Sudden Stop, Financial Factors and Economic Collpase in {Latin America}: Learning from {Argentina} and {Chile}",
 author = "Calvo, Guillermo A and Talvi, Ernesto",
 institution = "National Bureau of Economic Research",
 type = "Working Paper",
 series = "Working Paper Series",
 number = "11153",
 year = "2005"
}

@article{eichengreen1998exchange,
  title={Exchange rate stability and financial stability},
  author={Eichengreen, Barry},
  journal={Open Economies Review},
  volume={9},
  number={1},
  pages={569-608},
  year={1998},
  publisher={Springer}
}

@TechReport{hausmann1999financial,
type={IDB Publications (Working Papers)},
institution={Inter-American Development Bank},
author={Hausmann, Ricardo and Gavin, Michael and Pagés, Carmen and Stein, Ernesto H.},
title={Financial Turmoil and the Choice of Exchange Rate Regime},
year={1999},
number={1108}
}

@article{hopewell2017brics,
 author = {Hopewell, Kristen},
 journal = {International Affairs},
 number = {6},
 pages = {1377-1396},
 publisher = {[Royal Institute of International Affairs, Oxford University Press]},
 title = {The {BRICS}—merely a fable? {Emerging} power alliances in global trade governance},
 volume = {93},
 year = {2017}
}

@article{lubik2007central,
  title={Do central banks respond to exchange rate movements? {A} structural investigation},
  author={Lubik, Thomas A and Schorfheide, Frank},
  journal={Journal of Monetary Economics},
  volume={54},
  number={4},
  pages={1069-1087},
  year={2007},
  publisher={Elsevier}
}

@article{nasir2018implications,
  title={Implications of oil prices shocks for the major emerging economies: {A} comparative analysis of {BRICS}},
  author={Muhammad Ali Nasir and Lutchmee Naidoo and Muhammad Shahbaz and Nii Amoo},
  journal={Energy Economics},
  volume={76},
  pages={76-88},
  year={2018},
  publisher={Elsevier}
}

@article{stoica2016exchange,
  title={Exchange rate regimes and external financial stability},
  author={Stoica, Ovidiu and Ihnatov, Iulian},
  journal={Economic Annals},
  volume={61},
  number={209},
  pages={27-43},
  year={2016}
}

@article{taylor2001role,
  title={The role of the exchange rate in monetary-policy rules},
  author={Taylor, John B},
  journal={American Economic Review},
  volume={91},
  number={2},
  pages={263-267},
  year={2001}
}

@book{o2011growth,
  title={{The growth map: Economic opportunity in the BRICs and beyond}},
  author={O'Neill, Jim},
  year={2011},
  publisher={Penguin UK}
}

@book{hyndman2018forecasting,
  title={Forecasting: principles and practice},
  author={Hyndman, Rob J and Athanasopoulos, George},
  year={2018},
  publisher={OTexts}
}

@article{rumelhart1986learning,
  title={Learning representations by back-propagating errors},
  author={Rumelhart, David E and Hinton, Geoffrey E and Williams, Ronald J},
  journal={Nature},
  volume={323},
  number={6088},
  pages={533-536},
  year={1986},
  publisher={Nature Publishing Group UK London}
}

@article{box1970distribution,
 author = {G. E. P. Box and David A. Pierce},
 journal = {Journal of the American Statistical Association},
 number = {332},
 pages = {1509-1526},
 publisher = {[American Statistical Association, Taylor & Francis, Ltd.]},
 title = {Distribution of Residual Autocorrelations in Autoregressive Integrated Moving Average Time Series Models},
 volume = {65},
 year = {1970}
}

@article{salinas2020deepar,
  title={{DeepAR}: Probabilistic forecasting with autoregressive recurrent networks},
  author={Salinas, David and Flunkert, Valentin and Gasthaus, Jan and Januschowski, Tim},
  journal={International Journal of Forecasting},
  volume={36},
  number={3},
  pages={1181-1191},
  year={2020},
  publisher={Elsevier}
}

@article{faraway1998time,
  title={Time series forecasting with neural networks: {A} comparative study using the air line data},
  author={Faraway, Julian and Chatfield, Chris},
  journal={Journal of the Royal Statistical Society Series C: Applied Statistics},
  volume={47},
  number={2},
  pages={231-250},
  year={1998},
  publisher={Oxford University Press}
}

@article{koning2005m3,
  title={The {M3} competition: Statistical tests of the results},
  author={Koning, Alex J and Franses, Philip Hans and Hibon, Michele and Stekler, Herman O},
  journal={International Journal of Forecasting},
  volume={21},
  number={3},
  pages={397-409},
  year={2005},
  publisher={Elsevier}
}

@book{vovk2005algorithmic,
author = {Vovk, Vladimir and Gammerman, Alex and Shafer, Glenn},
title = {Algorithmic Learning in a Random World},
year = {2005},
publisher = {Springer}
}

@inproceedings{Oreshkin2020N-BEATS:,
title={{N-BEATS}: Neural basis expansion analysis for interpretable time series forecasting},
author={Boris N. Oreshkin and Dmitri Carpov and Nicolas Chapados and Yoshua Bengio},
booktitle={International Conference on Learning Representations},
year={2020}
}

@article{panja2023epicasting,
author = {Panja, Madhurima and Chakraborty, Tanujit and Kumar, Uttam and Liu, Nan},
year = {2023},
pages = {185–212},
title = {Epicasting: An Ensemble Wavelet Neural Network for forecasting epidemics},
volume = {165},
publisher={Elsevier BV},
journal = {Neural Networks}
}

@article{baker2016measuring,
  title={Measuring economic policy uncertainty},
  author={Baker, Scott R. and Bloom, Nicholas and Davis, Steven J.},
  journal={The Quarterly Journal of Economics},
  volume={131},
  number={4},
  pages={1593-1636},
  year={2016},
  publisher={Oxford University Press}
}

@article{granger1980introduction,
  title={An introduction to long-memory time series models and fractional differencing},
  author={Granger, Clive WJ and Joyeux, Roselyne},
  journal={Journal of Time Series Analysis},
  volume={1},
  number={1},
  pages={15-29},
  year={1980},
  publisher={Wiley Online Library}
}

@article{caldara2022measuring,
  title={Measuring geopolitical risk},
  author={Caldara, Dario and Iacoviello, Matteo},
  journal={American Economic Review},
  volume={112},
  number={4},
  pages={1194-1225},
  year={2022},
  publisher={American Economic Association 2014 Broadway, Suite 305, Nashville, TN 37203}
}

@book{weisberg1982residuals,
  title={Residuals and influence in regression},
  author={R. Dennis Cook and Sandford Weisberg},
  year={1982},
  publisher={Chapman \& Hall}
}

@article{ploberger1992cusum,
 author = {Werner Ploberger and Walter Krämer},
 journal = {Econometrica},
 number = {2},
 pages = {271-285},
 publisher = {[Wiley, Econometric Society]},
 title = {{The CUSUM test with OLS residuals}},
 volume = {60},
 year = {1992}
}

@article{granger1969investigating,
 author = {C. W. J. Granger},
 journal = {Econometrica},
 number = {3},
 pages = {424-438},
 publisher = {[Wiley, Econometric Society]},
 title = {Investigating Causal Relations by Econometric Models and Cross-spectral Methods},
 volume = {37},
 year = {1969}
}

@book{franses2000non,
  title={Non-linear time series models in empirical finance},
  author={Franses, Philip Hans and Van Dijk, Dick},
  year={2000},
  publisher={Cambridge University Press}
}

@article{firat2017setar,
  title={{SETAR (Self-Exciting Threshold Autoregressive)} non-linear currency Modelling in {EUR/USD}, {EUR/TRY} and {USD/TRY} parities},
  author={Firat, Emrah Hanifi},
  journal={Mathematics and Statistics},
  volume={5},
  number={1},
  pages={33-55},
  year={2017}
}

@article{bollerslev1986generalized,
  title={Generalized autoregressive conditional heteroskedasticity},
  author={Bollerslev, Tim},
  journal={Journal of Econometrics},
  volume={31},
  number={3},
  pages={307-327},
  year={1986},
  publisher={Elsevier}
}

@article{scott2014predicting,
  title={Predicting the present with {Bayesian} structural time series},
  author={Scott, Steven L and Varian, Hal R},
  journal={International Journal of Mathematical Modelling and Numerical Optimisation},
  volume={5},
  number={1-2},
  pages={4-23},
  year={2014},
  publisher={Inderscience Publishers Ltd}
}

@article{challu2023nhits,
  title={{NHITS}: {Neural Hierarchical Interpolation for Time Series Forecasting}},
  author={Challu, Cristian and Olivares, Kin G and Oreshkin, Boris N and Ramirez, Federico Garza and Canseco, Max Mergenthaler and Dubrawski, Artur},
  journal={Proceedings of the AAAI Conference on Artificial Intelligence},
  volume={37},
  number={6},
  pages={6989-6997},
  year={2023}
}

@misc{chen2303tsmixer,
  title={{TSM}ixer: An All-{MLP} Architecture for Time Series Forecasting},
  author={Chen, Si An and Li, Chun Liang and Arik, Sercan O and Yoder, Nathanael Christian and Pfister, Tomas},
  year={2023},
  note={Transactions on Machine Learning Research}
}

@misc{huang2023nonlinearity,
  title={A nonlinearity and model specification test for functional time series}, 
  author={Xin Huang and Han Lin Shang and Tak Kuen Siu},
  year={2023},
  howpublished={arXiv:2304.01558}
}

@article{prabowo2020performance,
  title={The performance of ramsey test, white test and Terasvirta test in detecting nonlinearity},
  author={Prabowo, Hendri and Suhartono, Suhartono and Prastyo, Dedy Dwi},
  journal={Inferensi},
  volume={3},
  number={1},
  pages={1-12},
  year={2020}
}

@article{SENGUPTA2025953,
  title   = {Forecasting {CPI} inflation under economic policy and geopolitical uncertainties},
  author  = {Sengupta, Shovon and Chakraborty, Tanujit and Singh, Sunny Kumar},
  journal = {International Journal of Forecasting},
  year    = {2025},
  volume  = {41},
  number  = {3},
  pages   = {953-981},
  publisher = {Elsevier}
}

@article{abid2020economic,
  title={Economic policy uncertainty and exchange rates in emerging markets: Short and long runs evidence},
  author={Abid, Abir},
  journal={Finance Research Letters},
  volume={37},
  pages={101378},
  year={2020},
  publisher={Elsevier}
}

@article{mueller2017exchange,
  title={Exchange rates and monetary policy uncertainty},
  author={Mueller, Philippe and Tahbaz-Salehi, Alireza and Vedolin, Andrea},
  journal={The Journal of Finance},
  volume={72},
  number={3},
  pages={1213-1252},
  year={2017},
  publisher={Wiley Online Library}
}

@article{christou2018role,
  title={The role of economic uncertainty in forecasting exchange rate returns and realized volatility: Evidence from quantile predictive regressions},
  author={Christou, Christina and Gupta, Rangan and Hassapis, Christis and Suleman, Tahir},
  journal={Journal of Forecasting},
  volume={37},
  number={7},
  pages={705-719},
  year={2018},
  publisher={Wiley Online Library}
}

@article{bartsch2019economic,
  title={Economic policy uncertainty and dollar-pound exchange rate return volatility},
  author={Bartsch, Zachary},
  journal={Journal of International Money and Finance},
  volume={98},
  pages={102067},
  year={2019},
  publisher={Elsevier}
}

@article{plakandaras2015forecasting,
  title={Forecasting daily and monthly exchange rates with machine learning techniques},
  author={Plakandaras, Vasilios and Papadimitriou, Theophilos and Gogas, Periklis},
  journal={Journal of Forecasting},
  volume={34},
  number={7},
  pages={560-573},
  year={2015},
  publisher={Wiley Online Library}
}

@article{ehm2016quanttiles,
    author = {Ehm, Werner and Gneiting, Tilmann and Jordan, Alexander and Krüger, Fabian},
    title = {Of Quantiles and Expectiles: Consistent Scoring Functions, Choquet Representations and Forecast Rankings},
    journal = {Journal of the Royal Statistical Society Series B: Statistical Methodology},
    volume = {78},
    number = {3},
    pages = {505-562},
    year = {2016},
    month = {05}
}

@article{hiemstra1994testing,
  title={Testing for linear and nonlinear Granger causality in the stock price-volume relation},
  author={Hiemstra, Craig and Jones, Jonathan D},
  journal={The Journal of Finance},
  volume={49},
  number={5},
  pages={1639-1664},
  year={1994},
  publisher = {Wiley Online Library}
}

@article{karemera2006assessing,
  title={Assessing the forecasting accuracy of alternative nominal exchange rate models: {The} case of long memory},
  author={Karemera, David and Kim, Benjamin JC},
  journal={Journal of Forecasting},
  volume={25},
  number={5},
  pages={369-380},
  year={2006},
  publisher={Wiley Online Library}
}

@article{ngan2013forecasting,
  title={{Forecasting foreign exchange rate by using ARIMA model: A case of VND/USD exchange rate}},
  author={Tran Mong Uyen Ngan},
  journal={Research Journal of Finance and Accounting},
  year={2016},
  volume={7},
  pages={38-44}
}

@article{galeshchuk2016neural,
  title={Neural networks performance in exchange rate prediction},
  author={Galeshchuk, Svitlana},
  journal={Neurocomputing},
  volume={172},
  pages={446-452},
  year={2016},
  publisher={Elsevier}
}

@techreport{davis2016index,
 title = "An Index of Global Economic Policy Uncertainty",
 author = "Davis, Steven J",
 institution = "National Bureau of Economic Research",
 type = "Working Paper",
 series = "Working Paper Series",
 number = "22740",
 year = "2016"
}

@techreport{baker2019policy,
 title = "Policy News and Stock Market Volatility",
 author = "Baker, Scott R and Bloom, Nicholas and Davis, Steven J and Kost, Kyle J",
 institution = "National Bureau of Economic Research",
 type = "Working Paper",
 series = "Working Paper Series",
 number = "25720",
 year = "2019"
}

@article{husted2020monetary,
  title={Monetary policy uncertainty},
  author={Husted, Lucas and Rogers, John and Sun, Bo},
  journal={Journal of Monetary Economics},
  volume={115},
  pages={20-36},
  year={2020},
  publisher={Elsevier}
}

@article{zhou2020can,
  title = {Can economic policy uncertainty predict exchange rate volatility? {New} evidence from the {GARCH-MIDAS} model},
  author = {Zhou, Zhongbao and Fu, Zhangyan and Jiang, Yong and Zeng, Ximei and Lin, Ling},
  journal = {Finance Research Letters},
  volume = {34},
  pages = {101258},
  year = {2020},
  publisher = {Elsevier}
}

@article{meese1983empirical,
  title={Empirical exchange rate models of the seventies: Do they fit out of sample?},
  author={Meese, Richard A and Rogoff, Kenneth},
  journal={Journal of International Economics},
  volume={14},
  number={1-2},
  pages={3-24},
  year={1983},
  publisher={Elsevier}
}

@article{rossi2013exchange,
  title={Exchange rate predictability},
  author={Rossi, Barbara},
  journal={Journal of Economic Literature},
  volume={51},
  number={4},
  pages={1063-1119},
  year={2013},
  publisher={American Economic Association 2014 Broadway, Suite 305, Nashville, TN 37203}
}

@article{pilbeam2015forecasting,
  title={Forecasting exchange rate volatility: {GARCH} models versus implied volatility forecasts},
  author={Pilbeam, Keith and Langeland, Kjell Noralf},
  journal={International Economics and Economic Policy},
  volume={12},
  pages={127-142},
  year={2015},
  publisher={Springer}
}

@article{balcilar2016does,
  title={Does economic policy uncertainty predict exchange rate returns and volatility? {Evidence} from a nonparametric causality-in-quantiles test},
  author={Balcilar, Mehmet and Gupta, Rangan and Kyei, Clement and Wohar, Mark E},
  journal={Open Economies Review},
  volume={27},
  pages={229-250},
  year={2016},
  publisher={Springer}
}

@article{benigno2012risk,
  title={Risk, monetary policy, and the exchange rate},
  author={Gianluca Benigno and Pierpaolo Benigno and Salvatore Nistico},
  journal={National Bureau of Economic Research Macroeconomics Annual},
  volume={26},
  number={1},
  pages={247--309},
  year={2012},
  publisher={University of Chicago Press Chicago, IL}
}

@article{colombo2013economic,
  title={Economic policy uncertainty in the {US}: Does it matter for the {Euro} area?},
  author={Colombo, Valentina},
  journal={Economics Letters},
  volume={121},
  number={1},
  pages={39-42},
  year={2013},
  publisher={Elsevier}
}

@article{juhro2018can,
  title={{Can Economic Policy Uncertainty Predict Exchange Rate and its Volatility? Evidence from ASEAN Countries}},
  author={Juhro, Solikin M and Phan, Dinh Hoang Bach},
  journal={Bulletin of Monetary Economics and Banking},
  volume={21},
  number={2},
  pages={251-268},
  year={2018},
  publisher={Bank Indonesia, Central Banking Research Department}
}

@article{sin2015economic,
  title={The economic fundamental and economic policy uncertainty of {Mainland China} and their impacts on {Taiwan} and {Hong Kong}},
  author={Sin, Chor Yiu},
  journal={International Review of Economics \& Finance},
  volume={40},
  pages={298-311},
  year={2015},
  publisher={Elsevier}
}

@article{istrefi2018subjective,
  title={Subjective interest rate uncertainty and the macroeconomy: {A} cross-country analysis},
  author={Istrefi, Klodiana and Mouabbi, Sarah},
  journal={Journal of International Money and Finance},
  volume={88},
  pages={296-313},
  year={2018},
  publisher={Elsevier}
}

@incollection{wieland2013forecasting,
title = {{Forecasting and Policy Making}},
editor = {Graham Elliott and Allan Timmermann},
series = {Handbook of Economic Forecasting},
publisher = {Elsevier},
volume = {2},
pages = {239-325},
year = {2013},
booktitle = {Handbook of Economic Forecasting},
author = {Volker Wieland and Maik Wolters},
}

@article{molodtsova2009out,
  title={Out-of-sample exchange rate predictability with Taylor rule fundamentals},
  author={Molodtsova, Tanya and Papell, David H},
  journal={Journal of International Economics},
  volume={77},
  number={2},
  pages={167-180},
  year={2009},
  publisher={Elsevier}
}

@article{kilian2003so,
  title={Why is it so difficult to beat the random walk forecast of exchange rates?},
  author={Kilian, Lutz and Taylor, Mark P},
  journal={Journal of International Economics},
  volume={60},
  number={1},
  pages={85-107},
  year={2003},
  publisher={Elsevier}
}

@article{kumar2024bayesian,
  title={Bayesian Markov switching model for {BRICS} currencies' exchange rates},
  author={Kumar, Utkarsh and Ahmad, Wasim and Uddin, Gazi Salah},
  journal={Journal of Forecasting},
  volume={43},
  number={6},
  pages={2322-2340},
  year={2024},
  publisher={Wiley Online Library}
}

@article{salisu2022exchange,
  title={Exchange rate predictability with nine alternative models for {BRICS} countries},
  author={Salisu, Afees A and Gupta, Rangan and Kim, Won Joong},
  journal={Journal of Macroeconomics},
  volume={71},
  pages={103374},
  year={2022},
  publisher={Elsevier}
}

@article{abir2024use,
author = {Abir, Shake and Shiam, Sarder and Zakaria, Rafi and Shimanto, Abid Hasan and Arefeen, S M Shamsul and Dolon, Md and Sultana, Nigar and Shoha, Shaharina},
year = {2024},
month = {12},
pages = {66-83},
title = {Use of {AI}-Powered Precision in Machine Learning Models for Real-Time Currency Exchange Rate Forecasting in {BRICS} Economies},
volume = {6},
journal = {Journal of Economics, Finance and Accounting Studies}
}

@article{date2025modelling,
  title={Modelling and forecasting of exchange rate pairs using the Kalman filter},
  author={Date, Paresh and Maunthrooa, Janeeta},
  journal={Journal of Forecasting},
  volume={44},
  number={2},
  pages={606-622},
  year={2025},
  publisher={Wiley Online Library}
}

@article{beckmann2020relationship,
  title={The relationship between oil prices and exchange rates: Revisiting theory and evidence},
  author={Beckmann, Joscha and Czudaj, Robert L and Arora, Vipin},
  journal={Energy Economics},
  volume={88},
  pages={104772},
  year={2020},
  publisher={Elsevier}
}

@article{chen2024dynamic,
  title={Dynamic analysis of the relationship between exchange rates and oil prices: {A} comparison between oil exporting and oil importing countries},
  author={Chen, Shiying and Chang, Bisharat Hussain and Fu, Hu and Xie, ShiQi},
  journal={Humanities and Social Sciences Communications},
  volume={11},
  number={1},
  pages={1-12},
  year={2024},
  publisher={Palgrave}
}

@article{andrieș2017relationship,
  title={The relationship between exchange rates and interest rates in a small open emerging economy: The case of {Romania}},
  author={Andrieș, Alin Marius and C{\u{a}}praru, Bogdan and Ihnatov, Iulian and Tiwari, Aviral Kumar},
  journal={Economic Modelling},
  volume={67},
  pages={261-274},
  year={2017},
  publisher={Elsevier}
}

@article{saracc2016impact,
  title={Impact of short-term interest rate on exchange rate: The case of {Turkey}},
  author={Sara{\c{c}}, Taha Bahad{\i}r and Karag{\"o}z, Kadir},
  journal={Procedia Economics and Finance},
  volume={38},
  pages={195-202},
  year={2016},
  publisher={Elsevier}
}

@techreport{trapletti1999ergodicity,
    author =  {Trapletti, Adrian and Leisch, Friedrich and Hornik, Kurt},
    title = {On the Ergodicity and Stationarity of the {ARMA}(1,1) Recurrent Neural Network Process},
    institution = {SFB Adaptive Information Systems and Modelling in Economics and Management Science, WU Vienna University of Economics and Business},
    year = {1999}
}

@article{trapletti2000stationary,
  title={Stationary and integrated autoregressive neural network processes},
  author={Trapletti, Alessandro and Leisch, Friedrich and Hornik, Kurt},
  journal={Neural Computation},
  volume={12},
  number={10},
  pages={2427--2450},
  year={2000},
  publisher={MIT Press}
}

@article{xu2018note,
journal={Journal of Theoretical Probability},
author={Lifeng Xu and Shaoyi Zhang and Ren Zhang},
title={A Note on {Foster–Lyapunov} Drift Condition for Recurrence of Markov Chains on General State Spaces},
year={2018},
pages={1923-1928},
volume={31},
number={4}
}

@inproceedings{zeng2023transformers,
author = {Zeng, Ailing and Chen, Muxi and Zhang, Lei and Xu, Qiang},
title = {Are transformers effective for time series forecasting?},
year = {2023},
publisher = {AAAI Press},
booktitle = {Proceedings of the Thirty-Seventh AAAI Conference on Artificial Intelligence and Thirty-Fifth Conference on Innovative Applications of Artificial Intelligence and Thirteenth Symposium on Educational Advances in Artificial Intelligence},
articleno = {1248},
numpages = {8}
}

@book{whittle1951hypothesis,
  title={Hypothesis Testing in Time Series Analysis},
  author={Whittle, Peter},
  series={Statistics Upsala},
  year={1951},
  publisher={Almqvist \& Wiksells}
}

@book{beran2017statistics,
  title={{Statistics for Long-Memory Processes}},
  author={Beran, Jan},
  year={2017},
  publisher={Routledge},
  edition   = {1st}
}

@article{hansen2011model,
author = {Hansen, Peter R. and Lunde, Asger and Nason, James M.},
title = {The Model Confidence Set},
journal = {Econometrica},
volume = {79},
number = {2},
pages = {453-497},
year = {2011}
}

@article{zivot1992test,
 author = {Eric Zivot and Donald W. K. Andrews},
 journal = {Journal of Business \& Economic Statistics},
 number = {3},
 pages = {251--270},
 publisher = {[American Statistical Association]},
 title = {{Further Evidence on the Great Crash, the Oil-Price Shock, and the Unit-Root Hypothesis}},
 volume = {10},
 year = {1992}
}

@article{Hosking1981fractional,
 author = {J. R. M. Hosking},
 journal = {Biometrika},
 number = {1},
 pages = {165--176},
 publisher = {[Oxford University Press, Biometrika Trust]},
 title = {{Fractional Differencing}},
 volume = {68},
 year = {1981}
}

@book{doukhan2002theory,
  title={{Theory and Applications of Long-Range Dependence}},
  author={Doukhan, P. and Oppenheim, G. and Taqqu, M.},
  year={2002},
  publisher={Birkh{\"a}user Boston}
}

@InProceedings{cont2005financial,
author="Cont, Rama",
editor="L{\'e}vy-V{\'e}hel, Jacques
and Lutton, Evelyne",
title="Long range dependence in financial markets",
booktitle="Fractals in Engineering",
year="2005",
publisher="Springer London",
address="London",
pages="159--179"
}

@article{kilian2003exchange,
title = {{Why is it so difficult to beat the random walk forecast of exchange rates?}},
journal = {Journal of International Economics},
volume = {60},
number = {1},
pages = {85-107},
year = {2003},
note = {Emperical Exchange Rate Models},
author = {Lutz Kilian and Mark P. Taylor}
}

@article{fernandes1998nonlinearity,
author = {Fernandes, Marcelo},
title = {Non-linearity and exchange rates},
journal = {Journal of Forecasting},
volume = {17},
number = {7},
pages = {497-514},
year = {1998}
}

@article{lee2013nonlinear,
title = {{The behavior of real exchange rate: Nonlinearity and breaks}},
journal = {International Review of Economics \& Finance},
volume = {27},
pages = {125-133},
year = {2013},
author = {Chia-Hao Lee and Pei-I Chou}
}

@ARTICLE{hochreiter1997lstm,
  author={Hochreiter, Sepp and Schmidhuber, Jürgen},
  journal={Neural Computation}, 
  title={{Long Short-Term Memory}}, 
  year={1997},
  volume={9},
  number={8},
  pages={1735-1780}
  }

@InProceedings{zhao2020memory,
  title = 	 {{Do {RNN} and {LSTM} have Long Memory?}},
  author =       {Zhao, Jingyu and Huang, Feiqing and Lv, Jia and Duan, Yanjie and Qin, Zhen and Li, Guodong and Tian, Guangjian},
  booktitle = 	 {Proceedings of the 37th International Conference on Machine Learning},
  pages = 	 {11365--11375},
  year = 	 {2020},
  editor = 	 {III, Hal Daumé and Singh, Aarti},
  volume = 	 {119},
  series = 	 {Proceedings of Machine Learning Research},
  month = 	 {13--18 Jul},
  publisher =    {PMLR}
}

@article{mun2008emv,
author = {Mun, Kyung-Chun},
year = {2008},
month = {01},
pages = {
77-102},
title = {{Effects of exchange rate fluctuations on equity market volatility and correlations: Evidence from the Asian financial crisis}},
volume = {47},
journal = {Quarterly Journal ofFinance and Accounting}
}

@article{baillie1996memory,
title = {Long memory processes and fractional integration in econometrics},
journal = {Journal of Econometrics},
volume = {73},
number = {1},
pages = {5-59},
year = {1996},
author = {Richard T. Baillie}
}

@article{hornik1989approximators,
title = {Multilayer feedforward networks are universal approximators},
journal = {Neural Networks},
volume = {2},
number = {5},
pages = {359-366},
year = {1989},
author = {Kurt Hornik and Maxwell Stinchcombe and Halbert White}
}

@INPROCEEDINGS{bengio1993dependence,
  author={Bengio, Y. and Frasconi, P. and Simard, P.},
  booktitle={IEEE International Conference on Neural Networks}, 
  title={The problem of learning long-term dependencies in recurrent networks}, 
  year={1993},
  volume={},
  number={},
  pages={1183-1188 vol.3}
  }

@article{li2020deeplearning,
author = {Li, Zhenwei and Han, Jing and Song, Yuping},
title = {{On the forecasting of high-frequency financial time series based on ARIMA model improved by deep learning}},
journal = {Journal of Forecasting},
volume = {39},
number = {7},
pages = {1081-1097},
year = {2020}
}

@Article{chapman2023macroeconomic,
journal={Forecasting},
author={James T. E. Chapman and Ajit Desai},
title={{Macroeconomic Predictions Using Payments Data and Machine Learning}},
year={2023},
month={November},
pages={1-32},
volume={5},
number={4}
}

\newpage

\section{Appendix}
\subsection{Proofs of the Theoretical Results}\label{prop_proof}
This section discusses the proofs of the theoretical results (Lemma~\ref{lemma_irreducible} and Theorem~\ref{theorem_ergodicity}) on geometric ergodicity and asymptotic stationarity of the NARFIMA model.
\subsubsection{Proof of Lemma~\ref{lemma_irreducible}}\label{proof_lemma_irreducible}
\begin{proof}
Fix $y \in \mathcal{R}$ and let  $O \subset \mathcal{R}$ be a nonempty open interval. Conditionally on $\left(y_{t-1}, e_{t-1}\right)=(y, e)$,
$$
y_t=f(y, e)+\varepsilon_t; \quad f(y, e):=\psi_1 y+\psi_2 e+g(y, e).
$$
By Assumption~\eqref{A4}, $\varepsilon_t$ has a strictly positive continuous density on $\mathcal{R}$. Therefore,
\begin{align*}
    \mathbb{P}\left(y_t \in O \mid y_{t-1} = y, e_{t-1}=e\right) =& \mathbb{P}\left(f(y_{t-1}, e_{t-1}) + \varepsilon_t \in O \mid y_{t-1}=y, e_{t-1}=e\right)\\
    =&\int_O f_{\varepsilon}\left(y^{\prime}-f(y, e)\right) d y^{\prime}>0, \quad \forall y^{\prime} \in \mathcal{R}.
\end{align*}
To show that the conditional one-step transition kernel in $y$ has a strictly positive density everywhere, we need to remove the conditioning on $e_{t-1}$. By the law of iterated expectations:
\begin{align*}
    \mathbb{P}\left(y_t \in O \mid y_{t-1} = y\right) =& \mathbb{E}(\mathbb{1}_{\{y \in O\}} | y_{t-1} = y) \\
     =& \mathbb{E}(\mathbb{E}(\mathbb{1}_{\{y \in O\}} | y_{t-1} = y, e_{t-1} = e) | y_{t-1} = y) \\ 
    =& \mathbb{E}(\mathbb{P}(y \in O\ | y_{t-1} = y, e_{t-1} = e) | y_{t-1} = y) \\
    =& \mathbb{E}\left[\int_O f_{\varepsilon}\left(y^{\prime}-f\left(y, e\right)\right) d y^{\prime} \mid y_{t-1}=y\right].
\end{align*}
By Assumption~\eqref{A4} the inner integral is strictly positive for every realization of $e_{t-1}$ since $f_{\varepsilon}>0$ everywhere. Hence, the conditional expectation is also positive. Therefore, from any $y$, every nonempty open set $O$ is reached in one step with positive probability. This establishes irreducibility. If we take $O$ to be any neighborhood of $y$, then $\mathbb{P}\left(y_t \in O \mid y_{t-1}=y\right)>0$, which rules out cyclic behavior and yields aperiodicity.
\end{proof}

\subsubsection{Proof of Theorem~\ref{theorem_ergodicity}}\label{proof_theorem_ergodicity}
\begin{proof}
In order to verify the geometric drift condition, we choose the general Lyapunov function $V(y)=1+y^2$. Then,
\begin{align*}
    \mathbb{E}\left[V\left(y_t\right) \mid y_{t-1} = y, e_{t-1} = e\right] =& 1+\mathbb{E}\left[\left(\psi_1 y+\psi_2 e+g(y, e)+\varepsilon_t\right)^2\right]\\
    =& 1+\left(\psi_1 y+\psi_2 e+g(y, e)\right)^2+\mathbb{E}(\varepsilon_t^2).
\end{align*}
For any $\eta>0,$ the bound $ (a+b+c)^2 \leq(1+\eta) a^2+\left(2+\frac{2}{\eta}\right)\left(b^2+c^2\right)$ holds for all $a,b,c\in\mathcal{R}.$ By Assumption~\eqref{A2}, $\exists A>0; \; |g(y, e)| \leq A$. Therefore, 
$$
\left(\psi_1 y+\psi_2 e+g\right)^2 \leq(1+\eta) \psi_1^2 y^2+\left(2+\frac{2}{\eta}\right)\left(\psi_2^2 e^2+A^2\right).
$$
Now, by Assumption~\eqref{A5}, $|\psi_1| < 1$. We may therefore choose $\eta > 0$ such that $(1+\eta)\psi_1^2 = 1-\delta$ for some $\delta \in (0,1)$; if $\psi_1 = 0$, this holds trivially for any $\delta \in (0,1)$. This yields, 
$$
\mathbb{E}\left[V\left(y_t\right) \mid y_{t-1} = y, e_{t-1} = e\right] \leq 1 +(1-\delta) y^2+\left[\left(2+\frac{2}{\eta}\right) A^2+\mathbb{E}(\varepsilon_t^2)\right]+\left(2+\frac{2}{\eta}\right) \psi_2^2 e^2.
$$
Notice that $1+(1-\delta) y^2 = (1-\delta) V(y)+\delta$. Hence, 
$$
\mathbb{E}\left[V\left(y_t\right) \mid y_{t-1} = y, e_{t-1} = e\right] \leq (1-\delta) V(y) + \underbrace{\left[\delta + \left(2+\frac{2}{\eta}\right) A^2+\mathbb{E}(\varepsilon_t^2)\right]}_{=: b_0}+\underbrace{\left(2+\frac{2}{\eta}\right) \psi_2^2}_{=: b_1} e^2.
$$
Finally, by Assumptions~\eqref{A1} and~\eqref{A4} (finite variance of $\varepsilon_t$), we have that
$$
\mathbb{E}\left[V\left(y_t\right) \mid y_{t-1}=y\right] \leq(1-\delta) V(y)+B, \quad B:=b_0+b_1e^2<\infty.
$$
This is a Foster-Lyapunov drift \citep{xu2018note}. This geometric drift condition, combined with irreducibility, results in geometric ergodicity. Thus, $\{y_t\}$ admits a unique invariant distribution and converges to it at a geometric rate. If the initial sample $y_0$ is drawn from the invariant law, the process is asymptotically stationary. 
\end{proof}

\subsection{Macroeconomic Datasets and Global Characteristics}
This section provides a brief overview of the uncertainty measures used in our analysis and describes the global features of all the macroeconomic variables.

\subsubsection{Overview Uncertainty Measures}\label{uncertainty_desc}
In this study, we consider four newspaper-based uncertainty measures, namely GEPU, US EMV, US MPU, and GPR index. The GEPU index, developed by \citealp{davis2016index}, reflects global economic uncertainty that impacts investment, trade, and financial markets. It is computed as a GDP-weighted average of national EPU indices from 21 countries, which collectively account for two-thirds of global output. The national EPU index measures uncertainty related to economic policy decisions, their timing, impact, and the economic consequences of non-economic events such as military actions \citep{baker2016measuring}. It is constructed by analyzing the frequency of terms related to economy, uncertainty, and policy in country-specific newspaper articles. The GEPU index, derived from these national EPU indices using PPP-adjusted GDP, effectively captures both global economic crises and geopolitical disruptions. The US EMV index, introduced by \citealp{baker2019policy}, quantifies the impact of wars, policy risks, commodity markets, and macroeconomic outlook on US equity returns and stock market volatility. It tracks the relative frequency of terms related to the economy, stock markets, and volatility across eleven major US newspapers. By leveraging electronic text search techniques, it helps quantify fluctuations in the CBOE Volatility Index and the realized volatility of S\&P 500 returns. This index reflects how policy uncertainty, market sentiment, and global economic shocks influence market fluctuations and investor behavior in the US financial markets. The US MPU index, proposed by \citealp{husted2020monetary}, specifically measures uncertainty surrounding Federal Reserve monetary policy actions and their implications. It bridges the gap between conventional and unconventional policy regimes by tracking the frequency of articles containing monetary policy, uncertainty, and Federal Reserve-related terms in the Washington Post, Wall Street Journal, and New York Times. Unlike GEPU and US EMV, which capture broader uncertainty sources such as fiscal policy, healthcare, and national security, the US MPU index focuses solely on the US monetary policy landscape, reflecting the time-varying influence of Federal Reserve communications. Alongside economic uncertainties, we also analyze the country-specific GPR index, developed by \citealp{caldara2022measuring}, which quantifies risks arising from adverse geopolitical events such as terrorism, political instability, violence, territorial conflicts, and wars. This country-specific index is constructed using electronic text searches of relevant geopolitical terms across ten prominent newspapers from the US, UK, and Canada. It captures global perceptions of geopolitical risks associated with a given country or its major cities. 

\subsubsection{Statistical Properties of the Macroeconomic Time Series}\label{stat_prop}
We compute the five-point summary, mean, standard deviation (SD), coefficient of variation (CoV), and entropy for all the macroeconomic variables and uncertainty measures used in our analysis. The values of these summary statistics, reported in Table \ref{tab:summary_statistics}, offer valuable economic insights. 
For instance, the CoV reveals substantial relative variability across most economic indicators, with notable differences between countries and variables. Generally, exchange rates exhibit moderate CoV values, whereas interest rate differentials and inflation differentials show much higher relative variability. 
Alongside the descriptive statistics, we analyze the global behavior of the exchange rate series and exogenous variables, and summarize the results in Table~\ref{Table_Global_characteristics}. The analysis includes key features such as skewness, kurtosis, nonlinearity, seasonality, stationarity, long-range dependence, and detected outliers, offering a comprehensive overview of the structural patterns across all time series. 

\begin{table}[h!]
\centering
\caption{Summary statistics of the datasets utilized in this analysis.}
\begin{adjustbox}{width=\textwidth}
\fontsize{5}{6}\selectfont
\begin{tabular}{ccccccccccc}
\hline
Country & Series & Min Value & Q1 & Median & Mean & SD & Q3 & Max Value & CoV & Entropy \\
\hline
\multirow{6}{*}{Brazil}
& Exchange Rate & 1.043 & 1.800 & 2.215 & 78.611 & 2.382 & 3.004 & 4.120 & 33.002 & 5.608 \\
& Short-term Interest Rate & 5.480 & 10.578 & 13.750 & 15.534 & 814.479 & 18.643 & 49.750 & 52.432 & 4.728 \\
& Short-term IRD & 3.650 & 8.798 & 11.720 & 13.291 & 721.781 & 15.523 & 44.680 & 54.306 & 5.425 \\
& CPI Inflation & 1.645 & 4.518 & 5.989 & 6.203 & 270.569 & 7.259 & 17.236 & 43.619 & 5.613 \\
& CPI Inflation Differential & -0.290 & 1.925 & 3.487 & 4.050 & 297.201 & 5.403 & 15.178 & 73.383 & 5.613 \\
& GPR & 0.003 & 0.026 & 0.039 & 0.048 & 3.365 & 0.059 & 0.225 & 70.107 & 5.613 \\
\hline

\multirow{6}{*}{Russia} 
& Exchange Rate & 5.629 & 27.624 & 29.899 & 34.673 & 1632.301 & 35.372 & 75.172 & 47.077 & 5.608 \\
& Short-term Interest Rate & 5.250 & 8.250 & 11.000 & 17.316 & 1632.830 & 21.000 & 150.000 & 94.296 & 3.432\\
& Short-term IRD & 4.670 & 6.368 & 9.600 & 15.073 & 1514.957 & 18.475 & 144.510 & 100.508 & 5.199 \\
& CPI Inflation & 2.197 & 6.853 & 10.232 & 15.022 & 1974.837 & 14.858 & 126.512 & 131.463 & 5.613 \\
& CPI Inflation Differential & -0.572 & 4.959 & 7.697 & 12.868 & 1977.207 & 13.297 & 124.367 & 153.653 & 5.613 \\
& GPR & 0.205 & 0.394 & 0.550 & 0.652 & 34.814 & 0.814 & 2.329 & 53.396 & 5.613\\
\hline

\multirow{6}{*}{India} 
& Exchange Rate & 35.747 & 43.924 & 46.787 & 50.983 & 992.537 & 60.935 & 73.561 & 19.468 & 5.613 \\
& Short-term Interest Rate & 5.400 & 6.000 & 6.500 & 7.161 & 142.468 & 8.188 & 12.000 & 19.895 & 2.301 \\
& Short-term IRD & 0.470 & 3.453 & 4.875 & 4.919 & 233.033 & 5.920 & 10.170 & 47.374 & 5.010 \\
& CPI Inflation & 0.000 & 4.167 & 5.932 & 6.675 & 334.077 & 8.824 & 19.672 & 50.049 & 5.506 \\
& CPI Inflation Differential & -2.622 & 1.444 & 3.932 & 4.521 & 381.554 & 6.785 & 18.124 & 84.396 & 5.613 \\
& GPR & 0.044 & 0.133 & 0.181 & 0.213 & 14.577 & 0.249 & 1.126 & 68.436 & 5.613\\
\hline

\multirow{6}{*}{China}
& Exchange Rate & 6.051 & 6.570 & 7.068 & 7.345 & 85.356 & 8.277 & 8.326 & 11.621 & 5.157 \\
& Short-term Interest Rate & 2.700 & 2.900 & 3.240 & 3.396 & 123.672 & 3.250 & 9.000 & 36.417 & 1.164 \\
& Short-term IRD & -5.260 & -2.118 & -0.270 & 187.761 & -0.904 & -0.090 & 3.810 & -207.700 & 4.639 \\
& CPI Inflation & -2.200 & 0.784 & 1.781 & 1.953 & 213.111 & 2.816 & 8.805 & 109.120 & 5.601 \\
& CPI Inflation Differential & -4.477 & -1.823 & -0.019 & -0.200 & 192.720 & 1.038 & 4.778 & -963.602 & 5.613 \\
& GPR & 0.098 & 0.296 & 0.389 & 0.456 & 23.725 & 0.540 & 1.521 & 52.028 & 5.613 \\
\hline

\multirow{6}{*}{Global} 
& GEPU & 51.141 & 76.741 & 104.319 & 117.880 & 5429.789 & 145.306 & 337.760 & 46.062 & 5.613 \\
& US EMV & 9.570 & 15.529 & 18.854 & 21.144 & 827.153 & 24.324 & 69.835 & 39.120 & 5.613 \\
& US MPU & 19.749 & 74.521 & 103.705 & 114.952 & 6343.051 & 134.119 & 407.365 & 55.180 & 5.613 \\
& Oil Price Growth Rate & -59.021 & -12.328 & 7.734 & 10.212 & 3650.524 & 31.315 & 144.605 & 357.474 & 5.613 \\
& US Short-term Interest Rate & 0.070 & 0.160 & 1.650 & 2.243 & 214.821 & 4.725 & 6.540 & 95.774 & 4.638 \\
& US CPI Inflation & -2.097 & 1.551 & 2.098 & 2.154 & 118.125 & 2.908 & 5.600 & 54.840 & 5.608 \\
\hline

\end{tabular}
\end{adjustbox}
\label{tab:summary_statistics}
\end{table}

\begin{table*}[ht]
\fontsize{5}{6}\selectfont
\centering
\caption{Global characteristics of the economic time series under study for BRIC countries.}
\begin{adjustbox}{width=\textwidth}
\begin{tabular}{ccccccccc} \hline
Countries & Series & Skewness & Kurtosis & Nonlinearity & Seasonality & Stationarity &  Hurst Exponent & Outlier(s) Detected\\\hline
\multirow{6}{*}{Brazil} & Exchange Rate & 0.417 & -0.689 & Non-linear & Non-seasonal & Non-stationary & 0.784 & 1 \\ 
 & Short-term Interest Rate & 1.820 & 3.970 & Non-linear & Non-seasonal & Non-stationary & 0.816 & 5 \\ 
 
 & Short-term IRD & 1.864 & 4.343 & Nonlinear & Non-seasonal & Non-stationary & 0.796 & 2 \\ 
 & CPI Inflation & 1.540 & 3.778 & Nonlinear & Non-seasonal & Stationary & 0.735 & 4 \\ 
 & CPI Inflation Differential & 1.233 & 1.842 & Linear & Non-seasonal & Stationary & 0.690 & 1 \\ 
 & GPR & 2.313 & 7.506 & Nonlinear & Non-seasonal & Stationary & 0.641 & 5\\ 
\hline
\multirow{6}{*}{Russia} 
 & Exchange Rate & 0.730 & -0.0378 & Nonlinear & Non-seasonal & Non-stationary & 0.824 & 0 \\ 
 & Short-term Interest Rate & 3.230 & 16.560 & Nonlinear & Non-seasonal & Non-stationary & 0.804 & 2\\ 
 & Short-term IRD & 3.552 & 20.162 & Nonlinear & Non-seasonal & Non-stationary & 0.793 & 2 \\ 
 & CPI Inflation & 4.042 & 16.931 & Nonlinear & Non-seasonal & Non-stationary & 0.745 & 8\\ 
 & CPI Inflation Differential & 4.048 & 16.957 & Nonlinear & Non-seasonal & Non-stationary & 0.743 & 8 \\ 
 & GPR & 2.056 & 5.412 & Nonlinear & Non-seasonal & Stationary & 0.622 & 4  \\ 
\hline

\multirow{6}{*}{India} 
 & Exchange Rate & 0.683 & -0.853 & Linear & Non-seasonal & Non-stationary & 0.848 & 1 \\ 
 & Short-term Interest Rate & 1.098 & 0.432 & Nonlinear & Non-seasonal & Non-stationary & 0.817 & 1\\ 
 & Short-term IRD & 0.100 & -0.604 & Nonlinear & Non-seasonal & Non-stationary & 0.826 & 1 \\ 
 & CPI Inflation & 0.921 & 0.938 & Nonlinear & Non-seasonal & Stationary & 0.768 & 1 \\ 
 & CPI Inflation Differential & 0.792 & 0.311 & Nonlinear & Non-seasonal & Stationary & 0.786 & 1 \\ 
 & GPR & 3.246 & 13.617 & Nonlinear & Non-seasonal & Non-stationary & 0.704 & 6\\ 
\hline

\multirow{6}{*}{China} 
& Exchange Rate & -0.0384 & -1.727 & Nonlinear & Non-seasonal & Non-stationary & 0.868 & 1\\ 
 & Short-term Interest Rate & 3.521 & 12.218 & Nonlinear & Non-seasonal & Non-stationary & 0.696 & 4\\ 
 & Short-term IRD & -0.469 & 0.193 & Nonlinear & Non-seasonal & Stationary & 0.730 & 1 \\ 
 & CPI Inflation & 0.617 & 0.569 & Linear & Non-seasonal & Non-stationary & 0.763 & 1 \\ 
 & CPI Inflation Differential & 0.0189 & -0.536 & Linear & Non-seasonal & Non-stationary & 0.807 & 1 \\ 
 & GPR & 1.568 & 2.757 & Linear & Non-seasonal & Non-stationary & 0.730 & 1 \\ 
\hline

\multirow{6}{*}{Global} 
 & GEPU & 1.374 & 1.884 & Nonlinear & Seasonal & Non-stationary & 0.806 & 1 \\ 
 & US EMV & 2.196 & 7.213 & Nonlinear & Seasonal & Non-stationary & 0.698 & 4 \\ 
 & US MPU & 1.874 & 4.761 & Nonlinear & Non-seasonal & Non-stationary & 0.706 & 3 \\ 
 & Oil Price Growth Rate & 0.600 & 0.556 & Nonlinear & Non-seasonal & Stationary & 0.723 & 1  \\ 
 & US Short-term Interest Rate & 0.573 & -1.235 & Nonlinear & Non-seasonal & Non-stationary & 0.830 & 1 \\ 
 & US CPI Inflation & -0.307 & 1.056 & Nonlinear & Non-seasonal & Non-stationary & 0.762 & 1 \\ 
 \hline
\end{tabular}
\label{Table_Global_characteristics}
\end{adjustbox}
\end{table*}

While the Hurst exponent provide preliminary indications of persistence, a more formal assessment of long memory is obtained by estimating the fractional differencing parameter $d$ under an ARFIMA specification. To this end, we employ the Whittle likelihood estimator, which fits a fractionally integrated process to the data and yields both an estimate of $d$ and a test of whether the estimate differs significantly from zero \citep{whittle1951hypothesis, beran2017statistics}. An estimate of $d$ significantly greater than zero indicates that the series exhibits long-range dependence rather than short memory. As reported in Table~\ref{Table_Whittle_Estimator}, the estimated values of $d$ are uniformly significant across all series, with $z$-statistics exceeding the critical threshold of 2.576 corresponding to the 1\% level and corresponding $p$-values below 0.0001. This provides decisive evidence against short memory for all macroeconomic indicators and confirms that a fractional integration component is essential for adequately capturing the persistence structure of the data.

\begin{table*}[ht]
\fontsize{5}{6}\selectfont
\centering
\caption{Estimates of the fractional differencing parameter $d$ obtained via Whittle likelihood for BRIC exchange rates and macroeconomic variables. Standard errors, $z$-statistics, and associated $p$-values are reported. The $z$-statistics are evaluated against the $1\%$ critical value of $2.576$. Rejection of the null hypothesis of short memory indicates the presence of long-range dependence.}
\begin{adjustbox}{width=\textwidth}
\begin{tabular}{cccccc} \hline
Countries & Series & Fractional Differencing Parameter ($\hat{d}$) & Standard Error & $z$-statistic & $p$-value \\\hline
\multirow{6}{*}{Brazil} & Exchange Rate & 0.490 & 0.047 & 20.906 & $< 0.0001$ \\ 
  & Short-term Interest Rate & 0.490 & 0.047 & 20.907 & $< 0.0001$ \\ 
  & Short-term IRD & 0.490 & 0.047 & 20.906 & $< 0.0001$ \\ 
  & CPI Inflation & 0.490 & 0.047 & 20.906 & $< 0.0001$ \\ 
  & CPI Inflation Differential & 0.490 & 0.047 & 20.906 & $< 0.0001$ \\ 
& GPR & 0.307 & 0.047 & 17.039 & $< 0.0001$\\ 
 \hline
\multirow{6}{*}{Russia} 
 & Exchange Rate & 0.490 & 0.047 & 20.907 & $< 0.0001$ \\ 
 & Short-term Interest Rate & 0.490 & 0.047 & 20.906 & $< 0.0001$ \\ 
 & Short-term IRD & 0.490 & 0.047 & 20.906 & $< 0.0001$ \\ 
 & CPI Inflation & 0.490 & 0.047 & 20.906 & $< 0.0001$ \\ 
 & CPI Inflation Differential & 0.490 & 0.047 & 20.906 & $< 0.0001$ \\ 
 & GPR & 0.479 & 0.047 & 20.669 & $< 0.0001$ \\
\hline

\multirow{6}{*}{India} 
 & Exchange Rate & 0.490 & 0.047 & 20.907 & $< 0.0001$ \\  
 & Short-term Interest Rate & 0.490 & 0.047 & 20.907 & $< 0.0001$ \\  
 & Short-term IRD & 0.490 & 0.047 & 20.907 & $< 0.0001$ \\ 
 & CPI Inflation & 0.490 & 0.047 & 20.906 & $< 0.0001$ \\ 
 & CPI Inflation Differential & 0.490 & 0.047 & 20.906 & $< 0.0001$ \\
 & GPR & 0.423 & 0.047 & 19.495 & $< 0.0001$ \\ 
\hline
\multirow{6}{*}{China} 
& Exchange Rate & 0.490 & 0.047 & 20.907 & $< 0.0001$ \\ 
 & Short-term Interest Rate & 0.490 & 0.075 & 13.125 & $< 0.0001$\\ 
 & Short-term IRD & 0.490 & 0.047 & 20.906 & $< 0.0001$ \\ 
 & CPI Inflation & 0.490 & 0.047 & 20.906 & $< 0.0001$  \\ 
 & CPI Inflation Differential & 0.490 & 0.047 & 20.907 & $< 0.0001$ \\ 
 & GPR & 0.490 & 0.047 & 20.907 & $< 0.0001$ \\ 
\hline

 \multirow{6}{*}{Global} 
  & GEPU & 0.490 & 0.047 & 20.907 & $< 0.0001$ \\ 
  & US EMV & 0.490 & 0.047 & 20.906 & $< 0.0001$ \\ 
  & US MPU & 0.453 & 0.047 & 20.126 & $< 0.0001$ \\ 
  & Oil Price Growth Rate & 0.490 & 0.047 & 20.907 & $< 0.0001$  \\ 
  & US Short-term Interest Rate & 0.490 & 0.047 & 20.907 & $< 0.0001$ \\ 
  & US CPI Inflation & 0.490 & 0.047 & 20.906 & $< 0.0001$ \\ 
  \hline
\end{tabular}
\label{Table_Whittle_Estimator}
\end{adjustbox}
\end{table*}

\subsection{Empirical Setup and Results}

This section provides an overview of the state-of-the-art forecasting models used for benchmarking,  the results of the MCS procedure used to statistically validate the superior predictive performance of the NARFIMA framework, and the empirical evidence regarding the nonlinearity and weak stationarity of ARFIMAx residuals.

\subsubsection{Overview of the Baseline Models}\label{Appendix_Baseline}

We evaluate the performance of the proposed NARFIMA model against several state-of-the-art models to demonstrate its forecasting effectiveness. The evaluation incorporates the following methods:

\begin{itemize}
\item \textit{Naïve} forecasting approach predicts future values by using the last observed value of a time series without any adjustments. Despite its simplicity, this stochastic model is often effective for economic time series and serves as a benchmark for evaluating the performance of more advanced forecasting techniques \citep{hyndman2018forecasting}.

\item \textit{Autoregressive} (AR) model assumes that future values of a time series are primarily influenced by its past observations. It closely resembles multiple linear regression, except that the target series is predicted using its own lagged values. This method is most suitable for modeling stationary time series datasets \citep{hyndman2018forecasting}.

\item \textit{Autoregressive Integrated Moving Average with exogenous variables} (ARIMAx) model is a widely used time series forecasting technique that captures linear dependencies by combining three key components: autoregressive (AR), differencing (I), and moving average (MA) \citep{box1970distribution}. In this framework, differencing of order $d_0$ ensures the stationarity of the input series, after which the AR component models the $p_0$ lagged values of the series, while the MA component incorporates $q_0$ lagged residuals. The parameters of ARIMAx are estimated by optimizing the Akaike information criterion (AIC).

\item \textit{Autoregressive Fractionally Integrated Moving Average with exogenous variables} (ARFIMAx) extends ARIMAx by allowing the differencing parameter to take fractional values in the range $(0, 0.5)$. This generalization enables the ARFIMAx$(\tilde{p},d,\tilde{q})$ model to effectively capture long-range dependencies in time series data, where correlations decay gradually \citep{granger1980introduction}. As shown in Table \ref{Table_Global_characteristics}, the exchange rate series exhibits long memory dynamics, justifying the use of the ARFIMAx model for accurate forecasting. 

\item \textit{Exponential Smoothing} (ETS) model decomposes a time series into three components: noise, trend, and seasonal patterns. Each component can be modeled using either an additive or multiplicative approach. ETS estimates its parameters by minimizing the sum of squared errors and is known for its flexibility in producing accurate forecasts across a wide range of time series data \citep{hyndman2018forecasting}. 

\item \textit{Self-exciting Threshold Autoregressive} (SETAR) model extends the traditional autoregressive approach by incorporating threshold effects. It partitions the time series into distinct regimes based on a threshold value, allowing different autoregressive processes in each regime. This structure enables SETAR to capture nonlinearities and model complex time series with regime-switching behavior \citep{firat2017setar}.

\item \textit{Trigonometric Box-Cox ARIMA Trend Seasonal} (TBATS) model is designed for time series with multiple seasonal patterns and nonlinear trends. TBATS enhances traditional models by incorporating Fourier terms, applying a Box-Cox transformation to stabilize variance, and using an ARIMA model to account for residual autocorrelation. This approach enables TBATS to handle complex seasonal structures that are often challenging for conventional models \citep{hyndman2018forecasting}.

\item \textit{Generalized Autoregressive Conditional Heteroscedasticity} (GARCH) model captures time-varying volatility in temporal data. GARCH$(p_g,q_g)$ incorporates \(p_g\) lagged values of the conditional variance and \(q_g\) lagged values of the squared residuals to model the volatility clustering commonly observed in financial time series. The parameter estimation is typically performed using maximum likelihood estimation. Due to its ability to forecast volatility, GARCH is widely used in econometric and financial modeling \citep{bollerslev1986generalized}.  

\item \textit{Bayesian Structural Time Series with exogenous variables} (BSTSx) model is a flexible forecasting approach that decomposes a time series into components including trend, seasonality, and the effects of external covariates. It relies on a state-space representation, where observations depend on hidden states that evolve over time. The Kalman filter is used to estimate these states, while the spike-and-slab prior helps identify relevant predictors. Markov Chain Monte Carlo methods are employed to generate samples from the model’s posterior distribution, and forecasts are derived by summarizing these samples \citep{scott2014predicting}.

\item \textit{Autoregressive Neural Network with exogenous variables} (ARNNx) extends conventional feed-forward neural networks to capture autoregressive time series patterns \citep{faraway1998time}. It processes $b$ lagged values in the input layer, passing them through $c$ neurons in the hidden layer. To ensure model stability and restrict overfitting, $c = \lfloor\frac{b + 1}{2}\rfloor$ is specified. The model is initialized with random weights and trained using the gradient descent backpropagation approach \citep{rumelhart1986learning}. This training process helps prevent overfitting and ensures a stable learning structure \citep{hyndman2018forecasting}. 

\item \textit{Deep learning-based Autoregressive} (DeepAR) model is an advanced neural network approach designed for time series forecasting. It leverages a recurrent neural network (RNN) architecture to capture temporal dependencies and predict future values in sequential data \citep{salinas2020deepar}.

\item \textit{Neural Basis Expansion Analysis for Time Series with exogenous variables} (NBeatsx) model is a specialized neural network architecture for time series forecasting. It consists of multiple stacked blocks, where each block has two key layers \citep{Oreshkin2020N-BEATS:}. The first layer captures patterns in the data, while the second layer refines the forecast by modeling residual errors. This iterative process enhances forecasting accuracy by progressively improving predictions.

\item \textit{Neural Hierarchical Interpolation for Time Series with exogenous variables} (NHiTSx) model builds on the NBeatsx framework by introducing a hierarchical approach to time series forecasting. Similar to NBeatsx, this framework utilizes a series of stacked blocks and enhances them with a hierarchical structure to better capture complex dependencies within the data \citep{challu2023nhits}.

\item \textit{Decomposition-based Linear model with exogenous variables} (DLinearx) employs a decomposition method to separate the input time series into trend and seasonal components using a moving average. Each component is then processed through a dedicated linear layer, and their sum generates the final forecast. By explicitly modeling trends, this method improves forecasting performance, making it more effective for capturing temporal patterns in the data \citep{zeng2023transformers}.

\item \textit{Normalization-based Linear model with exogenous variables} (NLinearx) applies a normalization technique to improve forecasting accuracy. It normalizes the input time series by subtracting the last observed value before passing the data through a linear layer. After processing, this value is added back to generate the final forecast. This approach helps mitigate distribution shifts between training and testing data, enhancing model robustness \citep{zeng2023transformers}.

\item \textit{Time Series Mixer with exogenous variables} (TSMixerx) is a neural network architecture designed for time series forecasting. It employs a sequence-mixing approach, where each layer mixes information across different time steps to capture dependencies within the data. The initial layers aggregate sequential information, while subsequent layers refine the learned patterns to improve forecasting accuracy \citep{chen2303tsmixer}.
\end{itemize}

\subsubsection{Model Confidence Set Analysis}\label{Appendix_MCS}
The Model Confidence Set (MCS) procedure of \citealp{hansen2011model} identifies a subset of models $\widehat{\mathcal{A}}^*_{1-\alpha}$ that contains the best forecasting model with confidence level $1-\alpha$. Starting from the full set of competitors $\mathcal{A}_0$, the procedure sequentially eliminates models whose forecast accuracy is significantly inferior relative to the remaining alternatives. At each step, pairwise comparisons are based on loss differentials $d_{ij,t} = \ell_{i,t} - \ell_{j,t}$, where $\ell_{i,t}$ denotes the forecast loss of model $i$ at time $t$. The procedure tests whether the expected loss differential is zero for all models under consideration using a bootstrap approximation to assess statistical significance. Models that consistently underperform are removed until only a set of statistically indistinguishable models remains. 
For our analysis, we implement the MCS procedure via the \texttt{MCS} package in \textbf{R}. Forecast accuracy is evaluated using the squared loss. 
Given the substantial number of competing models and the observed variability in loss differentials across countries, we adopt the range statistic ($T_R$) and approximate the bootstrap distribution using $B = 5000$ replications. The procedure is conducted at three significance levels $\alpha \in \{0.05, 0.10, 0.15\}$, allowing for a robust examination of model inclusion under decreasingly stringent confidence thresholds. Table~\ref{tab:MCS_transposed_final} summarizes the MCS results across all BRIC economies. A checkmark (\ding{51}) indicates that the model remains within the optimal set $\widehat{\mathcal{A}}^*_{1-\alpha}$, while a cross (\ding{55}) denotes elimination. Superscripts $^*$, $^\dagger$, and $^\ddagger$ correspond to exclusion at the 0.05, 0.10, and 0.15 levels, respectively. The results confirm that the proposed NARFIMA framework consistently outperforms the entire suite of baseline forecasters, as it is the only model to remain within the optimal set $\hat{\mathcal{A}}^*_{1-\alpha}$ across all countries, while all other competitors, including advanced deep learning architectures, are eliminated at one or more significance levels.

\begin{table*}[!htbp]
\caption{Model Confidence Set (MCS) results for all countries evaluated across the baseline forecasters (\ding{51}\ indicates the model belongs to the MCS set, while \ding{55}\ denotes elimination). Superscripts indicate the significance level at which a model is excluded: $^*$ $\alpha = 0.05$, $^\dagger$ $\alpha = 0.10$, and $^\ddagger$ $\alpha = 0.15$.}
\label{tab:MCS_transposed_final}
\centering
\begin{adjustbox}{width=1\textwidth}
    \small 
    \setlength{\tabcolsep}{2pt}
    \begin{tabular}{lccccccccccccccccc}
    \hline
    Country & Naïve & AR & ARIMAx & ARNNx & ARFIMAx & ETS & SETAR & TBATS & GARCH & BSTSx & DeepAR & NBeatsx & NHiTSx & DLinearx & NLinearx & TSMixerx & {\color{blue}NARFIMA} \\
    \hline
    Brazil & \ding{55}$^*$ & \ding{55}$^*$ & \ding{55}$^*$ & \ding{55}$^*$ & \ding{55}$^*$ & \ding{55}$^*$ & \ding{55}$^*$ & \ding{55}$^*$ & \ding{55}$^*$ & \ding{55}$^*$ & \ding{55}$^*$ & \ding{55}$^*$ & \ding{55}$^*$ & \ding{55}$^*$ & \ding{55}$^*$ & \ding{55}$^*$ & \ding{51} \\ \hline
    Russia & \ding{55}$^*$ & \ding{55}$^*$ & \ding{55}$^\dagger$ & \ding{55}$^\dagger$ & \ding{55}$^*$ & \ding{55}$^*$ & \ding{55}$^*$ & \ding{55}$^*$ & \ding{55}$^*$ & \ding{55}$^\ddagger$ & \ding{55}$^*$ & \ding{55}$^\dagger$ & \ding{55}$^*$ & \ding{55}$^\dagger$ & \ding{55}$^*$ & \ding{55}$^*$ & \ding{51} \\ \hline
    India  & \ding{55}$^*$ & \ding{55}$^*$ & \ding{55}$^*$ & \ding{55}$^*$ & \ding{55}$^*$ & \ding{55}$^*$ & \ding{55}$^*$ & \ding{55}$^*$ & \ding{55}$^*$ & \ding{51} & \ding{55}$^*$ & \ding{55}$^*$ & \ding{55}$^*$ & \ding{55}$^\dagger$ & \ding{55}$^\ddagger$ & \ding{55}$^*$ & \ding{51} \\ \hline
    China  & \ding{55}$^*$ & \ding{55}$^*$ & \ding{55}$^*$ & \ding{55}$^*$ & \ding{55}$^*$ & \ding{55}$^*$ & \ding{55}$^*$ & \ding{55}$^*$ & \ding{55}$^*$ & \ding{55}$^*$ & \ding{55}$^*$ & \ding{55}$^*$ & \ding{55}$^*$ & \ding{55}$^*$ & \ding{55}$^*$ & \ding{55}$^*$ & \ding{51} \\ \hline
    \end{tabular}
\end{adjustbox}
\end{table*}

\subsubsection{Empirical Evidence of Nonlinearity and Stationarity of ARFIMAx Residuals}\label{Appendix_Nonlinearity}
To justify the two-stage structure of the NARFIMA model, we first verify that the ARFIMAx component successfully captures the linear dependence in the data, leaving behind residuals that exhibit the nonlinear structure subsequently modeled by the feed-forward neural network component of NARFIMA. To this end, we apply the Teräsvirta and BDS tests to the residuals of the ARFIMAx model. The null hypothesis of linearity is rejected when the $p$-value falls below 0.05. As reported in Table~\ref{Residuals_Nonlinearity}, both tests reject the null hypothesis of linearity across all countries and forecast horizons, confirming that the ARFIMAx residuals contain substantial nonlinear structure.

In addition, the theoretical results on asymptotic stationarity of the NARFIMA process 
(Section~\ref{AS_GE}) require the innovation process to be weakly stationary. We therefore examine whether the ARFIMAx residuals satisfy this requirement using the Zivot-Andrews test, which is an extension of the Augmented Dickey-Fuller (ADF) test. The Zivot-Andrews test evaluates the null hypothesis of a unit root against the alternative of stationarity while allowing for a single endogenous structural break in the intercept \citep{zivot1992test}. 
The results, reported in Table~\ref{Residuals_Nonlinearity}, show that across all forecast horizons and all BRIC economies, the Zivot-Andrews test strongly rejects the unit root hypothesis. Test statistics fall substantially below the 1\% critical value of $-5.34$, with corresponding $p$-values below $0.0001$ in every case. These findings confirm that the ARFIMAx residuals are consistent with weak stationarity across all countries and horizons. Accordingly, the residual process satisfies the weak stationarity condition underlying the NARFIMA framework.

\begin{table}[ht]
\caption{Statistical properties of ARFIMAx residuals through Teräsvirta, BDS, and Zivot-Andrews tests across all forecast horizons for BRIC nations. For the Zivot-Andrews test, statistics falling below the $1\%$ critical value of $-5.34$ reject the unit root null in favor of stationarity.}
\centering
\begin{adjustbox}{width=\textwidth}
\begin{tabular}{cccccccc}
\hline
Country & Horizon & Terasvirta Test $p$-value & BDS Test $p$-value & Conclusion & Zivot-Andrews Test Statistic & Zivot-Andrews Test $p$-value & Conclusion \\ \hline
Brazil  & 1  & \( 1.53 \times 10^{-3} \)  & \( 7.59 \times 10^{-17} \) & Nonlinear Residuals & $-24.617$ & $< 0.0001$ & Stationary Residuals \\ 
        & 3  & \( 1.52 \times 10^{-3} \) & \( 1.58 \times 10^{-17} \) & Nonlinear Residuals & $-24.436$ & $< 0.0001$ & Stationary Residuals \\ 
        & 6  & \( 1.68 \times 10^{-3} \)  & \( 2.52 \times 10^{-16} \) & Nonlinear Residuals & $-24.318$ & $< 0.0001$ & Stationary Residuals \\ 
        & 12 & \( 1.64 \times 10^{-3} \)  & \( 2.07 \times 10^{-16} \) & Nonlinear Residuals & $-24.103$ & $< 0.0001$ & Stationary Residuals\\ 
        & 24 & \( 1.43 \times 10^{-3} \)  & \( 3.53 \times 10^{-21} \) & Nonlinear Residuals & $-23.732$ & $< 0.0001$ & Stationary Residuals\\ 
        & 48 & \( 1.93 \times 10^{-3} \)  & \( 2.72 \times 10^{-15} \) & Nonlinear Residuals & $-22.168$ & $< 0.0001$ & Stationary Residuals \\ \hline

Russia  & 1  & \( 2.56 \times 10^{-4} \)  & \( 3.58 \times 10^{-25} \) & Nonlinear Residuals & $-24.477$ & $< 0.0001$ & Stationary Residuals \\ 
        & 3  & \( 4.66 \times 10^{-4} \)  & \( 3.39 \times 10^{-23} \) & Nonlinear Residuals & $-24.338$ & $< 0.0001$ & Stationary Residuals\\ 
        & 6  & \( 4.55 \times 10^{-4} \)   & \( 1.83 \times 10^{-24} \) & Nonlinear Residuals & $-24.190$ & $< 0.0001$ & Stationary Residuals\\ 
        & 12 & \( 4.11 \times 10^{-2} \)  & \( 7.14 \times 10^{-42} \) & Nonlinear Residuals & $-23.111$ & $< 0.0001$ & Stationary Residuals\\
        & 24 & \( 9.14 \times 10^{-9} \)  & \( 4.90 \times 10^{-35} \) & Nonlinear Residuals & $-24.951$ & $< 0.0001$ & Stationary Residuals\\ 
        & 48 & \( 3.49 \times 10^{-4} \)  & \( 1.27 \times 10^{-30} \) & Nonlinear Residuals & $-22.370$ & $< 0.0001$ & Stationary Residuals\\ \hline
        
India   & 1  & \( 5.93 \times 10^{-6} \) & \( 1.04 \times 10^{-10} \) & Nonlinear Residuals & $-29.382$ & $< 0.0001$ & Stationary Residuals\\ 
        & 3  & \( 8.16 \times 10^{-6} \) & \( 2.82 \times 10^{-11} \) & Nonlinear Residuals & $-29.188$ & $< 0.0001$ & Stationary Residuals\\ 
        & 6  & \( 1.34 \times 10^{-5} \)   & \( 1.51 \times 10^{-11} \) & Nonlinear Residuals & $-28.890$ & $< 0.0001$ & Stationary Residuals\\ 
        & 12 & \( 1.88 \times 10^{-5} \) & \( 4.16 \times 10^{-10} \) & Nonlinear Residuals & $-28.388$ & $< 0.0001$ & Stationary Residuals\\ 
        & 24 & \( 5.38 \times 10^{-5} \) & \( 3.54 \times 10^{-10} \) & Nonlinear Residuals & $-27.732$ & $< 0.0001$ & Stationary Residuals\\ 
        & 48 & \( 5.36 \times 10^{-4} \)  & \( 1.26 \times 10^{-8} \)  & Nonlinear Residuals & $-26.084$ & $< 0.0001$ & Stationary Residuals\\ \hline

 China   & 1  & \( 2.8 \times 10^{-9} \)  & \( 3.19 \times 10^{-50} \) & Nonlinear Residuals & $-28.068$  & \(< 0.0001 \) & Stationary Residuals\\ 
         & 3  & \( 6.47 \times 10^{-9} \)  & \( 1.37 \times 10^{-44} \) & Nonlinear Residuals & $-28.300$  & \(< 0.0001 \) & Stationary Residualss\\ 
         & 6  & \( 3.35 \times 10^{-9} \)  & \( 2.83 \times 10^{-44} \) & Nonlinear Residuals & $-28.543$  & \(< 0.0001 \) & Stationary Residuals\\ 
         & 12 & \( 4.75 \times 10^{-10} \)  &  \(4.68 \times 10^{-63} \) & Nonlinear Residuals & $-28.107$  & \(< 0.0001 \) & Stationary Residuals\\ 
         & 24 & \( 5.75 \times 10^{-10} \)  & \(1.09 \times 10^{-50}\) & Nonlinear Residuals & $-29.890$  & \(< 0.0001 \) & Stationary Residuals\\ 
         & 48 & \( 1.14 \times 10^{-8} \)  & \( 0.00 \) & Nonlinear Residuals & $-19.453$  & \(< 0.0001 \) & Stationary Residuals\\ \hline        
\end{tabular}
\label{Residuals_Nonlinearity}
\end{adjustbox}
\end{table}

\end{document}